\documentclass[twocolumn]{aastex}
\usepackage{comment}
\usepackage{graphicx,floatrow,amsmath,gensymb,rotating}
\usepackage{subfigure}
\usepackage{natbib}
\usepackage{xcolor}
\usepackage{makecell}
\usepackage{multirow}
\usepackage{threeparttable,booktabs}

\begin{document}

\title{The Active Galactic Nuclei in the Hobby-Eberly Telescope Dark Energy Experiment Survey (HETDEX) II. Luminosity Function}

\author[0000-0001-5561-2010]{Chenxu Liu}
\affiliation{Department of Astronomy, The University of Texas at Austin, 2515 Speedway Boulevard, Austin, TX 78712, USA}
\correspondingauthor{Chenxu Liu}
\email{lorenaustc@gmail.com}

\author[0000-0002-8433-8185]{Karl Gebhardt}
\affiliation{Department of Astronomy, The University of Texas at Austin, 2515 Speedway Boulevard, Austin, TX 78712, USA}

\author[0000-0002-2307-0146]{Erin Mentuch Cooper}
\affiliation{Department of Astronomy, The University of Texas at Austin, 2515 Speedway Boulevard, Austin, TX 78712, USA}
\affiliation{McDonald Observatory, The University of Texas at Austin, Austin, TX 78712}

\author{Yechi Zhang}
\affiliation{Institute for Cosmic Ray Research, The University of Tokyo, 5-1-5 Kashiwanoha, Kashiwa, Chiba 277-8582, Japan}
\affiliation{Department of Astronomy, Graduate School of Science, the University of Tokyo, 7-3-1 Hongo, Bunkyo, Tokyo 113-0033, Japan}

\author{Donald P. Schneider}
\affiliation{Department of Astronomy \& Astrophysics, The Pennsylvania State University, University Park, PA 16802, USA}
\affiliation{Institute for Gravitation and the Cosmos, The Pennsylvania State University, University Park, PA 16802, USA}

\author[0000-0002-1328-0211]{Robin Ciardullo}
\affil{Department of Astronomy \& Astrophysics, The Pennsylvania State University, University Park, PA 16802, USA}
\affil{Institute for Gravitation and the Cosmos, The Pennsylvania State University, University Park, PA 16802, USA}

\author[0000-0002-8925-9769]{Dustin Davis}
\affiliation{Department of Astronomy, The University of Texas at Austin, 2515 Speedway Boulevard, Austin, TX 78712, USA}

\author[0000-0003-2575-0652]{Daniel J. Farrow}  
\affiliation{University Observatory, Fakult\"at f\"ur Physik, Ludwig-Maximilians University Munich, Scheiner Strasse 1, 81679 Munich, Germany}
\affiliation{Max-Planck Institut f\"ur extraterrestrische Physik, Giessenbachstrasse 1, 85748 Garching, Germany}

\author[0000-0001-8519-1130]{Steven L. Finkelstein}
\affiliation{Department of Astronomy, The University of Texas at Austin, 2515 Speedway, Austin, TX 78712, USA}

\author[0000-0001-6842-2371]{Caryl Gronwall}
\affiliation{Department of Astronomy \& Astrophysics, The Pennsylvania State University, University Park, PA 16802, USA}
\affiliation{Institute for Gravitation and the Cosmos, The Pennsylvania State University, University Park, PA 16802, USA}

\author[0000-0001-6717-7685]{Gary J. Hill}
\affiliation{McDonald Observatory, The University of Texas at Austin, 2515 Speedway Boulevard, Austin, TX 78712, USA}
\affiliation{Department of Astronomy, The University of Texas at Austin, 2515 Speedway Boulevard, Austin, TX 78712, USA}

\author[0000-0002-1496-6514]{Lindsay House}
\affiliation{Department of Astronomy, The University of Texas at Austin, 2515 Speedway Boulevard, Austin, TX 78712, USA}

\author[0000-0002-8434-979X]{Donghui Jeong}
\affiliation{Department of Astronomy \& Astrophysics, The Pennsylvania State University, University Park, PA 16802, USA}
\affiliation{Institute for Gravitation and the Cosmos, The Pennsylvania State University, University Park, PA 16802, USA}

\author{Wolfram Kollatschny}
\affiliation{Institut fuer Astrophysik, Friedrich-Hund-Platz, D-37077, Goettingen }

\author[0000-0002-6907-8370]{Maja Lujan Niemeyer}
\affiliation{Max-Planck-Institut f\"{u}r Astrophysik, Karl-Schwarzschild-Str. 1, 85741 Garching, Germany}

\author[0000-0002-7327-565X]{Sarah Tuttle}
\affiliation{Department of Astronomy, University of Washington, Seattle, Physics \& Astronomy Building, Seattle, WA, 98195, USA}

\author{(The HETDEX Collaboration)}

\begin{abstract}

We present the Ly$\alpha$ emission line luminosity function (LF) of the Active Galactic Nuclei (AGN) in the first release of the Hobby-Eberly Telescope Dark Energy Experiment Survey (HETDEX) AGN catalog (\citealt{Liu2022}, Paper I). The AGN are selected either by emission-line pairs characteristic of AGN or by single broad emission line, free of any photometric pre-selections (magnitude/color/morphology). The sample consists of 2,346 AGN spanning $1.88<z<3.53$, 
covering an effective area of 30.61 $\rm deg^2$. 
Approximately 2.6\% of the HETDEX AGN are not detected at $>5\sigma$ confidence at $r\sim26$ in the deepest $r$-band images we have searched.
The Ly$\alpha$ line luminosity ranges from $\sim10^{42.3}$ to~$10^{45.9}$~erg~s$^{-1}$. Our Ly$\alpha$ LF shows a turnover luminosity with opposite slopes on the bright end and the faint end: The space density is highest at $\rm L_{Ly\alpha}^*=10^{43.4}$~erg~s$^{-1}$.

We explore the evolution of the AGN LF over a broader redshift range ($0.8<z<3$); constructing the rest-frame ultraviolet (UV) LF with the 1450 \AA\ monochromatic luminosity of the power-law component of the continuum ($\rm M_{1450}$) from $\rm M_{1450}\sim-18$ to~$-27.5$. We divide the sample into three redshift bins ($z\sim$1.5, 2.1, and 2.6). 
In all three redshift bins, our UV LFs indicate that the space density of AGN is highest at the turnover luminosity $\rm M_{1450}^*$ with opposite slopes on the bright end and the faint end. The $\rm M_{1450}$ LFs in the three redshift bins can be well-fit with a luminosity-evolution-density-evolution (LEDE) model: the turnover luminosity ($\rm M_{1450}^*$) increases and the turnover density ($\rm \Phi^*$) decreases with increasing redshift. 

\end{abstract}


\section{Introduction}
\label{sec_intro}

Active Galactic Nuclei (AGN) are actively growing supermassive black holes (SMBHs) in the center of the galaxies. The strong observed correlation between the mass of the SMBHs and the velocity dispersion of their host galaxies implies that their evolution is correlated (\citealt{Gebhardt2000,Tremaine2002}, and see \citealt{Kormendy2013} for a detailed review). 
The AGN population is believed to be a significant contributor 
to the ultraviolet (UV) background and the re-ionization of the Universe \citep[e.g.][]{Haardt2015,Giallongo2015,Puchwein2019}. 
The AGN luminosity function (LF) describes the space density of AGN at a given luminosity. 
Accurate measurements of the AGN LF, especially at intermediate redshifts ($z\sim1-3$) when galaxies and SMBHs grow most rapidly, can not only characterize and help understand the strong evolution of the SMBHs, but also help to better picture the evolution of galaxies and the Universe.


The AGN LF is usually modeled with a double power-law profile with a steep bright-end slope and a shallow faint-end slope, with the characteristic break luminosity $\rm L^*$ and space density $\rm \Phi^*$ as shown by Equation \ref{e_dpl_L} \citep[e.g.][]{Richards2006,Hopkins2007,Bongiorno2007A&A,Ross2013,PD2016,Shen2020}.
\begin{equation}
  \rm  \Phi(L) = \frac{\Phi_L^*}{(L/L^*)^{-\alpha}+(L/L^*)^{-\beta}}.
\label{e_dpl_L}
\end{equation}
It can alternatively be re-written with the number density of AGN per unit magnitude as Equation \ref{e_dpl_M},
\begin{equation}
  \rm \Phi(M) = \frac{\Phi_M^*}{10^{0.4(\alpha+1)(M-M^*)+0.4(\beta+1)(M-M^*)}}.
\label{e_dpl_M}
\end{equation}

\subsection{The bright end}
The bright end of the AGN LF at intermediate redshifts ($1 \lesssim z \lesssim 3$) is well constrained by many large quasar (QSO) samples with consistent results. Most of these samples are pre-selected via their continuum brightness. For example, \cite{Richards2006} studied the AGN LF of 15,343 spectroscopically confirmed broad-line QSOs over 1622 deg$^2$ within $0<z<5$ defined from the first and the second generation of the Sloan Digital Sky Survey (SDSS-I/II) down to $i=19.1$ ($z\lesssim3$) and $i=20.1$ ($z\gtrsim3$). \cite{Ross2013} measured the AGN LF of 22,301 spectroscopically confirmed AGN over 2236 deg$^2$ between $2.2<z<3.5$ down to $i<21.8$ from the third generation of SDSS Baryon Oscillation Spectroscopic Survey (SDSS-III: BOSS) Data Release Nine (DR9). \cite{PD2016} studied the AGN LF of 13,876 AGN variability-selected from the fourth generation of SDSS (SDSS-IV) extended Baryon Oscillation Spectroscopic Survey (eBOSS) over 94.5 deg$^2$ in Stripe 82 at $0.68<z<4.0$ down to $g<22.5$. 

\subsection{Modeling the faint end}

Studies of the faint end of the AGN LF can help define the trigger mechanism(s) and the duty cycle of AGN. Hydrodynamical simulations show that mergers of gas-rich galaxies can lead to strong inflows of cold gas, feed central SMBHs, and thereby power bright AGN with high accretion rates \citep[e.g.][]{DiMatteo2005,Hopkins2008}. The feedback of AGN can in return expel enough gas and shut down the star formation in the host and leave the galaxy passive, red, and elliptical. The time scale of merger events is $\sim$ 1 Gyr, but for the majority of this time AGN are heavily obscured by surrounding dust. They can only shine as optical AGN for $\sim$ 100 Myr after enough dust is expelled. Low-luminosity AGN are more likely triggered by secular evolution of galaxies with low accretion rates \citep[see][for a detailed review of various triggering mechanisms of AGN]{Alexander2012}. For example, asymmetric structures, such as bars, can help remove the angular momentum of gas and trigger AGN activity. Whether massive SMBHs mainly consist of short-lived bright QSOs under near Eddington accretion, or of long-lived low-luminosity AGN with lower accretion rates remains uncertain. The number density of low-luminosity AGN is a key piece of information of the duty cycle of AGN triggered by the secular evolution with low accretion rates.

The faint end of the AGN LF is also important in understanding the evolution model of AGN which is usually described with the evolution of the break point ($\rm L^*$ and $\rm \Phi^*$) between the bright end and the faint end. The evolution of the AGN LF has been modeled with different models in various studies. \cite{Boyle2000} and \cite{Croom2004} found that the QSO sample ($b_J<20.85$, $0<z<2.5$) of the 2dF QSO Redshift Survey \citep[2QZ;][]{Croom2004} can be well fit by a pure luminosity evolution model (PLE, only $\rm L^*$ evolves with redshift while $\rm \Phi^*$ is constant). 
When going to higher redshifts, large QSO samples defined from SDSS show that the PLE model can only describe the QSO LF up to $z\sim2.2$, above which, besides the evolution of $\rm L^*$, $\rm \Phi^*$ decreases significantly with redshift \citep{Richards2006,Ross2013,PD2016}. They introduced the luminosity-evolution-density-evolution model (LEDE, $\rm L^*$ and $\rm \Phi^*$ evolve with redshift independently) at $z\gtrsim2.2$ in addition to the PLE model at $z\lesssim2.2$ to fit the evolution of the QSO LF. \cite{Bongiorno2007A&A} combined a faint AGN sample of 130 AGN ($\rm I_{AB}<24$; $0<z<5$) selected from the VIMOS-VLT Deep Survey (VVDS) \citep{Gavignaud2006} with the LF of bright QSOs of SDSS \citep{Richards2006}. They found that a luminosity-dependent-density-evolution model (LDDE), where $\rm \Phi^*$ evolves with redshift regulated by $\rm L^*$, can best describe the evolution of AGN LF. 

The faint end of the AGN LF at intermediate redshifts is still poorly constrained, especially in the UV/optical bands. 
X-ray observations are the most efficient method in identifying low-luminosity AGN as the X-ray emissions have strong penetration and low dilution from the host galaxies \citep[see][for a recent review]{Brandt2015}. \cite{Shen2020} studied the evolution of the AGN LF down to about three orders of magnitudes fainter than the break luminosity $\rm L^*$ ($\rm L_{bol}\sim10^{43}\ erg/s$ at $z\sim2$) from $z=0$ to $z=7$ with X-ray observations 
compiled by \cite{Hopkins2007} and updated by \cite{Shen2020}. However, X-ray observations are usually limited by small sky coverage ($\lesssim$ one or a few square degrees) and limited sample sizes ($\lesssim 10^3$). Infrared (IR) observations, especially mid-IR, provide a special approach of finding X-ray/optical obscured AGN 
and allow the study of the AGN LF down to $\sim 1-2$ orders of magnitude fainter than $\rm L^*$. 
However, X-ray, IR, and UV/optical AGN may represent different phases of AGN\null. Therefore, deep UV/optical surveys of AGN are still a unique way of describing the evolution of AGN despite the various studies of AGN LF in other bands.

\subsection{Early faint UV/optical AGN samples}
\label{sec_intro_faint_agn}

In the UV/optical bands, the measurements of the AGN LF fainter than the break magnitude ($\rm M^*$) are still limited: Deep surveys of AGN usually have small sample sizes and small sky coverage, and large samples are not deep enough to extend to the low luminosity region to constrain the faint end. \cite{Hunt2004} made the first direct measurement of the faint end of the AGN LF at $z\sim3$  down to $\rm M_{1450}=-21$ mag ($\sim 5$ mag fainter than $\rm M_{1450}^*$) using a sample of only 11 faint AGN discovered over 0.43 ${\rm deg}^2$ in a survey for Lyman-break galaxies \citep[LBG;][]{Steidel2003}. Their AGN are selected using multi-color selection criteria (the Lyman break technique) and spectroscopic follow-up. The 2dF–SDSS LRG and QSO survey \citep[2SLAQ;][]{Croom2009survey} provide measurements with a significantly larger (10,637 AGN) but much shallower ($\sim$ 1 mag fainter than the break magnitude) AGN sample covering 191.9 deg$^2$ \citep{Croom2009}. Their AGN sample is also pre-selected using a multi-color method which primarily selects UV-excess objects. The AGN LF derived from the VVDS survey used a sample of 130 AGN  (spanning 1.75 deg$^2$), which is larger than the LBG survey and deeper than the 2SLAQ survey ($\sim$ 2.5 mag fainter than the break magnitude) \citep{Bongiorno2007A&A}. Their sample is free from color selection, but is still a magnitude-limited sample which requires the continuum to be brighter than $\rm I_{AB}<24$ in their deep fields and $\rm I_{AB}<22.5$ in their wide fields.

\cite{Liu2022} (Paper I) selected 5,322 AGN at $0.25<z<4.2$ over 30.61 deg$^2$ from the Hobby-Eberly Telescope Dark Energy Experiment survey \citep[HETDEX;][]{Hill08, Gebhardt2021}. This AGN sample is purely emission-line selected free of any photometric pre-selection (broad-band magnitude/color/morphology). Our AGN selection is more effective in finding continuum faint AGN compared to the traditional photometric pre-selection with spectroscopic follow-ups. Cross matching with various imaging surveys shows that 2.6\% of the HETDEX AGN already reach the $5\sigma$ detection limit at $r\sim26$ of HSC which is the deepest imaging survey we searched. The median r-band magnitude of our AGN catalog is 21.6 mag, with 34\% having $r > 22.5$. Our AGN sample goes down to $\rm M_{1450}\sim-18$ mag. 
Both the sample size and the sky coverage of the HETDEX AGN are hundreds of times larger than those of the faint optical AGN samples of LBG and VVDS. 

\subsection{The $\rm Ly\alpha$ LF}
The two faint UV/optical AGN sample of the VVDS survey and the HETDEX survey introduced in Section \ref{sec_intro_faint_agn} are both identified by their emission line features. Besides the continuum LF, a more direct way to evaluate the number density of AGN is the emission-line LF. The $\rm Ly\alpha$ emission is a well studied emission feature that characterize the young ($\lesssim50\,Myr$), low-mass ($\lesssim10^{10}\ M_{\sun}$) galaxies with high star formation rates (SFR) of $1-100\ M_{\sun}\,yr^{-1}$ \citep[e.g.][]{Gawiser2006,Finkelstein2007,Finkelstein2009}. It is believed that the bright end ($L_{\text{Ly}\alpha}>10^{43.3}\ erg\,s^{-1}$) of the $\rm Ly\alpha$ LF is exclusively comprised of AGN \citep[e.g.][]{Spinoso2020}. 
Studying the $\rm Ly\alpha$ LF of AGN can help us better understand the contribution of AGN to each $\rm Ly\alpha$ luminosity. It is also interesting to search the possible connection between the emission line LF and the continuum LF.

In this paper, we present our measurements of the AGN LF based on the 5K emission-line selected AGN from the HETDEX survey. We briefly summarize the HETDEX survey and the AGN catalog in Section \ref{sec_catalog}. Section \ref{sec_completeness} describes the estimation of the completeness corrections. We introduce the $\rm V_{max}$ method for the LF calculation in Section \ref{sec_vmax}. We show the $\rm Ly\alpha$ LF of the AGN sample in Section \ref{sec_LyA_LF}. We explore the evolution of the AGN UV LF in Section \ref{sec_UV_LF_z}. In Section \ref{sec_Lbol}, we present the bolometric LF of our HETDEX AGN sample and compare it with the AGN LFs of other bands. 
We discuss the potential incompleteness of our AGN sample in the faint end and our efforts in correcting such selection effects in Section \ref{sec_discuss}.
We summarize this paper in Section \ref{sec_summary}. We use a flat $\rm\Lambda$CDM cosmology with $H_0=\rm 70\ km\ s^{-1}\ Mpc^{-1}$, $\rm \Omega_M=0.3$, $\rm \Omega_\Lambda=0.7$ throughout this paper. 


\section{The HETDEX AGN catalog}
\label{sec_catalog}

HETDEX \citep{Hill08, Gebhardt2021} is an ongoing spectroscopic survey (3500 \AA\ - 5500 \AA) without target pre-selection on the upgraded 10-m Hobby-Eberly Telescope (HET, \citealt{Ramsey98, Hill2021}). It uses the Visible Integral field Replicable Unit Spectrograph \citep[VIRUS;][]{Hill2021} to record spectra of every object falling within its field of view. A typical exposure contains 34,944 spectra, most of which capture sky or the background. The primary goal of this survey is to measure the cosmology at $z\sim3$ using $\rm Ly\alpha$ emitters (LAEs) as tracers of large scale structure. The HETDEX survey is expected to be active from 2017 to 2024, and eventually will cover 540 deg$^2$.

In this paper, we analyze the luminosity function of 5,322 AGN ($0.25<z<4.2$) identified from HETDEX observations from 2017-Jan-01 to 2020-Jun-26 covering an effective area of 30.61 $\rm deg^2$ (see Paper I for details of the HETDEX AGN sample). The AGN used for this analysis are identified solely based on their spectroscopic properties, free of any pre-selection based on imaging (magnitude, morphology, and color). Measurements with the Hyper Suprime-Cam (HSC) imager of the Subaru telescope from the HSC-HETDEX joint survey (HSC-DEX; S15A, S17A, S18A, PI: A. Schulze, and S19B, PI: S. Mukae) and the HSC Subaru Strategic Program (HSC-SSP; \citealt{Aihara2018,Aihara2019}), show that 2.6\% of the HETDEX AGN already reach their $5\sigma$ detection limit at $r\sim26$. The median r-band magnitude of our AGN catalog is 21.6 mag, with 34\% having $r > 22.5$.

There are two main steps in the sample selection: 
first, the HETDEX detection pipeline \citep{Gebhardt2021} consists of two detection algorithms, one for the identification of emission lines and the other for the detection of continuum sources. The 1.5 million $\rm S/N>4.5$ emission line candidates in the database were identified using a 3-dimensional grid search (2 dimensions in the spatial direction and 1 dimension in the wavelength dimension);  the 80,000 continuum candidates were selected via a detection threshold of 50 electron counts (roughly corresponds to $g\sim22.5$) in either a blue window (3700\,\AA\ - 3900\,\AA) or a red window (5100\,\AA\ - 5300\,\AA\null).  The union of these two datasets served as the basis of our AGN search.

The second step in our sample selection, AGN identification, is detailed in Paper~I.  To summarize:  AGN are found using two different methods: 2em selection, which identifies emission line pairs characteristic of AGN, and the sBL selection that searches for single broad ($\rm FWHM>1000$\,km/s) emission lines within the HETDEX wavelength range. There are 3,733 AGN with secure redshifts, either confirmed by emission line pairs or by matched spectral redshifts from literature, and flagged with $zflag=1$. The remaining 1,589 AGN are sBL selected broad-line AGN candidates provided with our best estimated redshifts and flagged with $zflag=0$. Our line identifications for the single broad emission lines are based on a combination of their line profiles, observed equivalent widths (EW), enhanced noise level at other expected lines, etc (see Paper I for more details). 

\section{Completeness Correction}
\label{sec_completeness}

The complexity of our AGN detection algorithm (see Paper I) means that there are two separate contributors to the completeness of the AGN sample: the completeness of the HETDEX detection algorithm, ($\rm C_{HETDEX}$ detailed in Section \ref{sec_c_het} and Farrow et al. in preparation), and the completeness of the AGN selection algorithm, ($\rm C_{AGN}$, described in Section \ref{sec_c_agn}).  In other words,
\begin{equation}
\rm C = C_{HETDEX} * C_{AGN}.
\label{e_c_tot}
\end{equation}

\subsection{Completeness of  HETDEX detection}
\label{sec_c_het}

As stated in Section \ref{sec_catalog}, the HETDEX detection algorithm identifies emission-line candidates and the continuum candidates separately. Therefore, the completeness of sources in the two candidate catalogs must be evaluated separately (Equation \ref{e_c_het_full}). 
\begin{equation}
\rm C_{HETDEX} = \begin{cases} 
                    \rm C_{HETDEX,line} &\text{emission line candidates}\\
                    \rm C_{HETDEX,cont} &\text{continuum candidates}
                 \end{cases}
\label{e_c_het_full}
\end{equation}
For the continuum candidates, we make a simple assumption that all continuum candidates would be automatically picked up by the pipeline, meaning $\rm C_{HETDEX,cont}\equiv1$.

\begin{figure*}[htbp]
\centering
\includegraphics[width=0.48\textwidth]{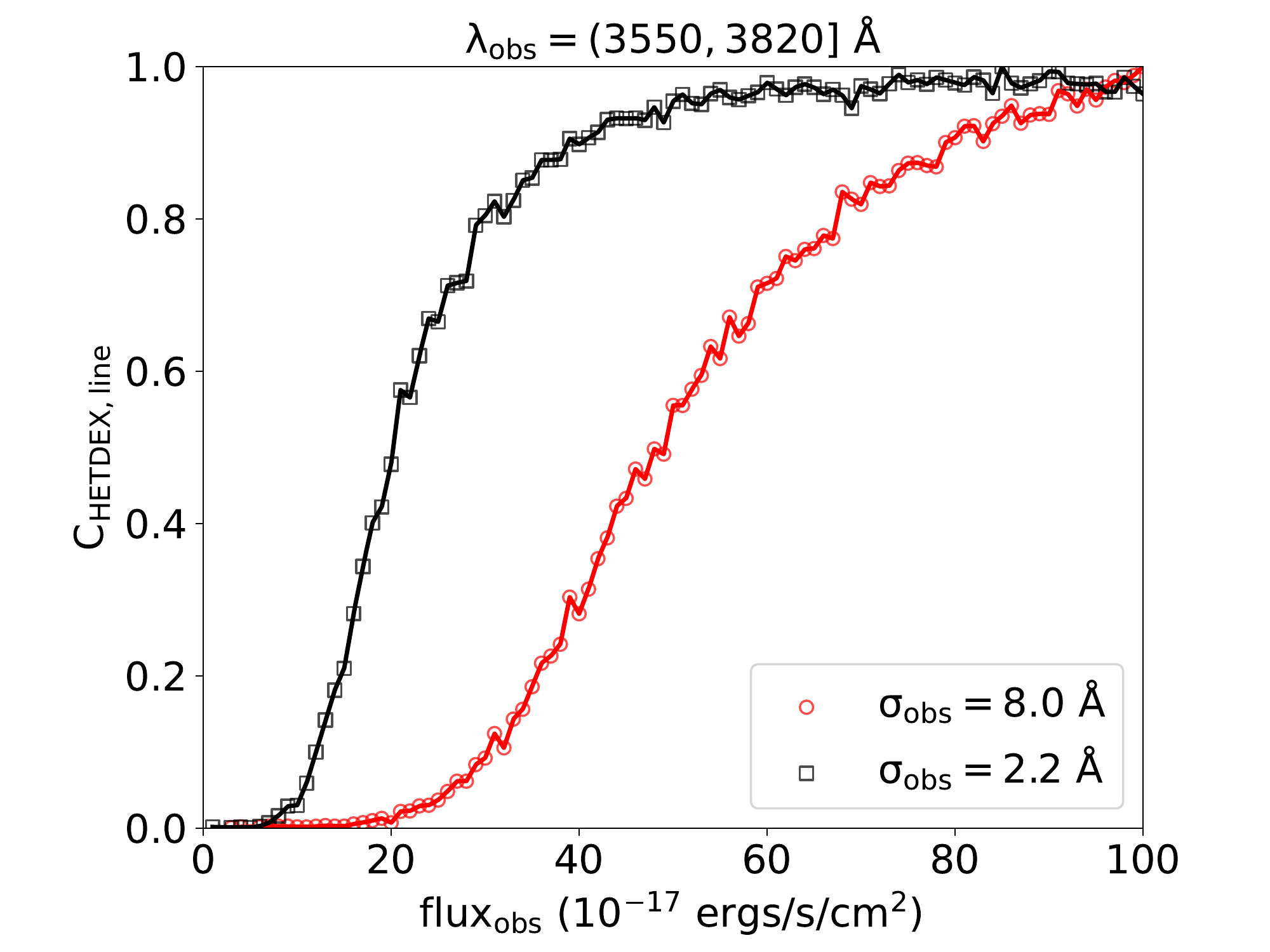}
\includegraphics[width=0.48\textwidth]{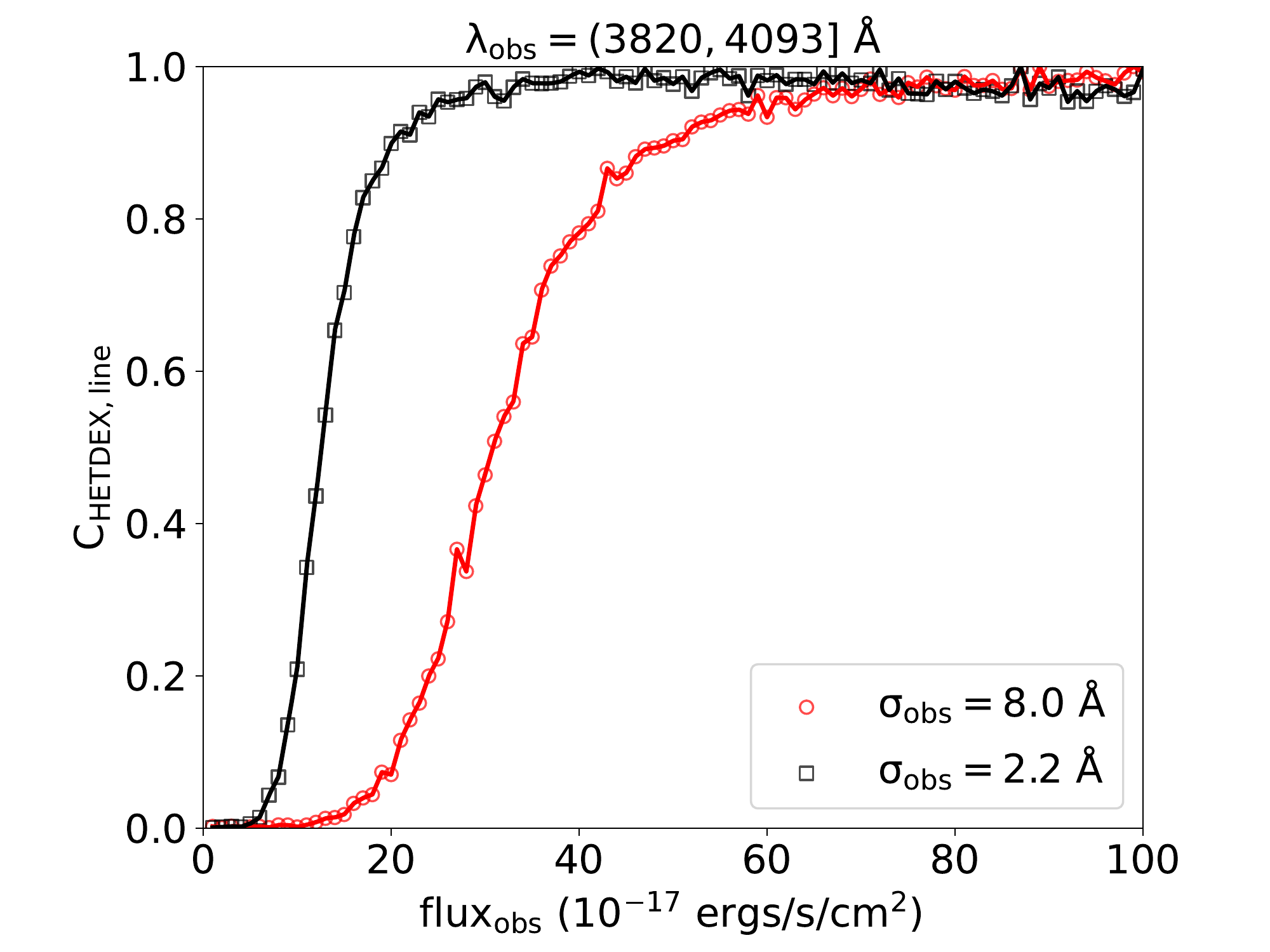}
\includegraphics[width=0.48\textwidth]{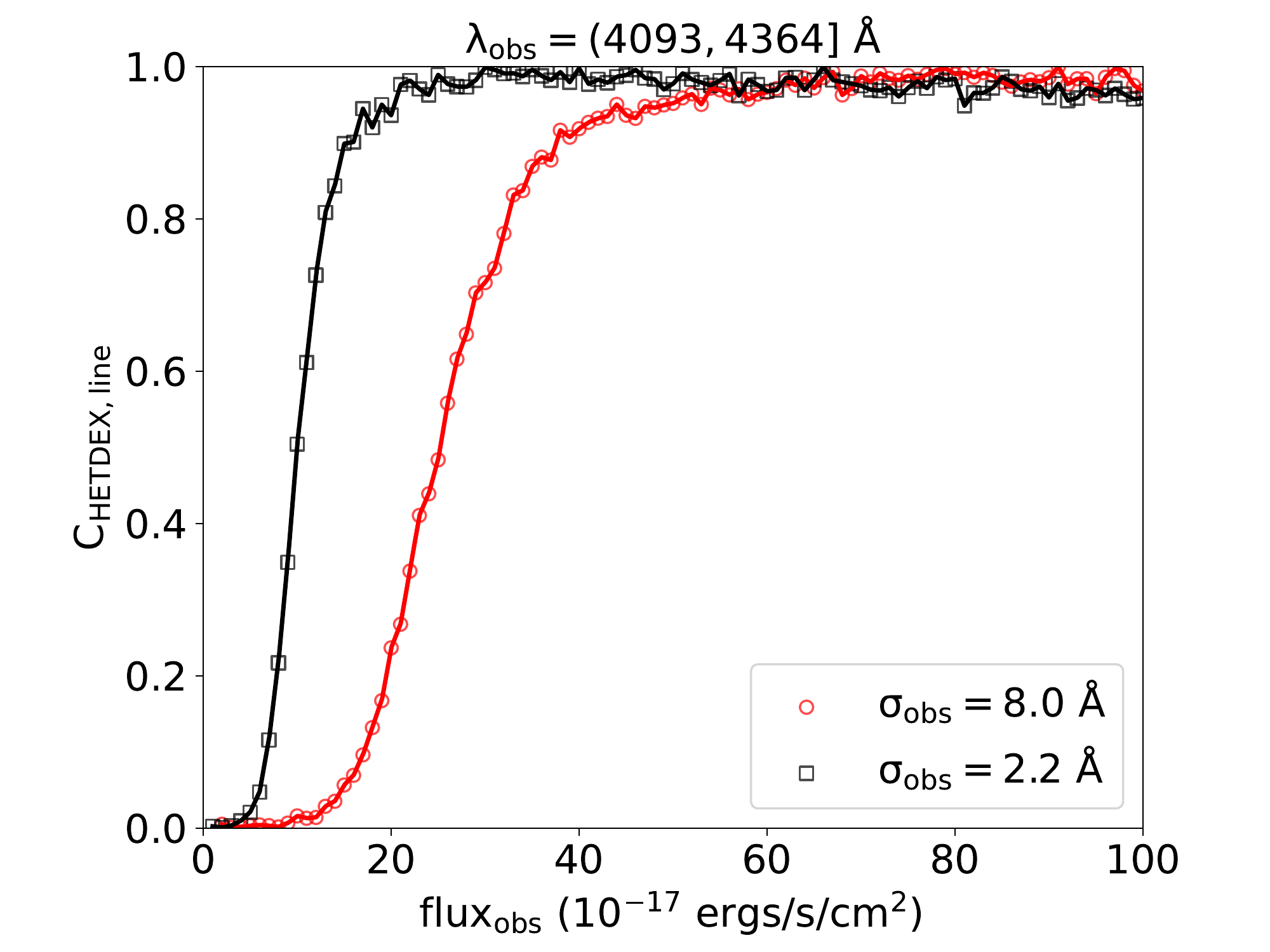}
\includegraphics[width=0.48\textwidth]{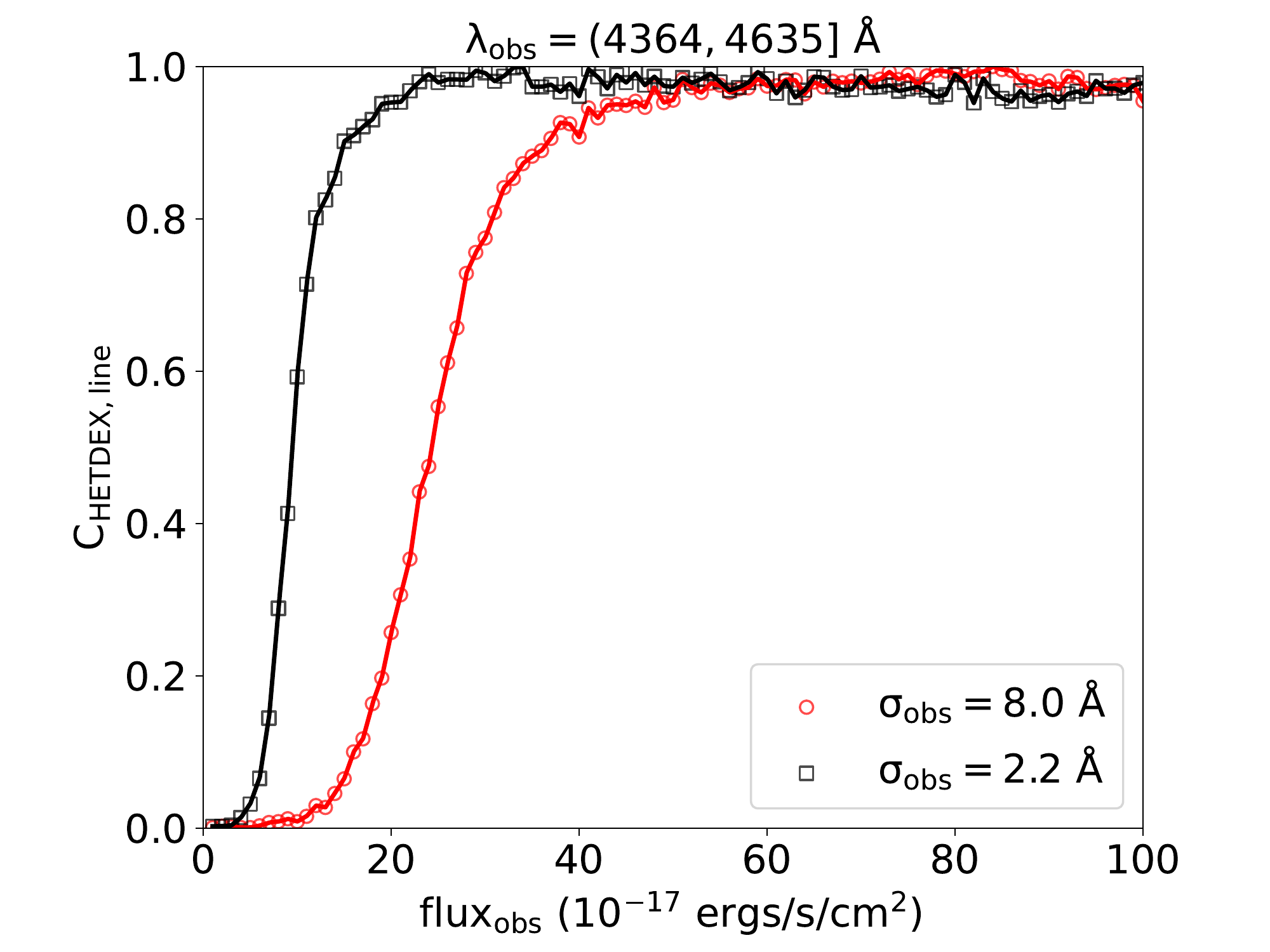}
\includegraphics[width=0.48\textwidth]{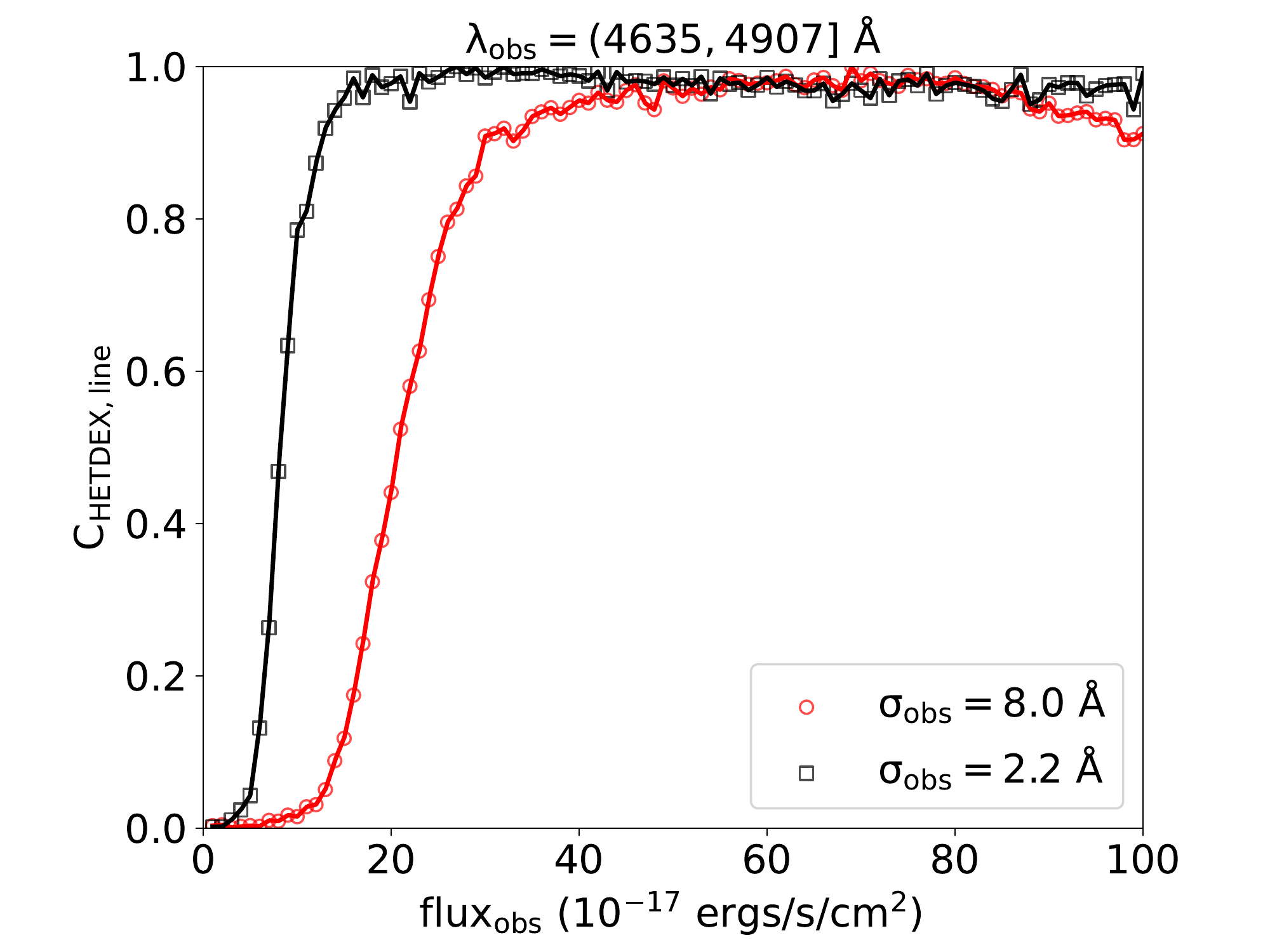}
\includegraphics[width=0.48\textwidth]{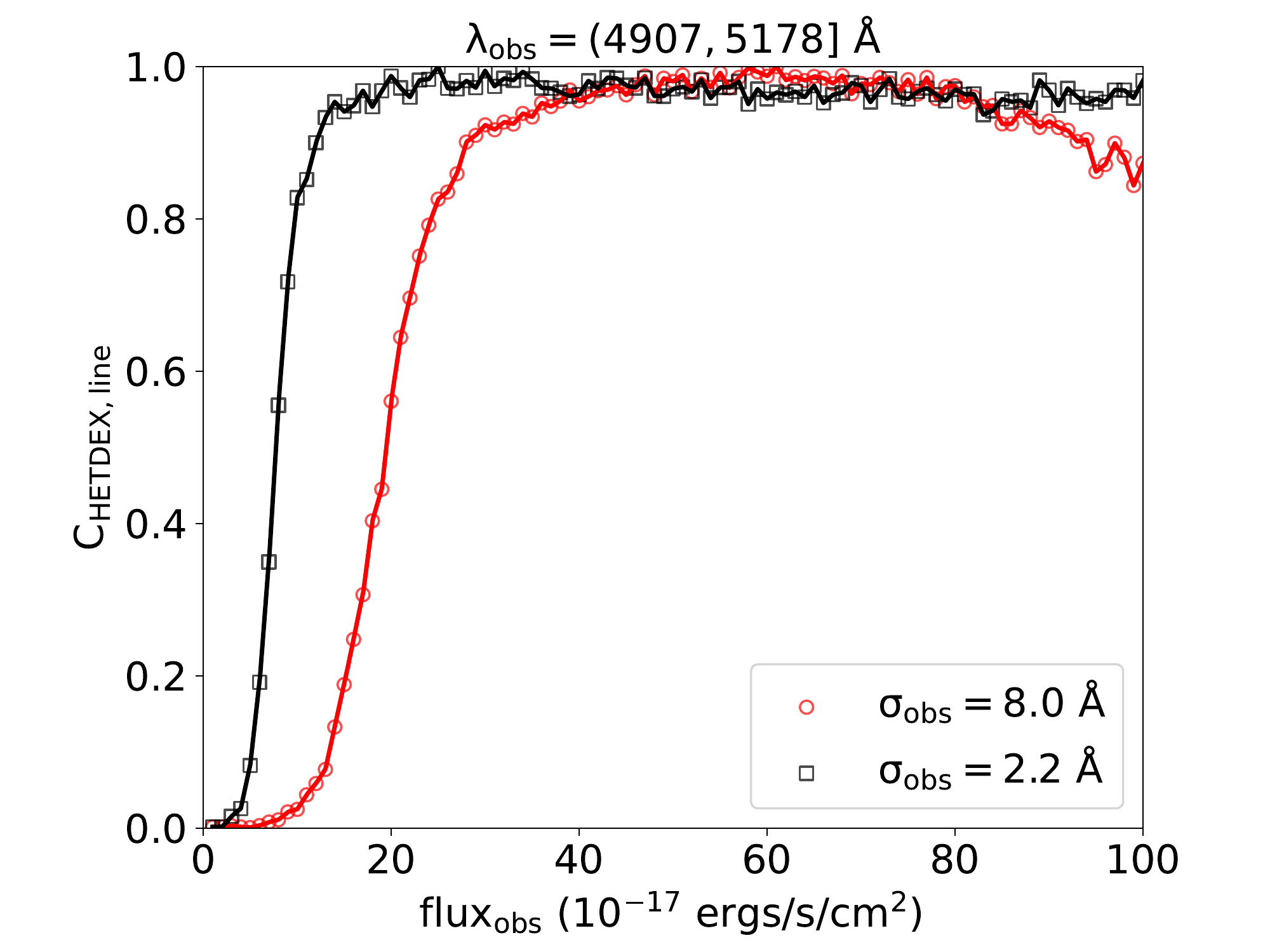}
\caption{The simulated completeness fraction of S/N$>4.8$ emission-line detections ($\rm C_{HETDEX,line}$) for the 13th exposure on the night of 2020-Jun-23 as a function of the observed emission line flux in different wavelength intervals. The red circles are for the broad emission lines, which have line widths of $\rm \sigma_{obs}=8$\,\AA, and the red curve is the spline interpolation to the red circles. The black squares show the reference completeness function for unresolved emission lines ($\rm\sigma_{obs}=2.2$\,\AA), and the black curve is the spline interpolation of the black squares.}
\label{f_c_het}
\end{figure*}

For the emission line candidates, the completeness simulations were first presented in \cite{Gebhardt2021}, but we review the methodology here. We employ a set of simulations to carefully evaluate the completeness of the HETDEX detection algorithm by inputting simulated emission lines with known observed wavelengths ($\rm \lambda_{obs}$), fluxes ($\rm flux_{obs}$), and line widths ($\rm \sigma_{obs}$) into each observed data cube. We then run the detection code and check how many of these artificial emission lines are recovered by the HETDEX detection algorithm. The detection rate is then the fraction of artificial sources recovered in the datacube.  This experiment demonstrates that the completeness of the HETDEX algorithm for an emission line is a function of observed wavelength, flux, and line width. It also varies with location on the sky via dependencies related to an object's placement on an IFU (edge fibers have lower completeness compared to fibers near the IFU center), and the observational conditions at the time of the exposure.  
The completeness of each detection from the HETDEX detection algorithm is then a function of six parameters as shown in Equation \ref{e_c_het},
\begin{equation}
  \rm  C_{HETDEX,line} = C(\lambda_{obs},\ flux_{obs},\ \sigma_{obs},\ R.A.,\ Dec.,\ S/N_{th}),
\label{e_c_het}
\end{equation}
where $\rm \sigma_{obs}$ is the line width of the emission fitted with a single Gaussian profile and S/N$_{\rm th}$ is the signal-to-noise ratio threshold for detections. We make use of flux limit routines in the HETDEX API \footnote{https://github.com/HETDEX/hetdex$\_$api} to evaluate Equation~\ref{e_c_het}, details and tests of the implementation will be presented in Farrow et al in preparation.

Figure \ref{f_c_het} shows the simulated completeness of the HETDEX detection algorithm for the observation taken by the 13th exposure in the night of 2020-Jun-23 as a typical example. The completeness increases as a function of the observed emission-line flux from 0 to about 1. The completeness of blue wavelengths is overall worse than that of the red wavelengths due to low throughput. The red line shows the simulation for a set of ``broad'' emission lines with $\rm \sigma_{obs}=8$\,\AA; the black line shows the completeness for unresolved emission with $\rm \sigma_{obs}=2.2$\,\AA\null.  In all cases, the curve giving the completeness versus line flux for broad emission line objects increases more slowly than that for narrow emission lines. This is because the HETDEX line-detection algorithm is tuned to select Ly$\alpha$ emitting galaxies (LAEs), whose emission lines are unresolved (or barely resolved) at HETDEX resolution ($\rm \sigma_{obs}=2.2$\,\AA\null).  The detection of broad emission lines is therefore less efficient, and that leads to a shallow slope to the completeness curve. 
Based on simulations with different $\rm \sigma_{obs}$, $\rm \lambda_{obs}$, and $\rm S/N_{th}$, we find that the difference between the completeness curve of ``broad'' emission lines with $\rm \sigma_{obs}$ and that of the unresolved emission lines is only related with $\rm \sigma_{obs}$ and $\rm S/N_{th}$. $\rm C_{HETDEX,line}$ of a broad emission line candidate with $\rm \sigma_{obs}$ can be estimated with that of the unresolved emission line with $\rm \sigma_{obs}=2.2$\,\AA\ by dividing its observed flux ($\rm flux_{obs}$) by $\rm \sqrt{\sigma_{obs}/2.2}*(15.09-S/N_{th})*(0.098-0.0004\sigma_{obs}+0.0004\sigma_{obs}^2)$.

\subsection{Completeness of AGN selection}
\label{sec_c_agn}

The AGN selection itself also has two cases: one for objects selected via the presence of two AGN emission lines (2em; Section \ref{sec_c_2em}) and the other for sources found via a single, broad emission line (sBL; Section \ref{sec_c_sBL}).  The completeness function for each must be evaluated separately as Equation \ref{e_c_tot2}, as is introduced in Section \ref{sec_catalog} (see Paper I for more details).
\begin{equation}
\rm C_{AGN} = \begin{cases}
                 \rm C_{2em} &\text{2em selected AGN}\\
                 \rm C_{sBL} &\text{sBL selected AGN}
              \end{cases}
\label{e_c_tot2}
\end{equation}

\subsubsection{Completeness of the 2em selection}
\label{sec_c_2em}

\begin{figure}[htbp]
\centering
\includegraphics[width=\textwidth]{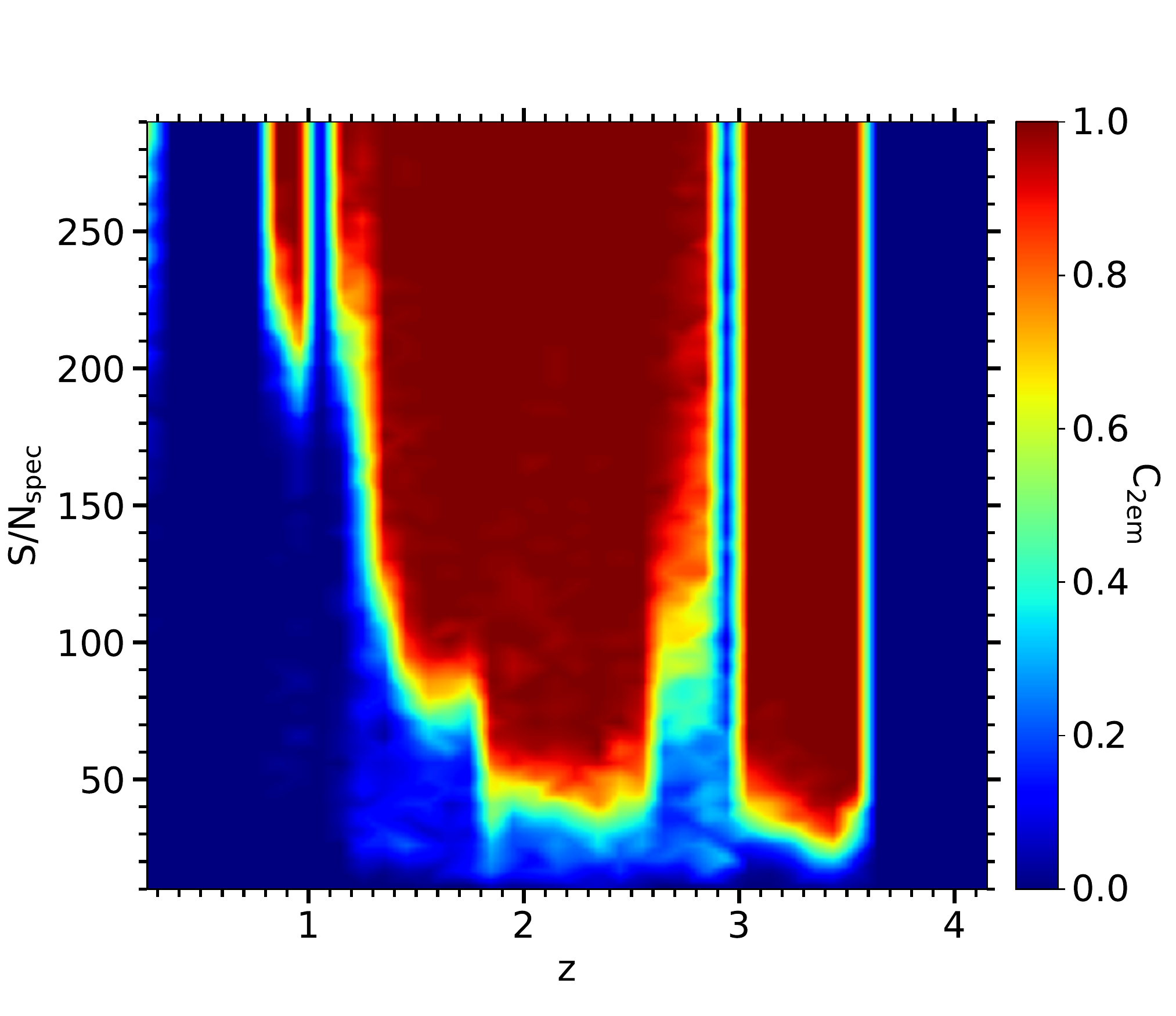}
\caption{The completeness of the 2em selection method in the S/N$\rm_{spec}$ - redshift space, where the S/N$\rm_{spec}$ is the signal-to-noise ratio across the full wavelength coverage 3500\,\AA\ - 5500\,\AA\ of the HETDEX spectra. All emission lines and continuum are considered as the signal in the calculation of S/N$\rm_{spec}$, so it can be as large as $\sim200$ for strong AGN.}
\label{f_c_2em}
\end{figure}

The 2em selection algorithm searched the HETDEX detection catalog for all emission line pairs characteristic of AGN\null. Such pairs include Ly$\alpha$+\ion{C}{4} $\lambda 1549$, \ion{C}{4} $\lambda 1549$ + \ion{C}{3}] $\lambda 1909$, \ion{C}{3}] $\lambda 1909$ + \ion{Mg}{2} $\lambda 2799$, etc.  The algorithm requires that the S/N of the strongest emission line be greater than 5 ($\rm S/N_{em,1st}>5$), and that of the second most significant emission feature have a S/N greater than 4 ($\rm S/N_{em,2nd}>4$). Thus, the probability of an AGN detection is related with both of $\rm S/N_{em,1st}$ and $\rm S/N_{em,2nd}$. To simplify the parameter space, we introduce the signal-to-noise ratio over the full 3500 \AA\ - 5500 \AA\ range ($\rm S/N_{spec}$) to take both $\rm S/N_{em,1st}$ and $\rm S/N_{em,2nd}$ into consideration with a single parameter. Since $\rm S/N_{spec}$ considers all emission lines and the continuum across 3500 \AA\ - 5500 \AA\ as the signal, it can be as large as $\sim 100 - 300$ for strong AGN.

The redshift is another key factor for calculating the completeness for the 2em selected AGN, as it regulates the observed wavelengths of the emission line pairs. At some redshifts, only one AGN emission line falls within the 3500\,\AA\ - 5500\,\AA\ wavelength range of the HETDEX survey, making it impossible to identify line pairs at these redshifts.  Consequently, the completeness function for 2em selected AGN is a function of two variables:  $\rm S/N_{spec}$ and the redshift. 

To calculate this function, we simulate a series of AGN spectra. We first add various redshifts to the rest-frame composite spectrum obtained from the HETDEX AGN catalog in Paper I and cut the composite spectra in the observed frames with the HETDEX wavelength coverage (3500\,\AA\ - 5500\,\AA). Noises are then added to the simulated observed AGN spectra with controlled $\rm S/N_{spec}$. We then run the 2em selection code and count how many of the simulated AGN would be identified as AGN in different ($\rm S/N_{spec}$, z) bins, i.e.,
\begin{equation}
  \rm  C_{2em} = C(S/N_{spec},\ z).
\label{e_c_2em}
\end{equation}
Since our simulated AGN spectra are based on the composite spectrum of the HETDEX AGN sample, they can reproduce the continuum and the relative line strength of different emission lines.

Figure \ref{f_c_2em} shows the completeness function for the 2em selected AGN in the $\rm S/N_{spec}$ - redshift space. The high completeness fractions near $z\sim 3.3$, 2.3, 1.6, 0.9, and 0.25 shows the preference for identifying the line pairs (\ion{O}{6} $\lambda 1034$ + Ly$\alpha$), (Ly$\alpha$ + \ion{C}{4} $\lambda 1549$), (\ion{C}{4} $\lambda 1549$ + \ion{C}{3}] $\lambda 1909$), (\ion{C}{3}] $\lambda 1909$ + \ion{Mg}{2} $\lambda 2799$), and (\ion{Mg}{2} $\lambda 2799$ + \ion{O}{2} $\lambda 3727$). Different line pairs require different levels of $\rm S/N_{spec}$ to be identified due to their different line strengths. For example, the (Ly$\alpha$ + \ion{C}{4} $\lambda 1549$) line pair is stronger than the  (\ion{C}{4} $\lambda 1549$ + \ion{C}{3}] $\lambda 1909$) line pair, so the high completeness region at $z\sim2.3$ starts from $\rm S/N_{spec}\sim 25$, while the high completeness region at $z\sim1.6$ starts from a higher $\rm S/N_{spec}$ level at $\sim 75$.
AGN at redshifts where the completeness is zero can only be identified by the single broad line selection method (Section \ref{sec_c_sBL}). 

\subsubsection{Completeness of the sBL selection}
\label{sec_c_sBL}

\begin{figure}[htbp]
\centering
\includegraphics[width=\textwidth]{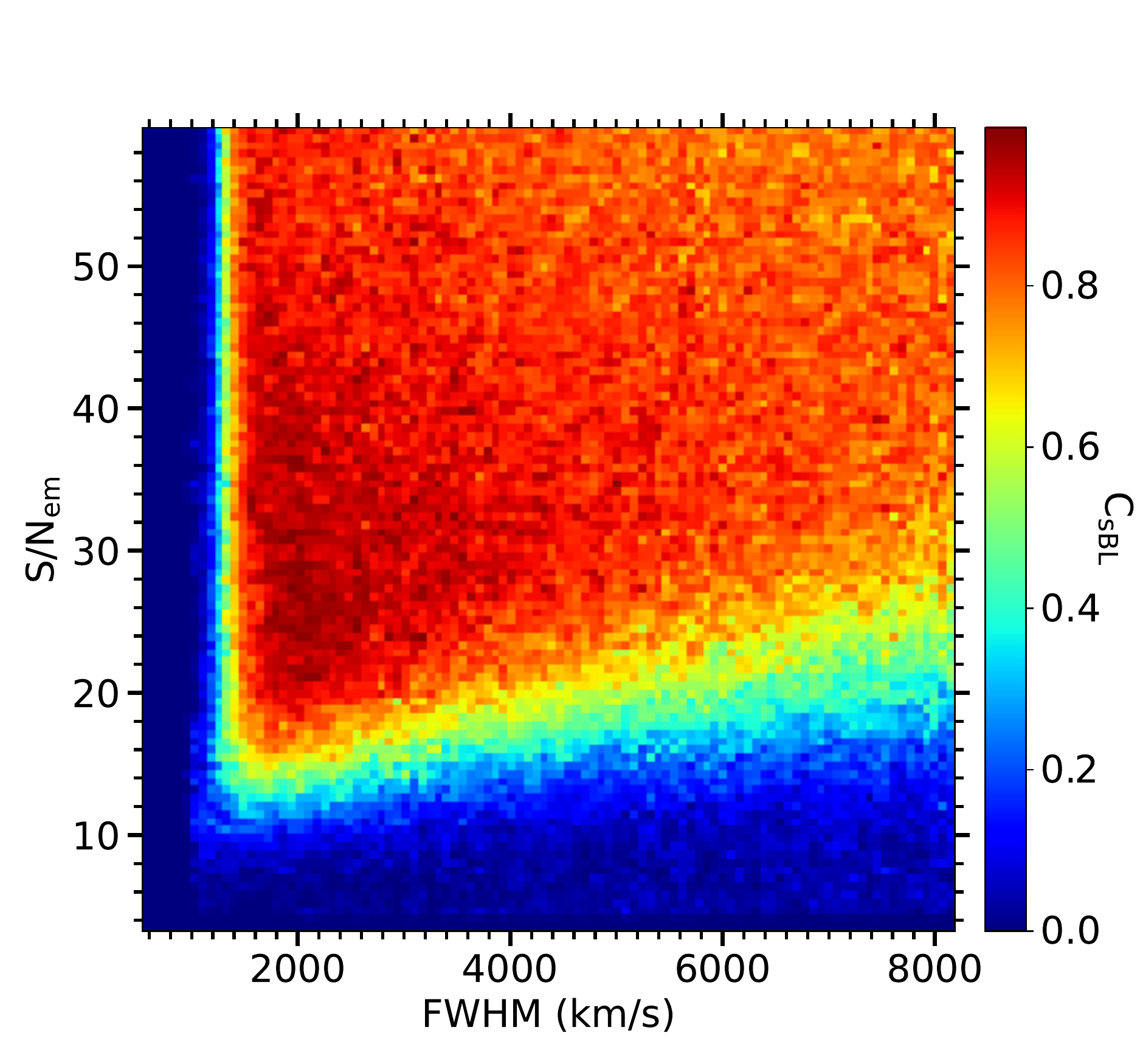}
\caption{The completeness of the sBL selection as a function of the $\rm S/N_{em}$ and full-width-at-half-maximum (FWHM) of simulated emission lines.}
\label{f_c_sBL}
\end{figure}

The sBL selection identifies AGN with a single broad emission line ($\rm FWHM>1000$\,km/s) within the wavelength range of the HETDEX survey as Type I AGN candidates. 
The single Gaussian fit for the emission lines of the HETDEX pipeline is primarily designed to search for the normal narrow emission line candidates, and is usually not suitable for the broad emission lines for AGN. We then fit the broad emission line candidates independently with multi-Gaussian profiles for our sBL search to get more accurate measurements of FWHMs (see Paper I for more details). The sBL selection requires the signal-to-noise ratio of the emission line from our multi-Gaussian fits ($\rm S/N_{em}$) to be greater than 3.8. The completeness of the sBL selection is a function of the line widths and the emission-line $\rm S/N_{em}$ (Equation \ref{e_c_sBL}).
\begin{equation}
  \rm  C_{sBL} = C(FWHM, S/N_{em})
\label{e_c_sBL}
\end{equation}
We again make simulate emission lines with known $\rm S/N_{em}$ and FWHM, run them through the sBL selection code, and plot the completeness fraction in the $\rm S/N_{em}$ - FWHM space, as is shown in Figure \ref{f_c_sBL}. As expected, the broader emission lines require a higher $\rm S/N_{em}$ to be identified. The region with intermediate line widths, $\rm 1000\,km/s \lesssim FWHM \lesssim 1500\,km/s$, also have lower completeness, as features of this width have a higher chance of being fit with FWHM $< 1000$\,km~s$^{-1}$ templates and therefore mis-identified as narrow emission line objects.

\subsection{Total Completeness}
\label{sec_c_tot}

\begin{figure}[htbp]
\centering
\includegraphics[width=0.48\textwidth]{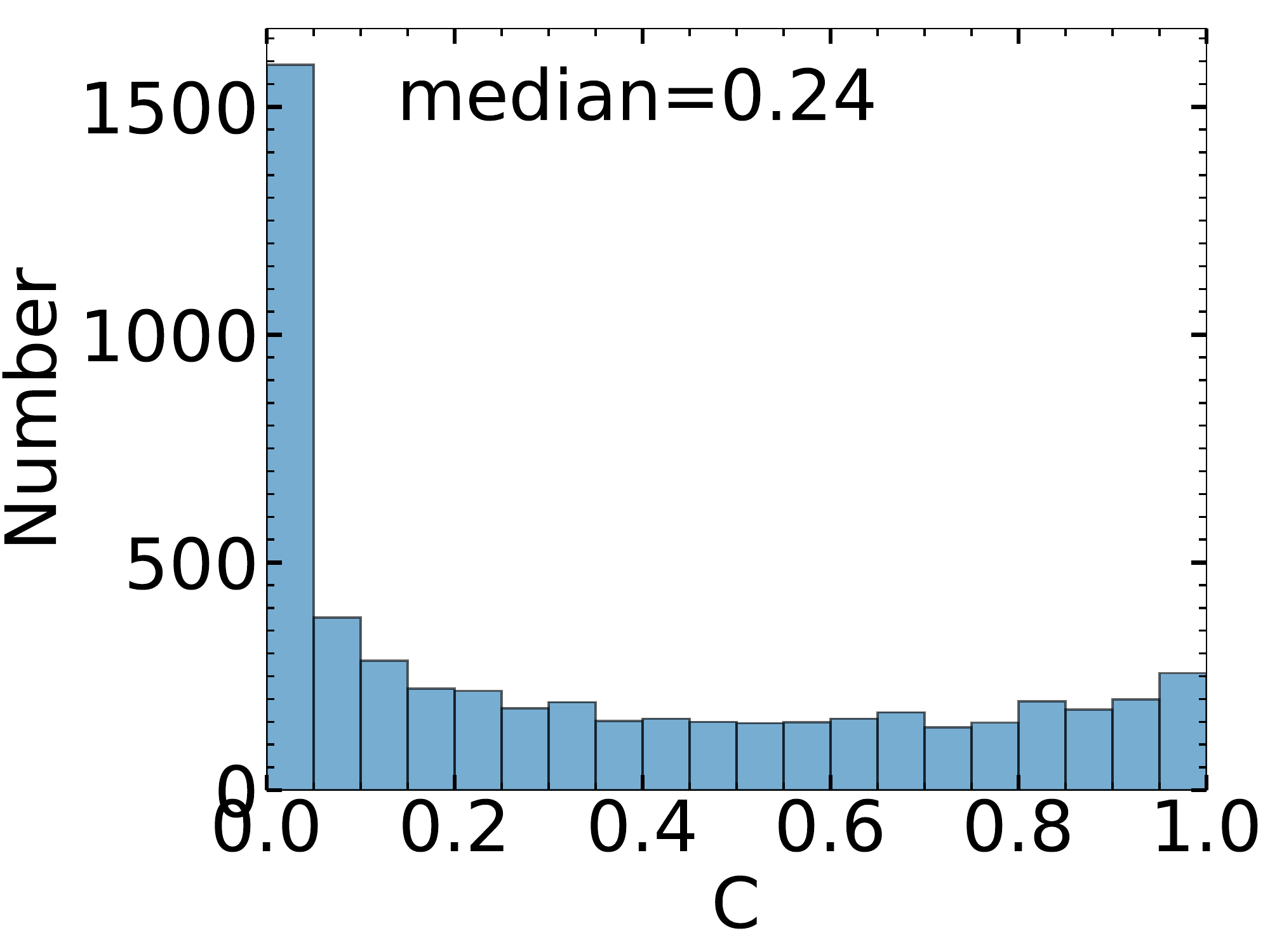}
\includegraphics[width=0.48\textwidth]{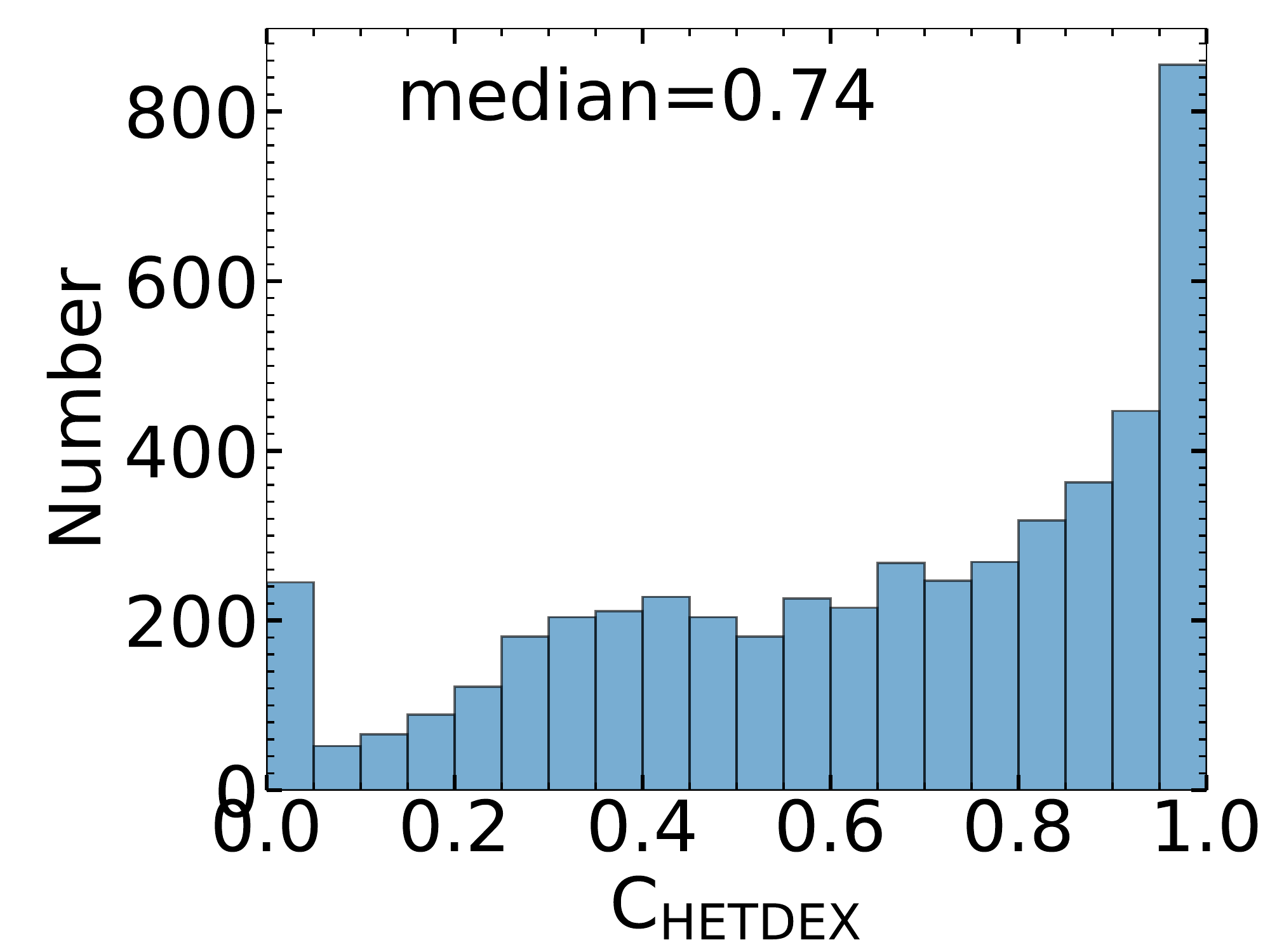}
\includegraphics[width=0.48\textwidth]{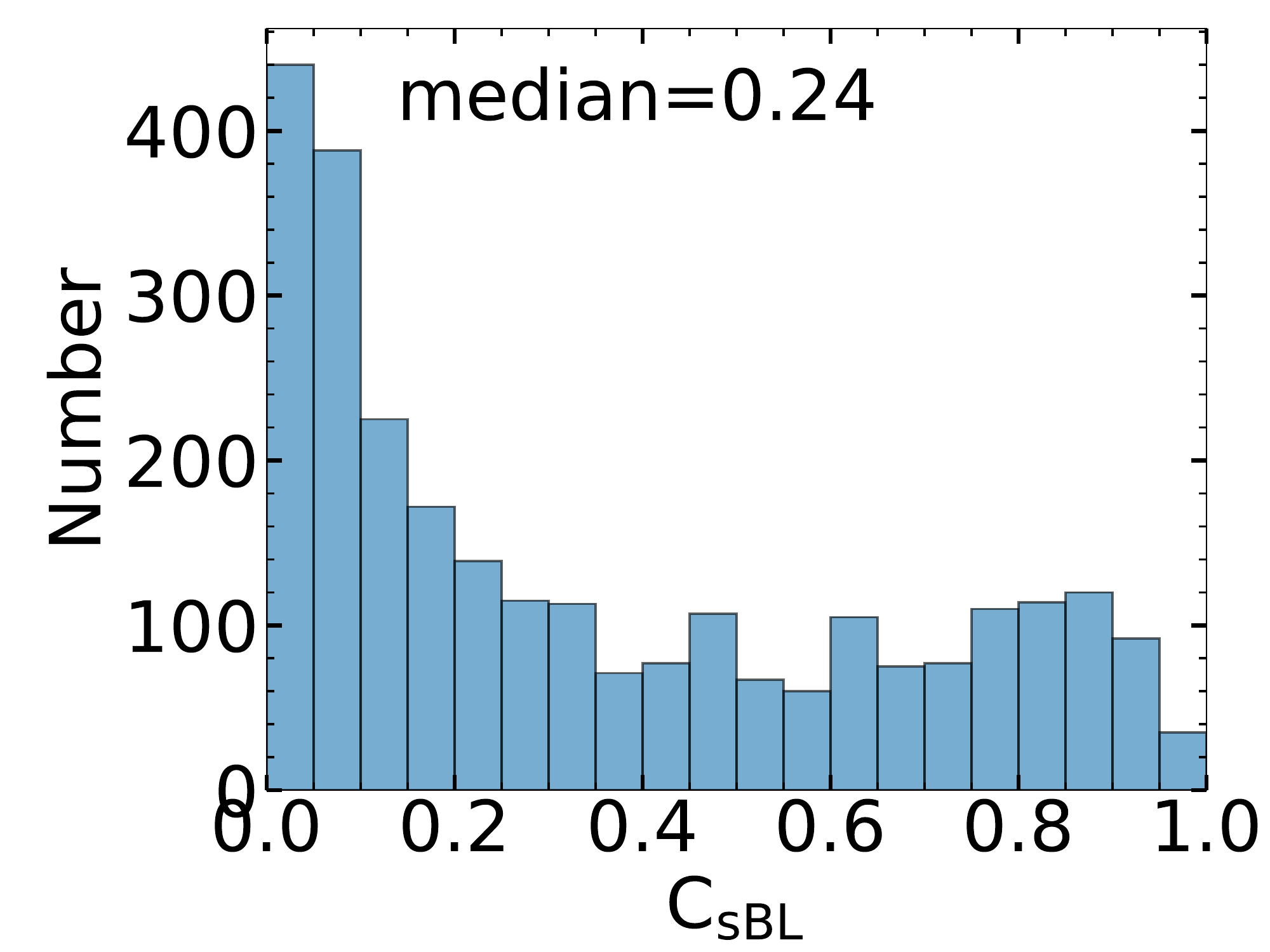}
\includegraphics[width=0.48\textwidth]{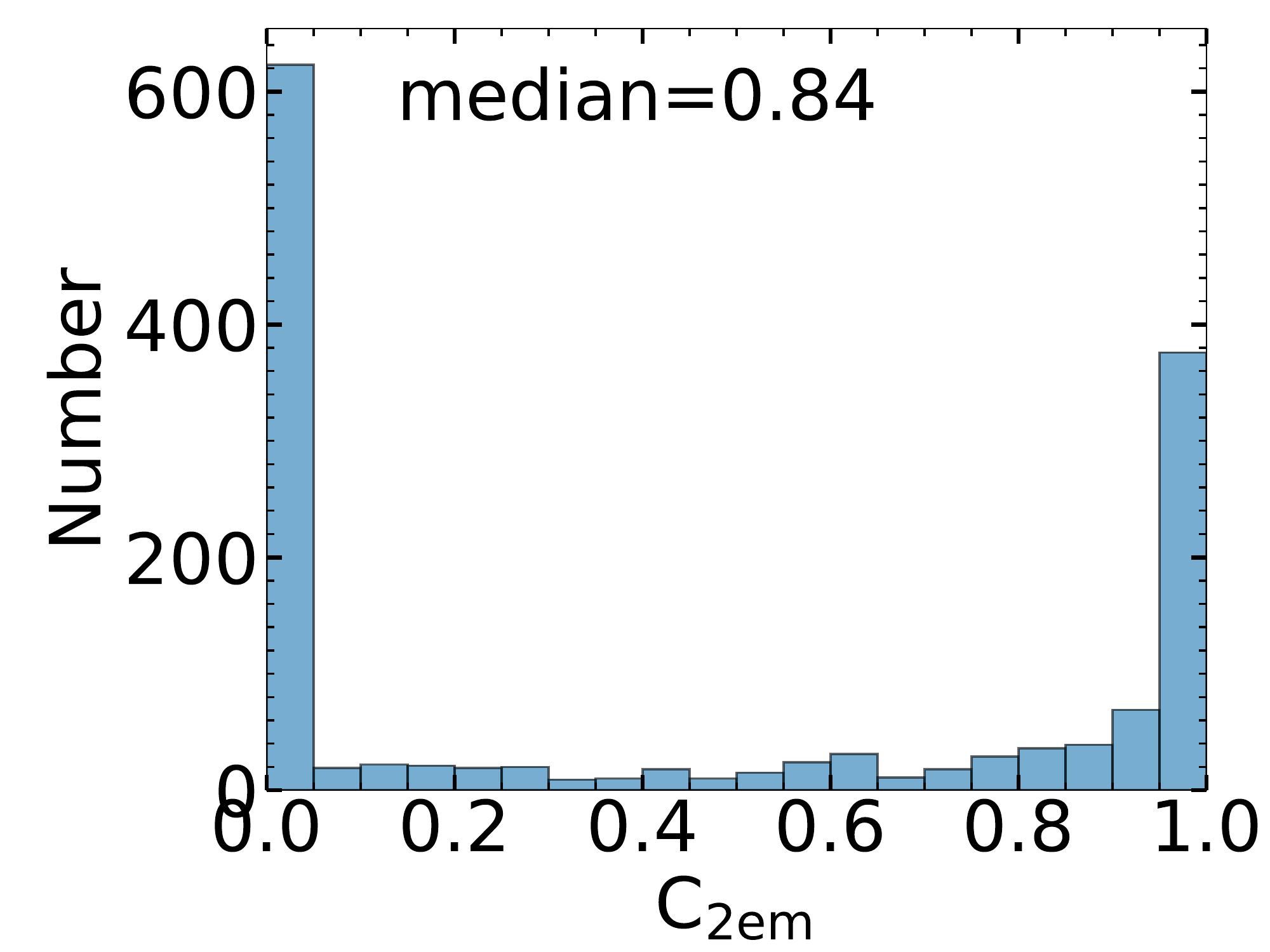}
\caption{Upper left: The distribution of the completeness of all HETDEX AGN\null. Upper right: Completeness distribution due to the HETDEX detection algorithm for all HETDEX AGN\null. Bottom left: Completeness distribution from the sBL algorithm for the sBL identified AGN\null. Bottom right:  Completeness distribution contributed from the 2em selection for the 2em identified AGN\null.}
\label{f_c_tot}
\end{figure}

We can combine the completeness curve of the HETDEX detection algorithm derived from simulations as described in Section \ref{sec_c_het}, with the AGN completeness functions of Section \ref{sec_c_agn} to compute the completeness applicable to every AGN in our catalog (i.e., Equation \ref{e_c_tot}). Figure \ref{f_c_tot} shows the distribution of the total completeness fraction in the upper left panel, and the distributions of various components of the completeness in the other three panels.  Note that in the upper left panel, there is a significant population of objects where completeness is smaller than 0.05. These are of lower confidence (larger completeness corrections) to be included in the statistics. Therefore, we require $\rm C>0.05$ in all LFs in the rest of this paper.



\section{The $\rm V_{max}$ method}
\label{sec_vmax}

All luminosity functions in this paper are calculated with the $\rm 1/V_{max}$ estimator \citep[e.g.][]{Schmidt1968,Blanc2011}. The space density in any luminosity bin can be calculated as:
\begin{equation}
  \Phi(\log_{10}{L}) = \frac{1}{\Delta \log_{10}{L}}\sum_{i}\frac{1}{\text{V}_{\text{max},i}},
\label{e_phi}
\end{equation}
where $\Delta \log_{10}{L}$ is the size of the bin, and $\text{V}_{\text{max},i}$ is the maximum comoving volume within which the $i$th object would still be detected by the survey.
\begin{equation}
\text{V}_{\text{max},i} = \omega\int_{z_{\text{min}}}^{z_{\text{max}}}\text{C}_i(z)\frac{d\text{V}}{dz}dz,
\label{e_vmax}
\end{equation}
where $\text{C}_i(z)$ is the completeness of the $i$th AGN estimated as Section \ref{sec_completeness}, $\omega$ is the total angular coverage of the sample which is 30.61 deg$^2$ for this paper, $\frac{dV}{dz}$ is the differential comoving volume element, and $z_{\text{min}}$ and $z_{\text{max}}$ are the minimum redshift and the maximum redshift.
The statistical uncertainty of $\Phi(\log_{10}{L})$ 
is estimated with the 2nd derivative of the poisson likelihood as follows,
\begin{equation}
  \sigma(\Phi(\log_{10}{L})) = \frac{1}{\Delta \log_{10}{L}}\sqrt{\sum_{i}\frac{1}{\text{V}_{\text{max},i}^2}}
\label{e_phi_er}
\end{equation}
\citep[e.g.][]{Johnston2011,Herenz2019}.

\section{The Emission-line LF of AGN}
\label{sec_LyA_LF}

\begin{figure}[htbp]
\centering
\includegraphics[width=\textwidth]{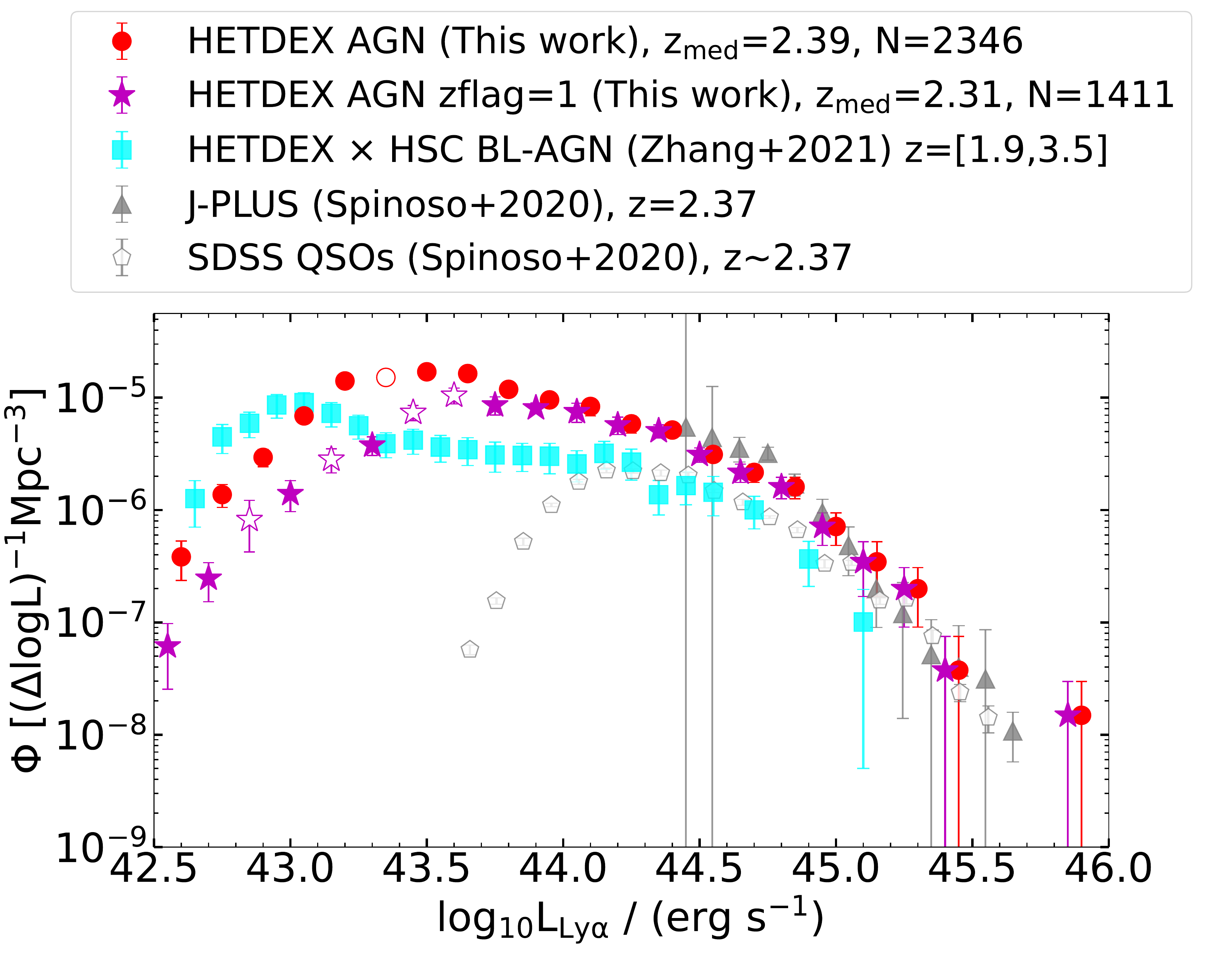}\\
\caption{The observed $\rm Ly\alpha$ LF for various AGN samples. Red circles: HETDEX AGN that have $\rm Ly\alpha$ emission within the wavelength range of the HETDEX survey ($1.88<z<3.53$). Magenta stars: HETDEX AGN with secure redshifts only ($zflag=1$) that have $\rm Ly\alpha$ emission within the experiment's wavelength range. The magenta stars are moved leftward by 0.05 dex for presentation purposes. The open red circles and the open magenta stars are the bins within which the completeness is lower than 0.15 (see Table \ref{t_LyA_LF_data} for the completeness of each luminosity bin).
Cyan squares: the $\rm Ly\alpha$ LF of the broad-line AGN selected from the overlap region between the HSC survey and the HETDEX survey in \cite{Zhang2021}. Black solid triangles: the $\rm Ly\alpha$ LF of the AGN in the J-PLUS survey 
at $z=2.37$ \citep{Spinoso2020}. Black open pentagons: the $\rm Ly\alpha$ LF of the SDSS DR14 QSOs \citep{Paris2018} at $z\sim 2.37$ estimated by \cite{Spinoso2020}. }
\label{f_LyA_LF}
\end{figure}

\begin{table*}[htbp]
\centering
\scriptsize
\begin{tabular}{c|cccccc|cccccc}
\hline\hline
 &  \multicolumn{6}{c|}{HETDEX AGN} & \multicolumn{6}{c}{HETDEX AGN $zflag=1$}\\\hline
$\rm \log_{10} L_{Ly\alpha}/(erg\,s^{-1})$ & $\rm \Phi$ & $\rm \sigma(\Phi)$ & N(AGN)$^\text{a}$ & C$^\text{b}$ & $\rm C_{HETDEX}^\text{c}$ & $\rm C_{AGN}^\text{d}$ & $\rm \Phi$ & $\rm \sigma(\Phi)$ & N(AGN)$^\text{a}$ & C$^\text{b}$ & $\rm C_{HETDEX}^\text{c}$ & $\rm C_{AGN}^\text{d}$ \\\hline
42.30 & 4.75e-07 & 2.42e-07 & 6 & 0.18 & 0.65 & 0.27 & 3.53e-07 & 2.25e-07 & 4 & 0.16 & 0.67 & 0.23 \\\hline
42.45 & 4.86e-08 & 3.44e-08 & 2 & 0.60 & 0.74 & 0.81 & 2.45e-08 & 2.45e-08 & 1 & 0.59 & 0.66 & 0.90 \\\hline
42.60 & 3.83e-07 & 1.47e-07 & 9 & 0.26 & 0.26 & 1.00 & 6.15e-08 & 3.61e-08 & 3 & 0.53 & 0.54 & 0.99 \\\hline
42.75 & 1.37e-06 & 3.16e-07 & 38 & 0.32 & 0.39 & 0.82 & 2.46e-07 & 9.38e-08 & 9 & 0.47 & 0.51 & 0.91 \\\hline
42.90 & 2.94e-06 & 5.13e-07 & 92 & 0.35 & 0.48 & 0.72 & 8.21e-07 & 3.98e-07 & 10 & 0.15 & 0.49 & 0.30 \\\hline
43.05 & 6.89e-06 & 9.70e-07 & 145 & 0.24 & 0.57 & 0.41 & 1.40e-06 & 4.29e-07 & 24 & 0.21 & 0.48 & 0.44 \\\hline
43.20 & 1.41e-05 & 1.53e-06 & 196 & 0.15 & 0.68 & 0.23 & 2.83e-06 & 6.84e-07 & 31 & 0.13 & 0.69 & 0.18 \\\hline
43.35 & 1.52e-05 & 1.47e-06 & 201 & 0.15 & 0.75 & 0.20 & 3.77e-06 & 7.08e-07 & 54 & 0.17 & 0.68 & 0.25 \\\hline
43.50 & 1.70e-05 & 1.62e-06 & 231 & 0.15 & 0.60 & 0.25 & 7.38e-06 & 1.13e-06 & 90 & 0.14 & 0.48 & 0.30 \\\hline
43.65 & 1.64e-05 & 1.86e-06 & 229 & 0.16 & 0.62 & 0.25 & 1.05e-05 & 1.69e-06 & 129 & 0.14 & 0.58 & 0.24 \\\hline
43.80 & 1.19e-05 & 1.67e-06 & 210 & 0.20 & 0.63 & 0.33 & 8.60e-06 & 1.58e-06 & 141 & 0.19 & 0.59 & 0.33 \\\hline
43.95 & 9.57e-06 & 9.44e-07 & 235 & 0.29 & 0.59 & 0.49 & 8.09e-06 & 8.71e-07 & 198 & 0.29 & 0.57 & 0.50 \\\hline
44.10 & 8.36e-06 & 1.47e-06 & 211 & 0.30 & 0.67 & 0.45 & 7.49e-06 & 1.46e-06 & 187 & 0.30 & 0.67 & 0.45 \\\hline
44.25 & 5.86e-06 & 1.02e-06 & 158 & 0.33 & 0.54 & 0.60 & 5.71e-06 & 1.02e-06 & 152 & 0.33 & 0.54 & 0.60 \\\hline
44.40 & 5.16e-06 & 6.71e-07 & 154 & 0.36 & 0.53 & 0.68 & 5.07e-06 & 6.70e-07 & 150 & 0.36 & 0.53 & 0.69 \\\hline
44.55 & 3.13e-06 & 4.55e-07 & 97 & 0.37 & 0.44 & 0.84 & 3.12e-06 & 4.54e-07 & 96 & 0.37 & 0.44 & 0.84 \\\hline
44.70 & 2.16e-06 & 4.02e-07 & 53 & 0.29 & 0.34 & 0.87 & 2.16e-06 & 4.02e-07 & 53 & 0.29 & 0.34 & 0.87 \\\hline
44.85 & 1.61e-06 & 3.47e-07 & 39 & 0.28 & 0.30 & 0.93 & 1.61e-06 & 3.47e-07 & 39 & 0.28 & 0.30 & 0.93 \\\hline
45.00 & 7.14e-07 & 2.30e-07 & 20 & 0.33 & 0.35 & 0.95 & 7.14e-07 & 2.30e-07 & 20 & 0.33 & 0.35 & 0.95 \\\hline
45.15 & 3.46e-07 & 1.76e-07 & 12 & 0.39 & 0.65 & 0.60 & 3.46e-07 & 1.76e-07 & 12 & 0.39 & 0.65 & 0.60 \\\hline
45.30 & 1.99e-07 & 1.09e-07 & 5 & 0.31 & 0.32 & 0.97 & 1.99e-07 & 1.09e-07 & 5 & 0.31 & 0.32 & 0.97 \\\hline
45.45 & 3.75e-08 & 3.75e-08 & 1 & 0.43 & 0.43 & 1.00 & 3.75e-08 & 3.75e-08 & 1 & 0.43 & 0.43 & 1.00 \\\hline
45.60 &    -- &      --  & 0 &   -- &   -- &   -- &       -- &      --  & 0 &   -- &   -- &   -- \\\hline
45.75 &    -- &      --  & 0 &   -- &   -- &   -- &       -- &      --  & 0 &   -- &   -- &   -- \\\hline
45.90 & 1.49e-08 & 1.49e-08 & 1 & 0.85 & 0.89 & 0.96 & 1.49e-08 & 1.49e-08 & 1 & 0.85 & 0.89 & 0.96 \\\hline
\hline
\end{tabular}
\caption{Binned $\rm Ly\alpha$ LF for HETDEX AGN and the sub-sample with secure redshifts only ($zflag=1$) in Figure \ref{f_LyA_LF}}
\begin{tablenotes}[flushleft]
\scriptsize
\item a. N(AGN) is the number of AGN in each luminosity bin. \\
\item b. C is the overall completeness in each luminosity bin: $\text{C}=\frac{\sum\limits_{i}{1/\text{V}_{\text{max},i,\text{uncor}}}}{\sum\limits_{i}{1/\text{V}_{\text{max},i}}}$, where $1/\text{V}_{\text{max},i}$ and $1/\text{V}_{\text{max},i,\text{uncor}}$ are the $\rm V_{max}$ of the $i$th AGN with and without completeness correction respectively. \\
\item c. $\rm C_{HETDEX}$ is the completeness in each luminosity bin contributed by the HETDEX pipeline (Section \ref{sec_c_het}).\\
\item d. $\rm C_{AGN}$ is the completeness in each luminosity bin contributed by the AGN selection (Section \ref{sec_c_agn}). 
\end{tablenotes}
\label{t_LyA_LF_data}
\end{table*}

In this section, we explore the observed LF of the Ly$\alpha$ emission-line associated with HETDEX AGN\null.  This line is the most abundant emission line at $z\sim2-3$. It is visible at $1.88<z<3.53$ within the survey's wavelength range (i.e., 3500\,\AA\ - 5500\,\AA). The median redshift of this AGN sample is 2.39.

\begin{figure*}[htbp]
\centering
\includegraphics[width=0.8\textwidth]{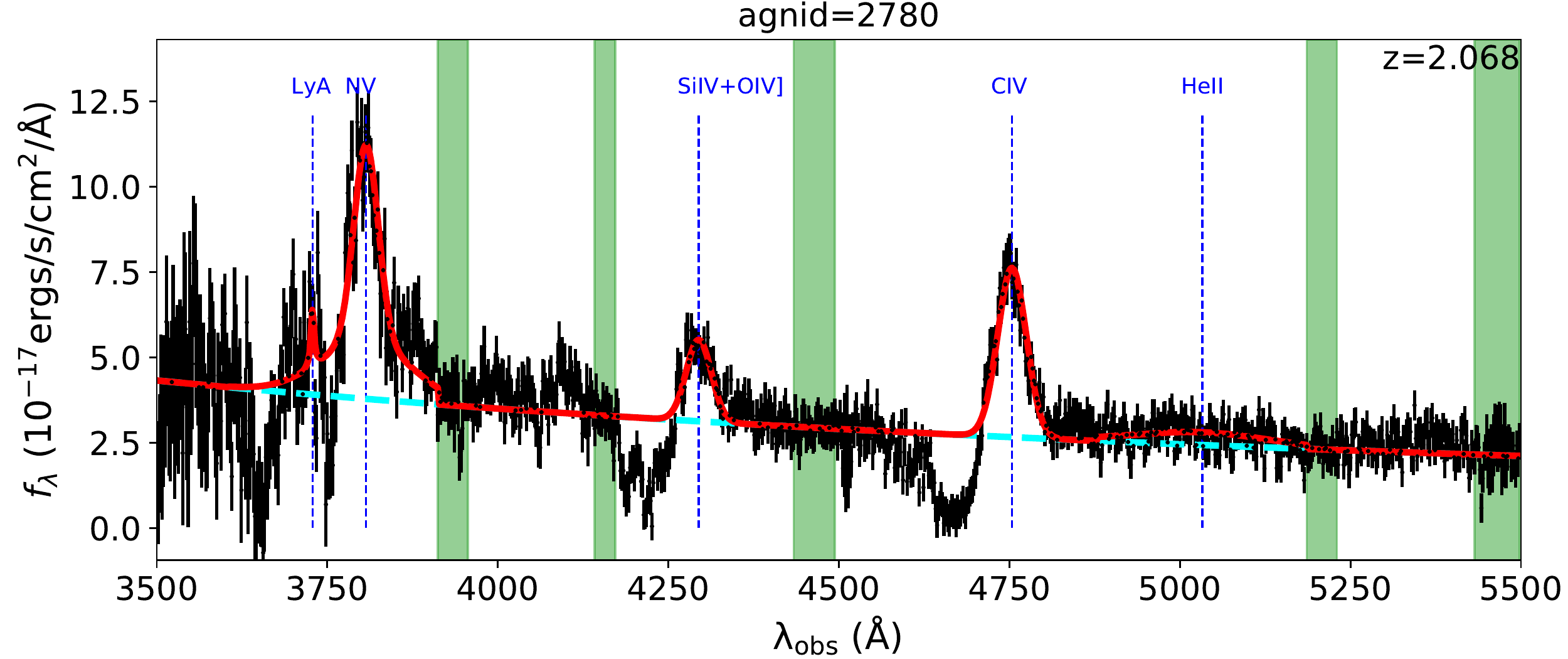}\\
\caption{An example of a 2em selected AGN ($agnid=2780$) with low observed $\rm Ly\alpha$ luminosity $\rm \log_{10} L_{Ly\alpha}/(erg\,s^{-1}) = 42.59$, but high completeness from our AGN selection ($\rm C_{2em}=1$ ) due to its high $\rm S/N_{spec}$ at 143.4. The low observed $\rm Ly\alpha$ luminosity is caused by heavy foreground absorption. The black data points with error bars are the observed flux and the associated error. The red solid line shows our best-fit model to the observed spectrum. The green shaded areas are the continuum windows used to fit the continuum. The cyan dashed line is our best-fit continuum model.}
\label{f_example}
\end{figure*}

Figure \ref{f_LyA_LF} shows the observed $\rm Ly\alpha$ LF of the HETDEX AGN (red circles) and that of the sub-sample with secure redshifts only (magenta stars) as well as those in other studies. 
Details of the $\rm Ly\alpha$ LF for HETDEX AGN in Figure \ref{f_LyA_LF}, the space density, the number of AGN, and the completeness correction of each luminosity bin, are listed in Table \ref{t_LyA_LF_data}. The completeness does not decrease with the observed $\rm Ly\alpha$ luminosity monotonically, as it is a combination of $\rm C_{HETDEX}$ and $\rm C_{AGN}$. $\rm C_{HETDEX}$ is not necessarily high for higher $\rm Ly\alpha$ luminosities as it not only increases with the emission line flux, but also decreases with the line widths (Figure \ref{f_c_het}, more details can be found in Section \ref{sec_c_het}). Lines with higher fluxes usually have larger line widths. $\rm C_{AGN}$ is also not necessarily high for AGN with higher $\rm Ly\alpha$ luminosity. For the sBL selected AGN, similar to $\rm C_{HETDEX}$, $\rm C_{sBL}$ is higher for narrower emission lines as shown in Figure \ref{f_c_sBL}. For 2em selected AGN, the situation is more complicated as an AGN with low observed $\rm Ly\alpha$ luminosity can have high $\rm S/N_{spec}$ because the $\rm Ly\alpha$ emission line is heavily contaminated by foreground absorption while $\rm S/N_{spec}$ catches the other strong emission lines and the strong continuum as the signal across 3500\,\AA\ - 5500\,\AA. Figure \ref{f_example} shows a typical example: $agnid=2780$ has the observed $\rm Ly\alpha$ luminosity measured 
as $\rm \log_{10} L_{Ly\alpha}/(erg\,s^{-1}) = 42.59$, while its $\rm S/N_{spec}$ is 143.4, which is high enough to locate it in the $\rm C_{2em}=1$ region in Figure \ref{f_c_2em}. It is hard to reliably recover the intrinsic $\rm Ly\alpha$ luminosity for such AGN. We decide to use their observed $\rm Ly\alpha$ luminosity in our $\rm Ly\alpha$ LF study.

Figure \ref{f_LyA_LF} shows that our $\rm Ly\alpha$ LF of the HETDEX AGN (red circles) agrees well with that of the AGN in the Javalambre Photometric Local Universe Survey \citep[J-PLUS;][]{Cenarro2019} (black solid triangles) at the bright end ($\rm Ly\alpha \gtrsim 10^{44.5}$ erg/s) .
\cite{Spinoso2020} extended the $\rm Ly\alpha$ LF at $2.2\lesssim z\lesssim3.3$ above $\rm Ly\alpha\sim 10^{44}$ erg/s for the first time with J-PLUS.
\cite{Spinoso2020} also estimated the $\rm Ly\alpha$ LF of the SDSS DR14 QSOs \citep{Paris2018} (black open pentagons) by performing synthetic photometry of SDSS QSOs in J-PLUS footprint with J-PLUS filters for comparison.  The LF of the SDSS QSOs shows a strong decline of the densities for fainter QSOs at $\rm Ly\alpha \sim 10^{43.6} - 10^{44.1}$ erg/s, while the space density of the HETDEX AGN continues to increase for AGN within this luminosity range. This might be caused by the over-estimated completeness of the SDSS QSOs in this luminosity range \citep[see][for related discussions]{Kulkarni2019}.

\cite{Zhang2021} combined the HETDEX spectra with the HSC $r$-band imaging and identified broad-line AGN independently. The overlap region between the two surveys is only 11.4 deg$^2$ (less than 40\% of the present study). Moreover, \cite{Zhang2021}  extracted spectra from the HETDEX database at the positions of HSC $r$-band sources detected at more than $5\sigma$ significance; the emission line signals were then identified with $\rm S/N>5.5$ and characterized as narrow-line emitters or broad-line AGN\null. Our AGN selection is purely based on blind spectroscopic observations, with no requirements for the continuum strengths based on photometric surveys, so our catalog is more complete for the AGN with faint continuum levels. In addition, there is no cut on the line widths in our 2em selection, so our AGN catalog additionally contains a number of narrow-line AGN.

There is a strong discrepancy between the AGN LF of this work (red circles) and that of \cite{Zhang2021} (cyan squares) at intermediate luminosities ($\rm Ly\alpha \sim 10^{43}-10^{44}$ erg/s) in Figure \ref{f_LyA_LF}: the $\rm Ly\alpha$ LF from \cite{Zhang2021} has a dip compared to that of ours. 
This discrepancy can be explained by our improved modelling of the completeness, which includes the effects of the AGN selection. When we remove the completeness correction from our AGN selection, and only use the completeness from the HETDEX pipeline, our $\rm Ly\alpha$ LF agrees well with that of \cite{Zhang2021} in this intermediate luminosity region. 

At $\rm Ly\alpha \lesssim 10^{43}$ erg/s, our LF is slightly lower compared to that of \cite{Zhang2021}.
This could result from the different ways of measuring FHWMs of emission lines. \cite{Zhang2021} measure the FWHM using a single Gaussian model while we measure the FWHM with our best-fitting multi-Gaussian model (Paper I). We find that the FWHM of $\rm Ly\alpha$ can be easily over-estimated with a single Gaussian model as the $\rm Ly\alpha$ emission region is usually complicated due to strong foreground narrow absorption lines, asymmetric profiles, blends with the nearby \ion{N}{5} $\lambda1241$ emission, etc. Our multi-Gaussian fit can better handle these effects. In addition, there is a significant fraction of low-luminosity AGN identified by \cite{Zhang2021} that have FWHMs lower than 1000 km/s in our multi-Gaussian fits.  
Since \cite{Zhang2021} focused on bridging between normal LAEs and bright AGNs,  as long as there is strong emission line in the spectrum, it did not matter whether the object was called a broad-line AGN or a normal LAE. These low line-width objects are rejected in our AGN selection. 
Finally, the \cite{Zhang2021} analysis assumes all broad, single-line detections in the HETDEX survey to be Ly$\alpha$.  Their luminosity function may therefore suffer contamination from lower-redshift \ion{C}{4} $\lambda 1549$, \ion{C}{3}] $\lambda 1909$ emitters or \ion{Mg}{2} $\lambda 2799$ emitters.
 


\begin{table}[htbp]
\centering
\begin{tabular}{c|c|c|c|c}
\hline\hline

sample    & $L_{\rm F}^*$ & $\Phi^*_{L,\rm F}$  & $L_{\rm B}^*$ & $\Phi^*_{L,\rm B}$ \\\hline
all AGN   & 2.37E+43  & 2.14E-5         & 3.06E+44  & 4.57E-6       \\\hline
$zflag=1$ & 3.86E+43  & 1.23E-5         & 4.32E+44  & 3.26E-6       \\\hline
\hline
\end{tabular}
\caption{The best-fit turnover points of the triple-power law profile for the HETDEX AGN.}
\label{t_LyA_LF}
\end{table}

We additionally fit the $\rm Ly\alpha$ LF of our HETDEX AGN sample at $z\sim2$ with a triple-power law model.
There are two turnover points in the triple-power fit: the break point between the bright end and the intermediate luminosities ($L_{\rm B}^*$, $\Phi^*_{L,\rm B}$) and the turnover point between the intermediete luminosities and the faint end ($L_{\rm F}^*$, $\Phi^*_{L,\rm F}$). We list our best-fit turnover points in Table \ref{t_LyA_LF}. In both LFs, the space density of AGN is highest at $L_{\rm F}^*$ 
with opposite slopes on both sides. We will further discuss the potential under-estimated incompleteness 
of the decreased space density for the faintest AGN ($L<L_{\rm F}^*$) in Section \ref{sec_discuss}.



\begin{figure}[htbp]
\centering
\includegraphics[width=\textwidth]{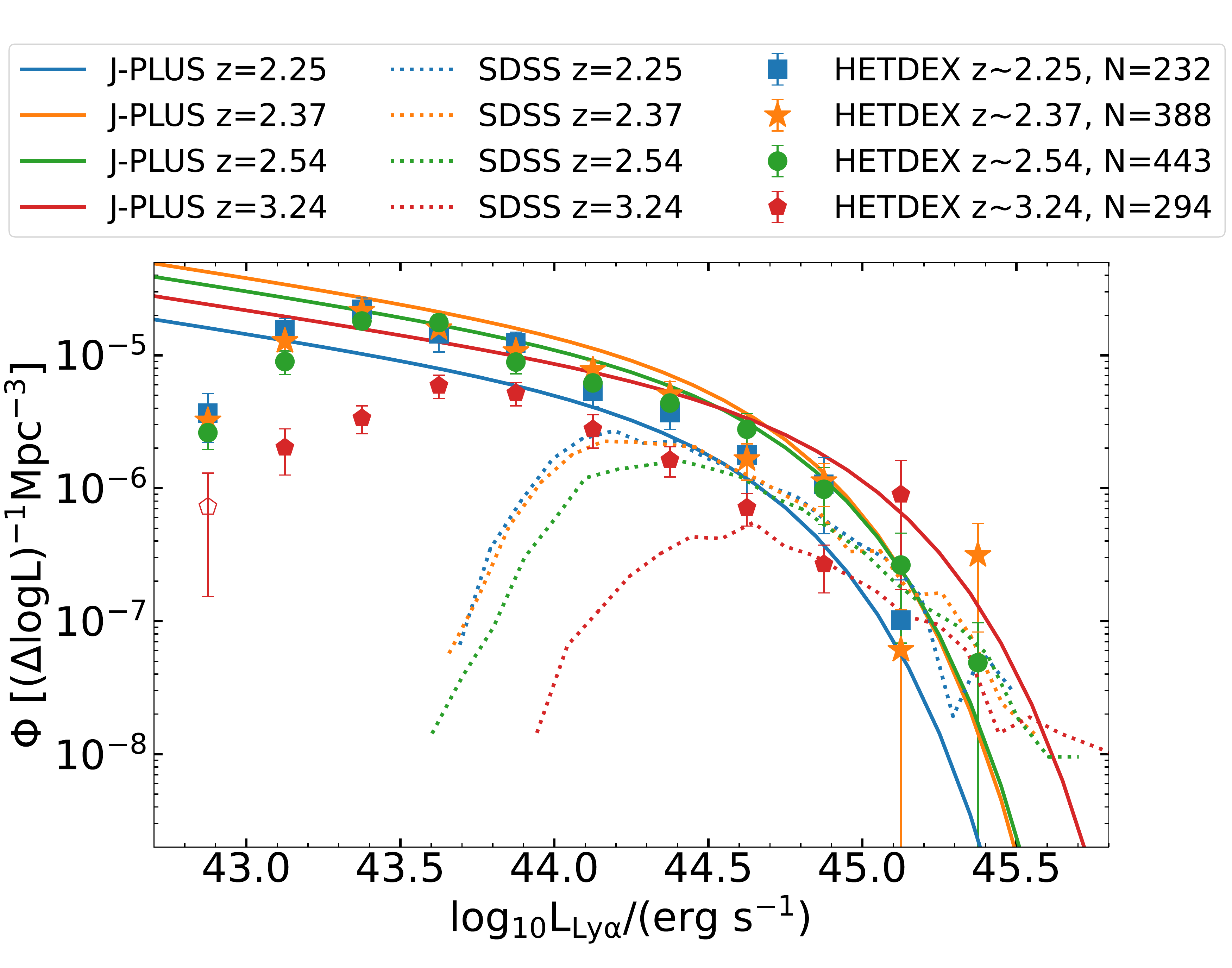}\\
\caption{The $\rm Ly\alpha$ luminosity function at $z\sim$ 2.25 (blue), 2.37 (orange), 2.54 (green), and 3.24 (red). Data points shown as blue squares, orange stars, green circles, and red pentagons are for the HETDEX AGN at $z\sim$2.25, 2.37, 2.54, and 3.24, respectively. Open data points are again for the luminosity bins within which the completeness is lower than 0.15. The solid lines are the best-fit model for the AGN in the J-PLUS survey \citep{Spinoso2020}. The dotted lines are for QSOs in the SDSS survey. The LFs of the SDSS QSOs are also taken from the estimation in \cite{Spinoso2020}.}
\label{f_LyA_LF_z}
\end{figure}

In Figure \ref{f_LyA_LF_z} we show the evolution of the $\rm Ly\alpha$ LF of the HETDEX AGN at $z\sim$ 2.25 (blue), 2.37 (orange), 2.54 (green), and 3.24 (red) and compare our results with those of the J-PLUS and SDSS QSOs surveys \citep{Spinoso2020}. The four redshifts have been chosen to facilitate comparison with the four bluest narrow bands of the J-PLUS survey. There are 232, 388, 443, and 294 HETDEX AGN in the redshift bins of [2.20, 2.29), [2.29 - 2.46), [2.46 - 2.70), and [2.95 - 4.25).
The J-PLUS AGN LFs only have data brighter than $\rm Ly\alpha \sim 10^{43.3}$ erg~s$^{-1}$, and in this range, our HETDEX AGN LFs agrees well with their LFs at all four redshift bins (within $1\sigma$ confidence regions of their LFs, see Figure 14 in \cite{Spinoso2020} for their confidence regions). At fainter luminosities ($\rm Ly\alpha \lesssim 10^{43.3}$~ergs~s$^{-1}$), there is a strong discrepancy between the HETDEX and J-PLUS LFs. This is not surprising since for these luminosities, 
the J-PLUS LFs are only an extrapolation to their best-fit model and contain no actual observations. Therefore, the J-PLUS survey may not properly reflect the faint end of the AGN LF\null.  Similarly, the SDSS QSOs also display positive faint end slopes at all four redshifts, but a brighter inflection point ($\rm L^*$). This could easily be caused by an overestimate of the completeness for the lower luminosity SDSS QSOs.

The three lower redshifts ($z\sim$ 2.25, 2.37, 2.54) span only $\sim 0.2$~Gyr of cosmic time, so the lack of strong evolution among the bins does not necessarily mean a lack of the evolution in the LF\null. In fact, in the the highest redshift ($z\sim$ 3.24) bin, the space density of AGN is significantly lower than that at the other three redshifts, especially at the faint end. A similar trend can also be found with the SDSS QSOs. This might indicate that the AGN density is lower at higher redshifts. The possible evolution of the LFs of the AGN should be better explored with more separated redshifts in a wider redshift range.



\section{UV LF as a function of redshift}
\label{sec_UV_LF_z}


$\rm Ly\alpha$ is only visible within a relatively small redshift range  ($1.88<z<3.53$), as the wavelength coverage of HETDEX only extends from 3500\,\AA\ to 5500\,\AA\null.  However, the full HETDEX AGN catalog covers $0.25<z<4.2$ via the detection of other emission lines.  For these objects, we can measure the luminosity function of the AGN UV continuum.  In this paper, we choose the monochromatic luminosity of the power-law (PL) continuum at 1450\,\AA\ ($\rm M_{1450}$) to study the evolution of the AGN LF. For AGN with the rest-frame 1450\,\AA\ out of the wavelength range of the HETDEX survey, we extrapolate the best-fit PL continuum and get an estimation of $\rm M_{1450}$. 

\begin{figure}[htbp]
\centering
\includegraphics[width=\textwidth]{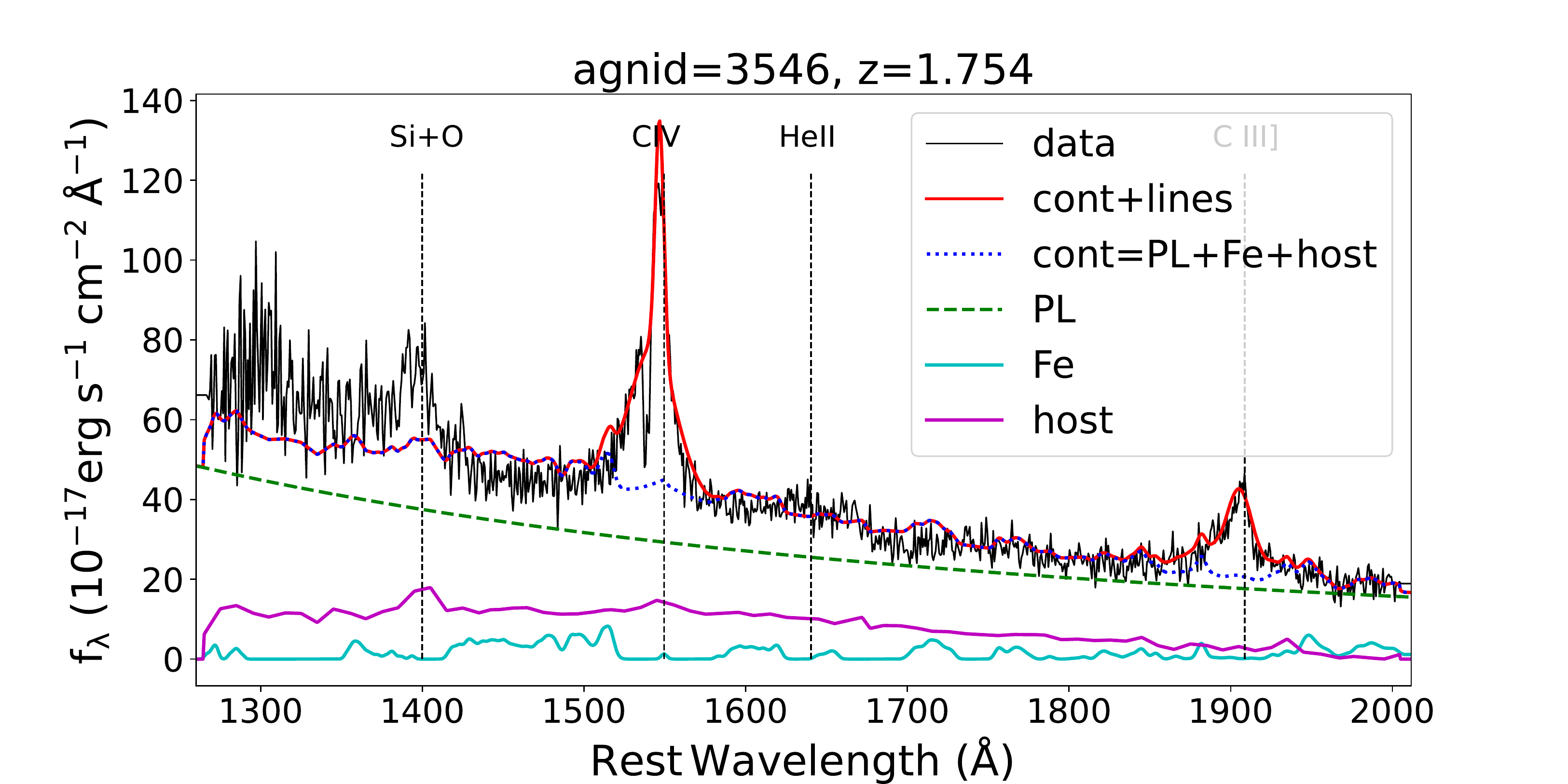}
\caption{An example of the decomposition of the continuum using {\tt PyQSOFit}. The black spectrum is the HETDEX spectrum of {\tt agnid}\,$=3546$. The continuum (blue dotted line) has three components: the PL continuum contributed by AGN (green dashed line); the continuum of the host galaxy (magenta solid line), and the \ion{Fe}{2} emission (cyan solid line). The final fitted spectrum by {\tt PyQSOFit} is shown by the red solid spectrum including both emission lines and the continuum.}
\label{f_pyqsofit}
\end{figure}

\begin{figure*}[htbp]
\centering
\includegraphics[width=0.48\textwidth]{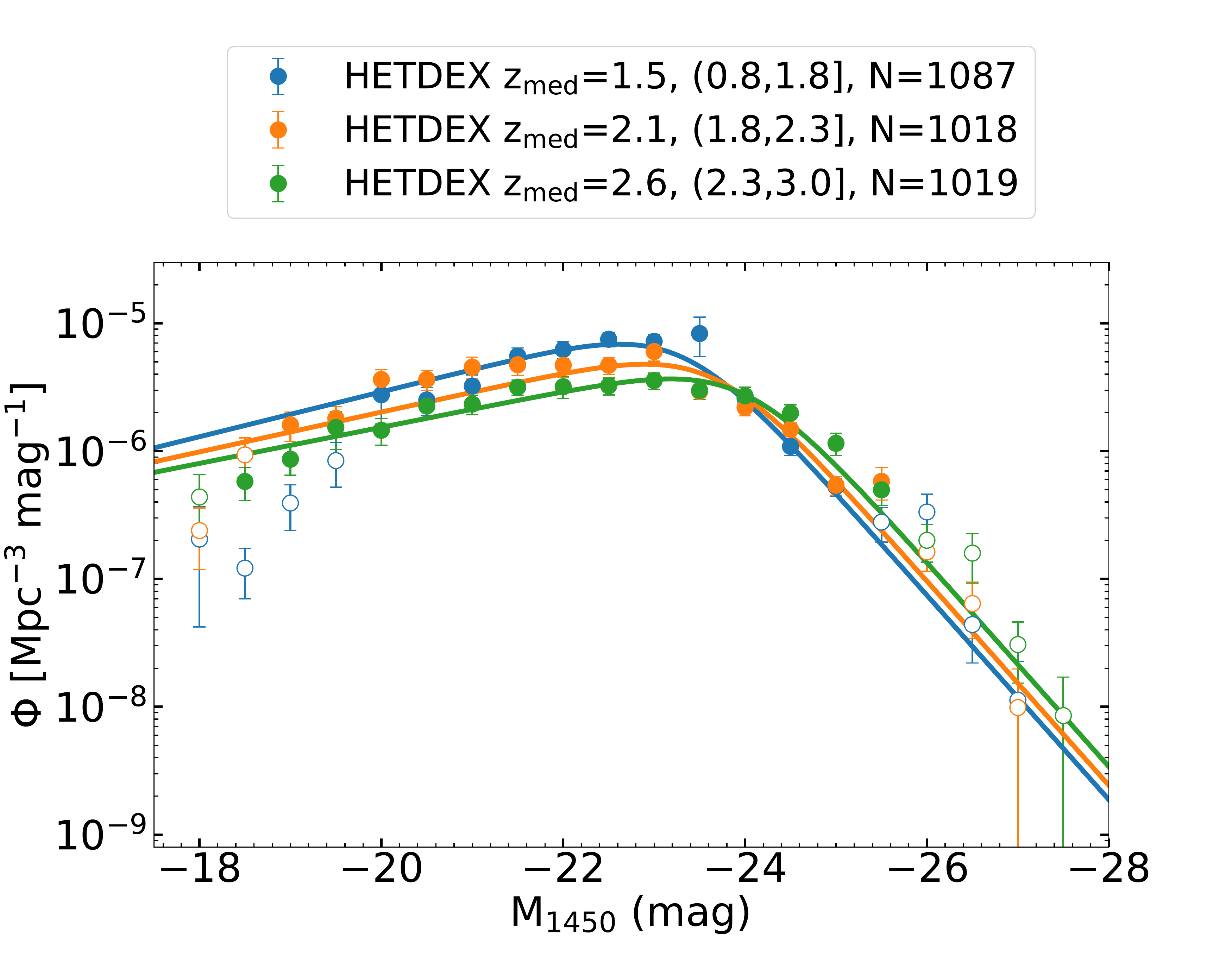}
\includegraphics[width=0.48\textwidth]{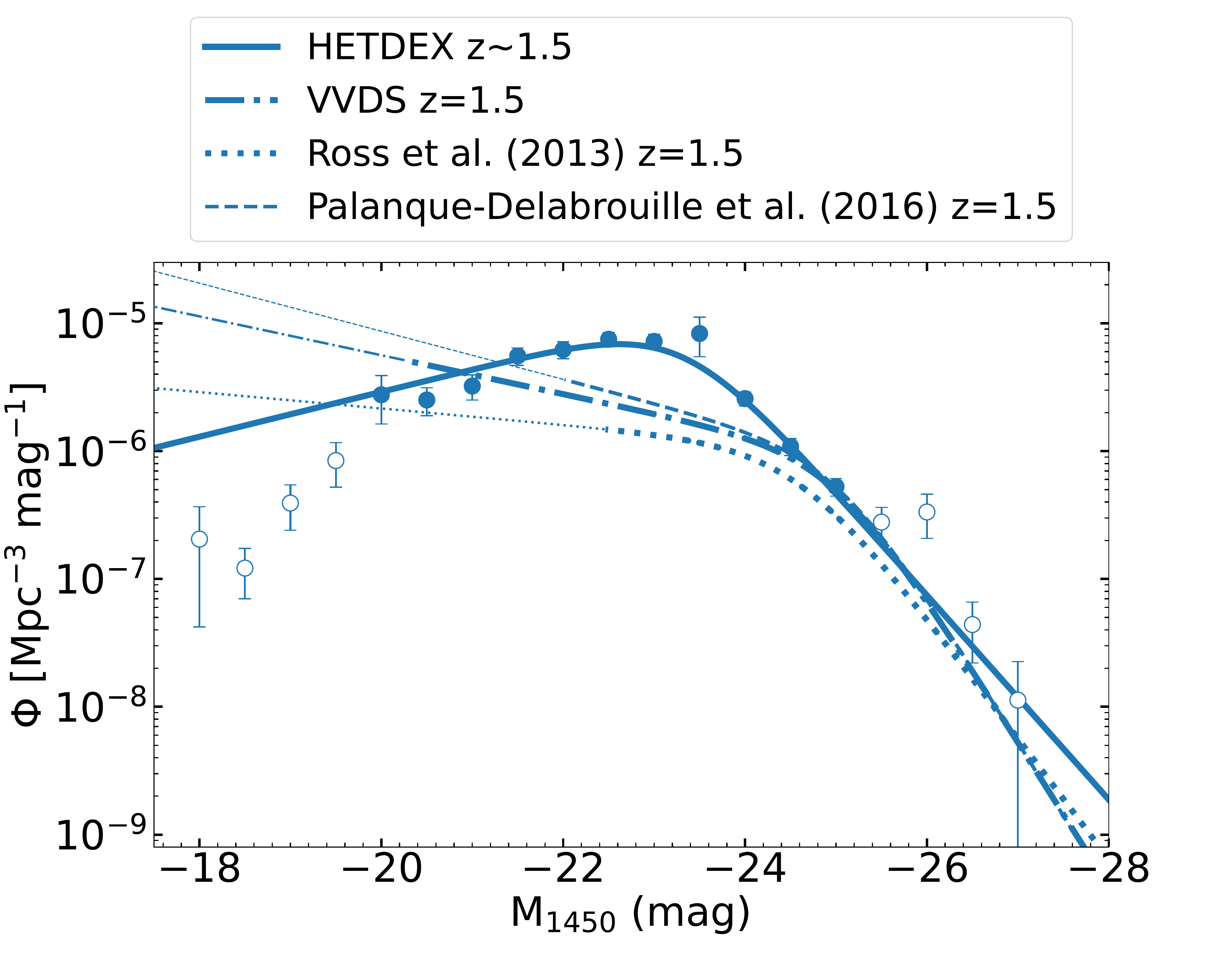}
\includegraphics[width=0.48\textwidth]{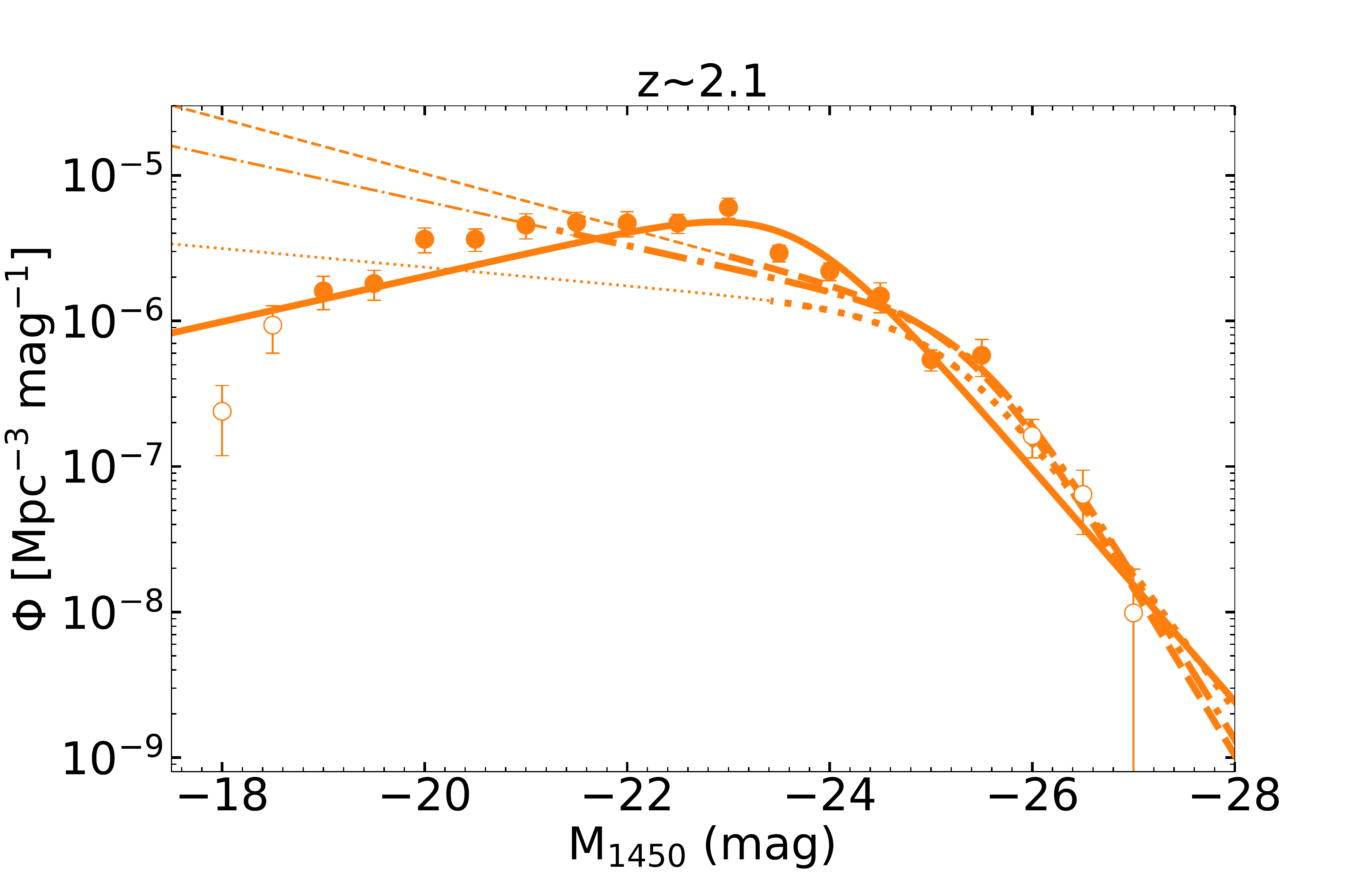}
\includegraphics[width=0.48\textwidth]{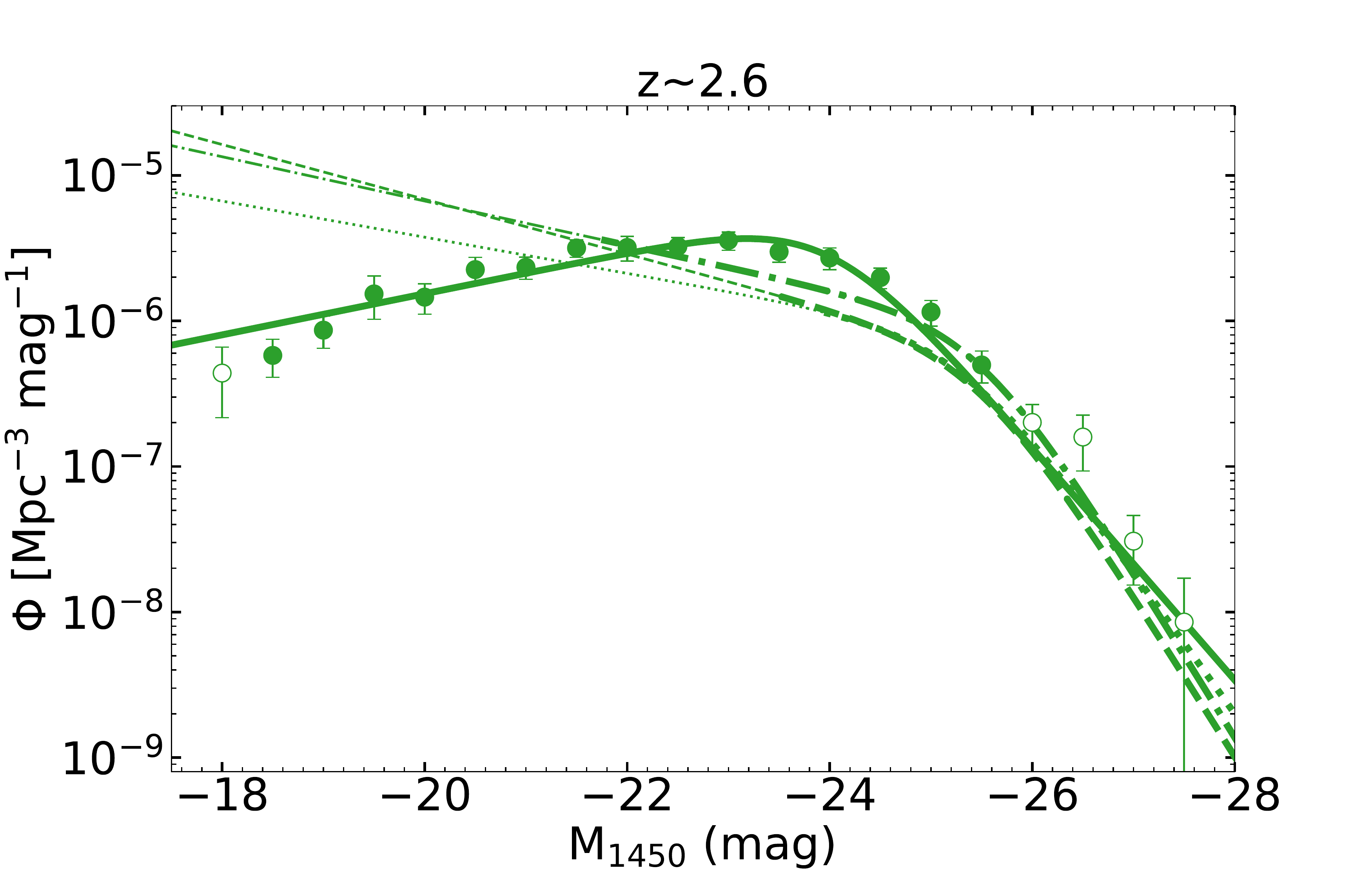}
\caption{The UV LF of AGN as measured from the power-law component of the continuum at 1450\,\AA\ ($\rm M_{1450}$). Upper left: The data from the HETDEX AGN in three redshift bins: $0.8<z\leq1.8$ (blue), $1.8<z\leq2.3$ (orange), and $2.3<z\leq3.0$ (green). There are 1087, 1018, and 1019 AGN in each redshift interval, with median redshifts of $z=$ 1.5, 2.1, and 2.6, respectively. Points containing fewer than 20 AGN or the binned completeness lower than 0.15 are shown as open dots and are not included in our fit. The three lines show our best-fit LEDE model.  Upper right: Blue data points and the blue solid line are the same with those in the upper left panel. The dotted line is the best-fit evolution model for color selected $z=1.5$ QSOs in SDSS DR9 \citep{Ross2013}. The lines are drawn thicker in regions covered by their data, and the lines are drawn thin for the extrapolations of the best-fit model. The dashed line is similar to the dotted line, but for the variability-selected AGN from SDSS DR9 \citep{PD2016}. The dash-dotted line is also similar to the dotted line, but for the AGN of the VVDS survey \citep{Bongiorno2007A&A}. The bottom left and the bottom right panels are similar to the upper right panel, but at $z=$ 2.1 and 2.6, respectively. We note that both \cite{Ross2013} and \cite{PD2016} fitted the evolution of the AGN LF with the PLE model at $z<2.2$ and switched to the LEDE model at $z>2.2$. 
}
\label{f_M1450_z}
\end{figure*}

\begin{table}[htbp]
\centering
\scriptsize
\begin{tabular}{c|cccccc}

\hline

\hline
 \multicolumn{7}{c}{$z\sim1.5$} \\\hline 
 
 \hline
$\rm M_{1450}$ & $\rm \Phi$ & $\rm \sigma(\Phi)$ & N(AGN) & C & $\rm C_{HETDEX}$ & $\rm C_{AGN}$ \\\hline
-27.0 & 1.13e-08 & 1.13e-08 & 1 & 1.00 & 1.00 & 1.00 \\\hline
-26.5 & 4.40e-08 & 2.20e-08 & 4 & 1.00 & 1.00 & 1.00 \\\hline
-26.0 & 3.34e-07 & 1.26e-07 & 12 & 0.37 & 0.38 & 0.98 \\\hline
-25.5 & 2.78e-07 & 8.43e-08 & 18 & 0.62 & 0.66 & 0.94 \\\hline
-25.0 & 5.28e-07 & 8.34e-08 & 46 & 0.83 & 0.88 & 0.94 \\\hline
-24.5 & 1.09e-06 & 1.64e-07 & 76 & 0.76 & 0.83 & 0.92 \\\hline
-24.0 & 2.58e-06 & 3.20e-07 & 137 & 0.52 & 0.66 & 0.79 \\\hline
-23.5 & 8.32e-06 & 2.85e-06 & 173 & 0.19 & 0.62 & 0.32 \\\hline
-23.0 & 7.22e-06 & 9.70e-07 & 172 & 0.24 & 0.59 & 0.41 \\\hline
-22.5 & 7.51e-06 & 9.70e-07 & 129 & 0.21 & 0.59 & 0.36 \\\hline
-22.0 & 6.23e-06 & 9.46e-07 & 96 & 0.19 & 0.54 & 0.34 \\\hline
-21.5 & 5.56e-06 & 8.66e-07 & 81 & 0.15 & 0.54 & 0.28 \\\hline
-21.0 & 3.23e-06 & 7.20e-07 & 47 & 0.17 & 0.50 & 0.35 \\\hline
-20.5 & 2.51e-06 & 6.22e-07 & 34 & 0.16 & 0.45 & 0.35 \\\hline
-20.0 & 2.76e-06 & 1.13e-06 & 30 & 0.15 & 0.47 & 0.33 \\\hline
-19.5 & 8.44e-07 & 3.21e-07 & 12 & 0.20 & 0.56 & 0.36 \\\hline
-19.0 & 3.93e-07 & 1.52e-07 & 10 & 0.25 & 0.58 & 0.43 \\\hline
-18.5 & 1.22e-07 & 5.16e-08 & 6 & 0.49 & 0.59 & 0.83 \\\hline
-18.0 & 2.05e-07 & 1.63e-07 & 3 & 0.15 & 0.67 & 0.23 \\\hline

\hline
 \multicolumn{7}{c}{$z\sim2.1$} \\\hline
 
 \hline
-27.0 & 9.86e-09 & 9.86e-09 & 1 & 1.00 & 1.00 & 1.00 \\\hline
-26.5 & 6.42e-08 & 3.01e-08 & 5 & 0.83 & 0.83 & 1.00 \\\hline
-26.0 & 1.63e-07 & 4.79e-08 & 13 & 0.86 & 0.86 & 0.99 \\\hline
-25.5 & 5.80e-07 & 1.66e-07 & 25 & 0.47 & 0.56 & 0.84 \\\hline
-25.0 & 5.42e-07 & 8.97e-08 & 42 & 0.83 & 0.83 & 1.00 \\\hline
-24.5 & 1.48e-06 & 3.43e-07 & 69 & 0.52 & 0.63 & 0.83 \\\hline
-24.0 & 2.20e-06 & 3.07e-07 & 104 & 0.50 & 0.66 & 0.76 \\\hline
-23.5 & 2.93e-06 & 3.85e-07 & 121 & 0.44 & 0.65 & 0.67 \\\hline
-23.0 & 6.01e-06 & 9.51e-07 & 128 & 0.24 & 0.57 & 0.42 \\\hline
-22.5 & 4.70e-06 & 7.11e-07 & 86 & 0.21 & 0.60 & 0.34 \\\hline
-22.0 & 4.70e-06 & 9.17e-07 & 74 & 0.17 & 0.62 & 0.28 \\\hline
-21.5 & 4.74e-06 & 8.44e-07 & 64 & 0.15 & 0.56 & 0.27 \\\hline
-21.0 & 4.55e-06 & 8.90e-07 & 71 & 0.17 & 0.59 & 0.29 \\\hline
-20.5 & 3.64e-06 & 6.40e-07 & 68 & 0.21 & 0.64 & 0.33 \\\hline
-20.0 & 3.64e-06 & 7.05e-07 & 56 & 0.17 & 0.56 & 0.31 \\\hline
-19.5 & 1.81e-06 & 4.21e-07 & 41 & 0.26 & 0.57 & 0.45 \\\hline
-19.0 & 1.61e-06 & 4.13e-07 & 25 & 0.18 & 0.45 & 0.39 \\\hline
-18.5 & 9.35e-07 & 3.35e-07 & 17 & 0.21 & 0.61 & 0.34 \\\hline
-18.0 & 2.39e-07 & 1.21e-07 & 8 & 0.38 & 0.75 & 0.51 \\\hline

\hline
 \multicolumn{7}{c}{$z\sim2.6$} \\\hline 
 
\hline
-27.0 & 3.07e-08 & 1.54e-08 & 4 & 0.96 & 1.00 & 0.96 \\\hline
-26.5 & 1.59e-07 & 6.63e-08 & 10 & 0.49 & 0.51 & 0.96 \\\hline
-26.0 & 2.01e-07 & 6.54e-08 & 15 & 0.62 & 0.66 & 0.93 \\\hline
-25.5 & 4.98e-07 & 1.23e-07 & 29 & 0.50 & 0.55 & 0.91 \\\hline
-25.0 & 1.15e-06 & 2.30e-07 & 51 & 0.37 & 0.51 & 0.73 \\\hline
-24.5 & 1.98e-06 & 3.20e-07 & 80 & 0.37 & 0.49 & 0.75 \\\hline
-24.0 & 2.71e-06 & 4.61e-07 & 102 & 0.36 & 0.60 & 0.60 \\\hline
-23.5 & 2.99e-06 & 4.59e-07 & 88 & 0.28 & 0.56 & 0.50 \\\hline
-23.0 & 3.57e-06 & 5.11e-07 & 109 & 0.26 & 0.68 & 0.38 \\\hline
-22.5 & 3.24e-06 & 4.91e-07 & 90 & 0.21 & 0.75 & 0.29 \\\hline
-22.0 & 3.19e-06 & 6.17e-07 & 83 & 0.21 & 0.77 & 0.27 \\\hline
-21.5 & 3.17e-06 & 4.36e-07 & 94 & 0.23 & 0.71 & 0.33 \\\hline
-21.0 & 2.33e-06 & 4.03e-07 & 71 & 0.24 & 0.77 & 0.31 \\\hline
-20.5 & 2.25e-06 & 4.80e-07 & 61 & 0.21 & 0.68 & 0.31 \\\hline
-20.0 & 1.46e-06 & 3.44e-07 & 45 & 0.23 & 0.68 & 0.33 \\\hline
-19.5 & 1.53e-06 & 5.05e-07 & 26 & 0.16 & 0.72 & 0.21 \\\hline
-19.0 & 8.64e-07 & 2.15e-07 & 30 & 0.26 & 0.64 & 0.40 \\\hline
-18.5 & 5.79e-07 & 1.69e-07 & 24 & 0.31 & 0.67 & 0.46 \\\hline
-18.0 & 4.38e-07 & 2.22e-07 & 6 & 0.10 & 0.74 & 0.14 \\\hline

\hline
\end{tabular}
\caption{Binned $\rm M_{1450}$ LF for HETDEX AGN in Figure \ref{f_M1450_z}}
\begin{tablenotes}[flushleft]
\scriptsize
\item Note: N(AGN), C, $\rm C_{HETDEX}$, and $\rm C_{AGN}$ are calculated in the same way as Table \ref{t_LyA_LF_data}. 
\end{tablenotes}
\label{t_UV_LF_data}
\end{table}

The UV continuum of an AGN contains PL emission from the AGN itself, starlight from the AGN's host galaxy, and broad \ion{Fe}{2} emission from the AGN's broad-line region. In most studies, the continuum of bright QSOs is dominated by the AGN component, so the broad-band magnitude can be directly adopted to represent the AGN luminosity without decomposition. However, the AGN in our study are emission-line selected, their continua may be much fainter than those of continuum-selected objects.  For these low-luminosity AGN, the contribution from the host galaxy cannot be ignored.  To determine the host galaxy contribution, we fit each AGN's spectrum with the publicly available code \texttt{PyQSOFit} \citep{Guo2018}. This program fits the continuum from the host galaxy using the stellar library of \cite{BC03}, and \ion{Fe}{2} emission is modeled using a series of templates for different rest-frame wavelength: \cite{Vestergaard2001} for 1000\,\AA\ - 2200\,\AA, \cite{Salviander2007} for 2200\,\AA\ - 3090\,\AA, \cite{Tsuzuki06} for 3090\,\AA\ - 3500\,\AA, and \cite{Boroson1992} for the longer optical wavelengths. Figure \ref{f_pyqsofit} shows one example of a {\tt PyQSOFit} modeled spectrum at $z=1.754$.

For reliable measurement of the AGN UV continuum, we require the AGN sample used in this study to be between $0.83<z<3.04$. At redshifts higher than $z=3.04$, there are not enough data points within the continuum window redward of $\rm Ly\alpha$ emission, and the windows blueward of $\rm Ly\alpha$ emission are usually heavily contaminated by foreground absorptions. Our low redshift cut of $z=0.83$ is set by the visibility of the \ion{C}{3}] $\lambda1909$ emission line within the wavelength coverage of the HETDEX survey. Below this redshift, AGN are identified in Paper I by single broad \ion{Mg}{2} feature which may be broadened by strong stellar winds from  star formation rather than AGN activities.  In our $0.83<z<3.04$ redshift range, there are 3,124 AGN that satisfy our $\rm C>0.05$ criterion. 
To guarantee that there are enough AGN ($\rm N\sim$\,1000) within each redshift interval, we define three bins: $0.83<z\le1.80$, $1.80<z\le2.33$, and $2.33<z\le3.04$.

\begin{table*}[htbp]
\centering
\begin{tabular}{c|c|c|c|c|c|c|c|c}
\hline\hline
$\alpha_{z=0} $ & A & $\beta$ & $M^*_{z=0}$ & $k_{1m}$ & $k_{2m}$ & $\Phi^*_{z=0}$ & $k_{1p}$ & $k_{2p}$ \\\hline
-0.45 & -0.078 & -3.00 & -22.95 & 0.000 & 0.062 & 2.72E-05 & -0.266 & 0.000 \\\hline
\hline
\end{tabular}
\caption{The  parameters of the LEDE model.}
\label{t_LEDE}
\end{table*}

Figure \ref{f_M1450_z} shows the $\rm M_{1450}$ LF of the HETDEX AGN in three redshift bins in the upper left panel. Table \ref{t_UV_LF_data} lists details of each data points. Similar with our explanations for Table \ref{t_LyA_LF_data} in Section \ref{sec_LyA_LF}, the completeness does not necessarily decrease monotonically with the emission line luminosity. And continuum faint AGN are not always emission line weak (see some examples in Figure 12 of Paper I). Therefore, the completeness does not decrease monotonically for continuum faint AGN in Table \ref{t_UV_LF_data}.

Points containing fewer than 20 AGN or the binned completeness lower than 0.15 are shown as open dots in Figure \ref{f_M1450_z}. In all three redshift bins, our AGN sample is not sensitive to the traditional break point between the bright end and the intermediate luminosities at $\rm M_{1450}^*\sim -26$ for $z\sim2$ AGN because we don't have enough reliable solid LF data points brighter than $\rm M_{1450}^*\sim -26$. Thus, we decide to only fit a simple double power law profile as described in Equation \ref{e_dpl_M}: $\rm M^*$ and $\rm \Phi^*$ fit the turnover point where the space density is highest, with 
$\alpha$ and $\beta$ fitting the slope of the faint end and that of intermediate luminosities.

A strong evolution can be seen: as the redshift increases from $z\sim1.5$ to $z\sim2.6$, the turnover luminosity increases while the turnover space density decreases.  We fit this evolution with a simple luminosity-evolution-density-evolution model (LEDE), using second-order polynomials for both $\rm M^*$ and $\rm \Phi^*$, as shown in Equations \ref{e_Mz} and \ref{e_pz}. Besides the evolution of the turnover point, the faint end slope also changes significantly as a function of the redshift and we fit the evolution with a simple linear form (Equation \ref{e_az}).
\begin{equation}
M^*(z) = M^*_{z=0} - 2.5\cdot ( k_{1m}\cdot z + k_{2m}\cdot z^2 )
\label{e_Mz}
\end{equation}
\begin{equation}
\Phi^*(z) = \Phi^*_{z=0}\cdot 10^{k_{1p}\cdot z + k_{2p}\cdot z^2}
\label{e_pz}
\end{equation}
\begin{equation}
\alpha(z) = \alpha_{z=0} + A\cdot z
\label{e_az}
\end{equation}
The best-fit LEDE model is shown by the solid lines in the upper left panel of Figure \ref{f_M1450_z}. Only solid data points are used in the fit. The best-fit parameters are listed in Table \ref{t_LEDE}.

The remaining three panels of Figure \ref{f_M1450_z} compare our  LEDE model with the  models of the VVDS survey \citep{Bongiorno2007A&A} and the SDSS survey \citep{Ross2013,PD2016}.  At the bright end ($\rm M_{1450}\lesssim-25$ mag), our best-fit solution for all three redshift bins agree well with the results of the VVDS and SDSS surveys. 
Our LFs do show a decrease in the space density of faint AGN ($M>M^*$), while the LFs in the VVDS survey and the SDSS survey both show continuous no such turnover. We note that the SDSS AGN LFs mainly cover the bright end and the brightest magnitudes of the faint end. The faint end shown by the thin dotted lines and the thin dashed lines are only extrapolations to their  models based on the bright luminosity bins. Both \cite{Ross2013} and \cite{PD2016} use the QSOs in the SDSS survey, but \cite{PD2016} used variability-selected QSOs, which can go fainter compared to the color-selected QSOs. The VVDS survey only covers the faint end, their bright end LF is also fitted with the QSO LF based on the SDSS survey, but with an earlier QSO catalog \citep{Richards2006}. The VVDS survey covers a very wide redshift range of $0<z<5$. \cite{Bongiorno2007A&A} broke their 130 AGN sample into seven redshift bins, and further into detailed luminosity bins. For most of their luminosity bins, the number of AGN is less than 10 or even 5. We note that the original AGN LFs of \cite{Bongiorno2007A&A}, \cite{Ross2013}, and \cite{PD2016} are reported with $M_\text{B}$, $M_i(z=2)$, and $M_g(z=2)$, respectively. We adopt the K-corrections in \cite{Ross2013} to convert these magnitudes into $\rm M_{1450}$. The  evolution models in both \cite{Ross2013} and \cite{PD2016} has two models separated at $z=2.2$: Below this redshift the evolution is pure luminosity evolution (PLE); Above this redshift the evolution is LEDE.

\begin{figure}[htbp]
\centering
\includegraphics[width=\textwidth]{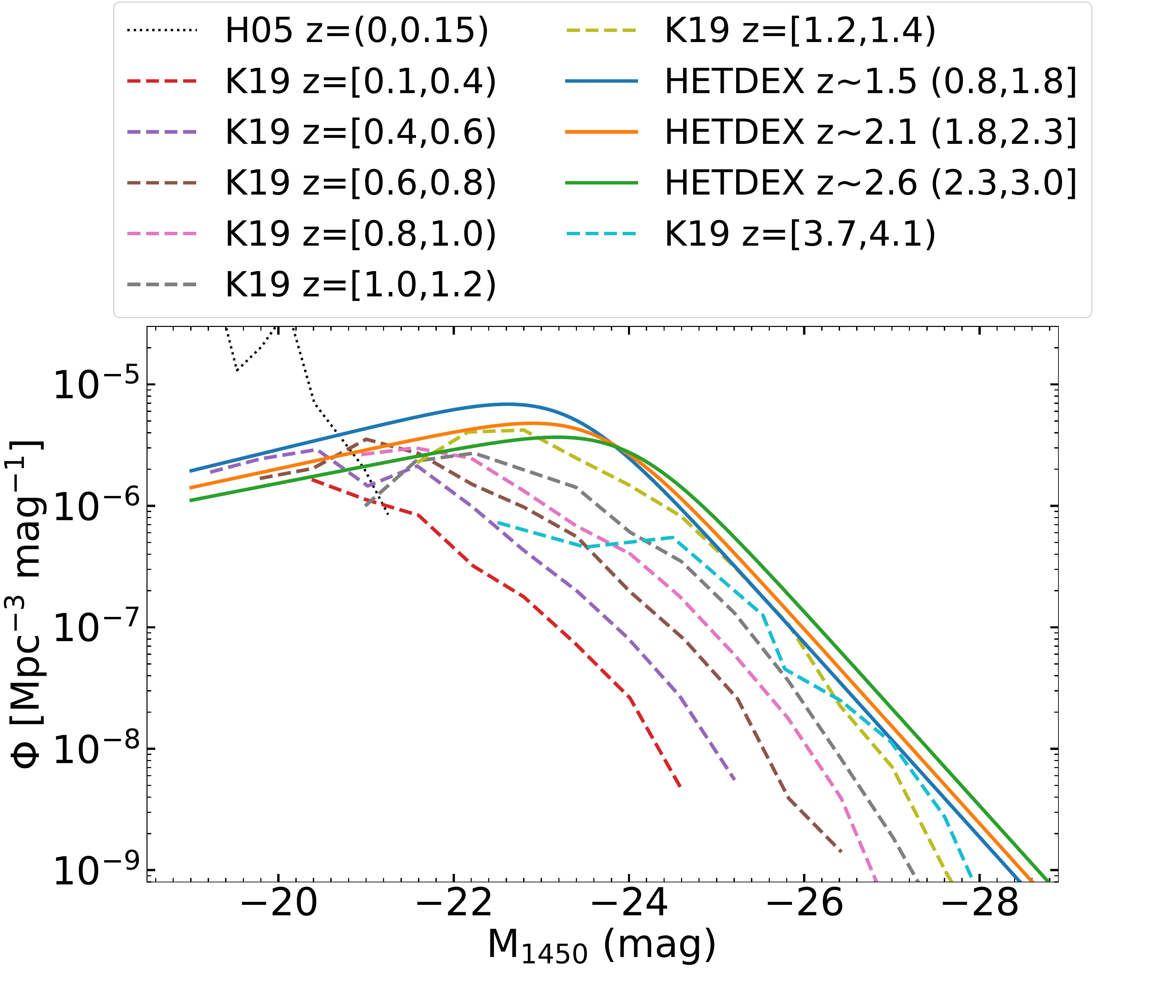}
\caption{The $\rm M_{1450}$ LF of AGN from $z\sim0$ to $z\sim4$. The solid lines are the  LEDE model to our HETDEX AGN catalog. The dashed lines are taken from \cite{Kulkarni2019} (K19). The black dotted line is from \cite{Hao2005} (H05).}
\label{f_M1450_z_comp}
\end{figure}

Figure \ref{f_M1450_z_comp} compares our AGN $\rm M_{1450}$ LF to those of the previous studies. \cite{Kulkarni2019}, hereafter K19, compiled large optical AGN catalogs and presented the LFs spanning over $0<z<7$ with the largest optical AGN catalog. At $0<z\lesssim2.1$, the K19 AGN sample exclusively consists of the SDSS DR7 QSOs \citep{Schneider2010} and the AGNs of the 2SLAQ survey \citep{Croom2009}. At $2.1\lesssim z<3.5$, the K19 sample only contains the BOSS DR9 color-selected QSOs \citep{Ross2013}. They provide the $\rm M_{1450}$ LF in fine redshift intervals with a bin size of $0.2\sim0.4$. For the redshift bins lower than $z=1.5$, the faint ends of the K19 AGN LFs consists exclusively of the AGNs of the 2SLAQ survey. For the redshift interval of $3.7\le z<4.1$ (the cyan dashed line), the faint end of the K19 AGN LF only consists of the AGN sample of \cite{Glikman2011}. We plot the data points of their LFs in different redshift bins in Figure \ref{f_M1450_z_comp} with dashed lines of different colors. We do not compare with their model because they exclude all data points at the faint end that shows decreased space density for fainter AGN during their fitting, as they believe these trends are suspicious and could be caused by over-estimated completeness. These data points are excluded by eye, not based on cuts such as the number of AGN per bin or the completeness.
Our HETDEX AGN LFs have more solid data points in the faint end and are therefore more reliable in the faint end than the bright end (see the upper left panel of Figure \ref{f_M1450_z} and the related caption, and Table \ref{t_UV_LF_data}). Similar with the K19 AGN LFs in all other redshift intervals, our HETDEX AGN LFs at $z\sim$ 1.5, 2.1, and 2.6 also show decreased space density for fainter AGN at the faint ends. This might again suggest that this trend is real, and there is probably a turnover luminosity ($ L^*$) that is favored by the AGN at different redshifts, below which it is more difficult for the AGN activity to be triggered.




\section{$\rm L_{bol}$ LF}
\label{sec_Lbol}

\begin{figure}[htbp]
\centering
\includegraphics[width=\textwidth]{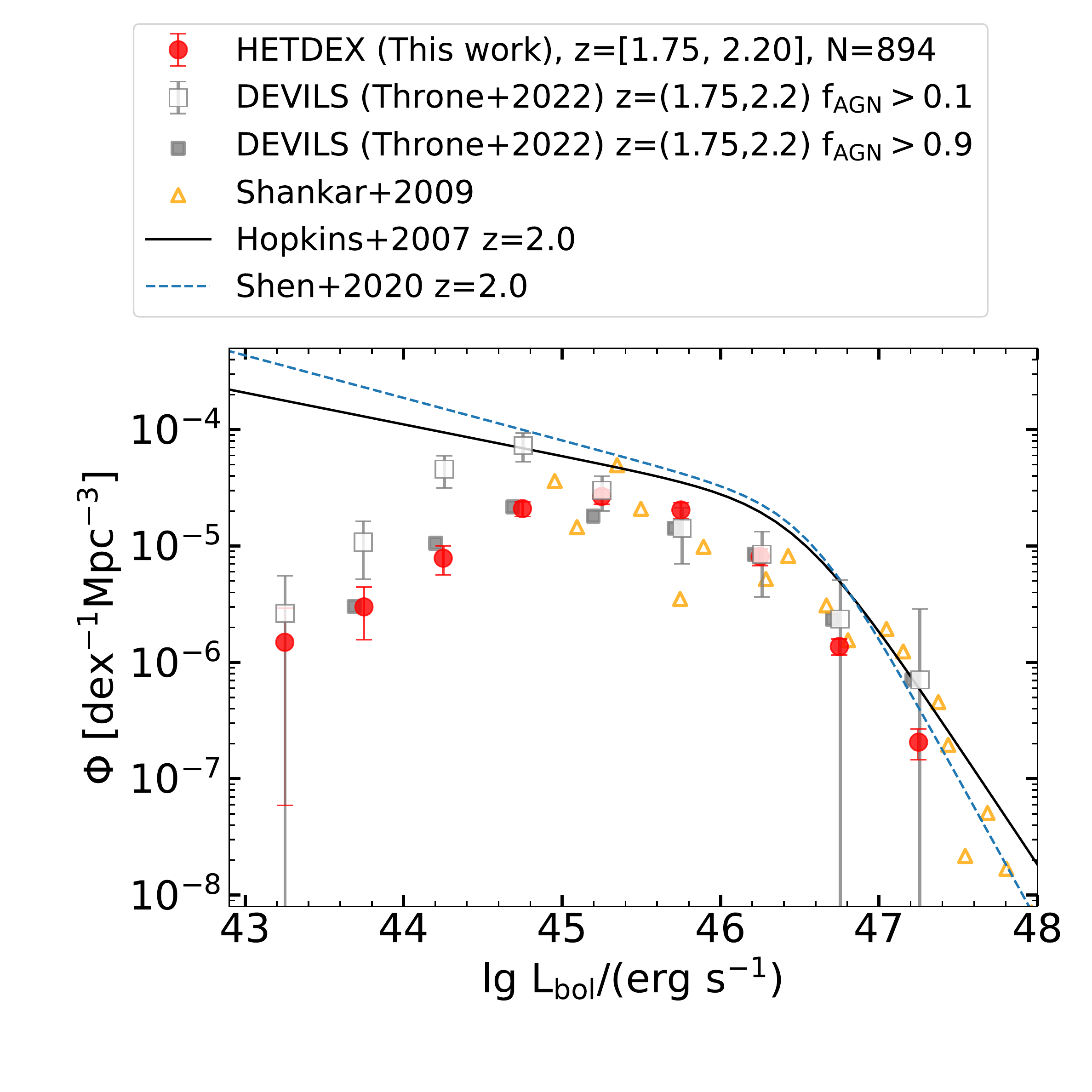}\\
\caption{The bolometric luminosity function of AGN at $z\sim2$. The red circles are the optical spectroscopically identified HETDEX AGN in Paper I. The grey squares are the SED identified AGN from the DEVILS survey \citep{Thorne2022}. $\rm f_{AGN}$ is the fraction of the mid-IR luminosity contributed by AGN. The open squares are for the AGN with $\rm f_{AGN}>0.1$, and the solid squares are for the AGN with $\rm f_{AGN}>0.9$. The solid grey squares are moved leftward by 0.05 dex for presentation purpose. The open orange triangles are taken from \cite{Shankar2009}, which compiled AGN selected from multi-bands and studied the evolution of the AGN LF. The black solid line shows the AGN LF based on the  evolution model of \cite{Hopkins2007}. The blue dashed lines shows the  evolution model of \cite{Shen2020}.}
\label{f_Lbol}
\end{figure}

There are many studies that compile AGN selected by different bands over a wide redshift range and explore the evolution of the bolometric LF of AGN, such as \cite{Hopkins2007}, \cite{Shankar2009}, \cite{Shen2020}, etc. The faint end, especially at $\rm L_{bol}\lesssim10^{45}$ erg/s, consists exclusively of the X-ray identified AGN and the mid-IR identified AGN in these studies. We show their  bolometric LF of AGN at $z=2.0$ and compare with that of the HETDEX AGN (red circles) in Figure \ref{f_Lbol}. The bolometric luminosity of the HETDEX AGN are estimated with their 1450\,\AA\ monochromatic luminosity following the correction in \cite{Shen2020}. Both the HETDEX AGN and the evolution models are plotted at $z=2.0$, so that they can be compared with the LFs of the AGN of the Deep Extragalactic VIsible Legacy Survey \citep[DEVILS;][]{Davies2018} (grey squares). \cite{Thorne2022} identified AGN from the DEVILS survey with the spectral energy distribution (SED) fitting to 22 broad bands spanning from far-ultraviolet to far-infrared (1500\,\AA\ – 500 $\mu$m). There are strong differences between our LF and the LFs in \cite{Hopkins2007} and \cite{Shen2020} at the faint end, where our LF shows a strong decline of the space density while their LFs show moderate increment of the space density for fainter AGN. This might suggest that AGN selected by different bands represent different phases of the AGN activity. X-ray, mid-IR, and optical AGN can have very different space density. The optical phase may have a shorter time scale compared to the X-ray phase and the mid-IR phase, especially for the fainter AGN, which are not powerful enough to expel the surrounding dust quickly and allow the AGN observed in optical.

Similar with our LF based on emission line identified AGN from HETDEX, the LF of the SED identified AGN from the DEVILS survey also shows decreased space density for fainter AGN at the faint end ($\rm L_{bol}\lesssim10^{45}$ erg/s). 
$\rm f_{AGN}$ is the fraction of the mid-IR luminosity contributed by AGN.
Our LF agrees well with their LF for the $\rm f_{AGN}>0.9$ AGN, while it shows lower space density compared to their LF for the $\rm f_{AGN}>0.1$ AGN at the faint end. This might suggest that objects with low AGN fraction can only be observed by mid-IR or X-ray observations, and hard to be observed as optical AGN.

\section{Discussion}
\label{sec_discuss}

In Section \ref{sec_LyA_LF} and Section \ref{sec_UV_LF_z}, we show that there is a turnover luminosity ($L^*$) where the space density is highest for our pure emission line identified AGN from the HETDEX survey in both the emission line LF and the continuum LF. At the faint end, $L<L^*$, the space density decreases for fainter AGN. 
One major concern about the observed low space density at the faint end is that fainter AGN are more heavily influenced by selection effects. In this section, we discuss the potential selection effects of our AGN identification and explain our efforts in correcting such selection effects with the completeness corrections in terms of 
EW (Section \ref{sec_ew}), line width (Section \ref{sec_fwhm}), and host galaxy contamination (Section \ref{sec_host}).


\subsection{Equivalent Width}
\label{sec_ew}
For the emission line identified sample, there is a general concern that the low-EW population whose emission-line are not strong enough to make it into the sample can be missed by the selection, as an emission-line identified sample is similar to the narrow-band drop-out selected sample. The missing low-EW population problem of the narrow-band drop-out selection can make the space density of the faint end under-estimated compared to the continuum selected sample (see Figure 8 of \citealt{Gronwall2007} for an example). However, the emission-line identified sample and the narrow-band drop-out selected sample are not identical: The emission line identification relies more on the S/N of emission lines, while the narrow-band drop-out method relies more on the equivalent widths. 
 We will discuss whether our emission line identified sample is biased against low-EW sources in this sub-section using the $\rm Ly\alpha$ emission line as an example. After similar analysis on other emission lines, we found that the following results are general and can also be applied to other lines.

We check whether our HETDEX AGN sample is biased against low-EW sources by comparing the distribution of the rest-frame EW of the $\rm Ly\alpha$ emission of our HETDEX AGN and that of the latest SDSS QSOs \citep{Paris2018, Rakshit2020} in Figure 14 of Paper I. The median $\rm EW_{Ly\alpha,rest}$ of our HETDEX AGN sample (124\,\AA) is higher than that of the SDSS QSO sample (106\,\AA), but there is no systematic bias towards higher $\rm EW_{Ly\alpha,rest}$. In fact, our HETDEX AGN sample contains more low-EW AGN (see Figure \ref{f_example} for one example). The higher median $\rm EW_{Ly\alpha,rest}$ of our HETDEX AGN sample is the result of our higher sensitivity to the extremely high EW AGN, which have strong emission lines and weak continua (see Figure 12 of Paper I for some examples) instead of missing low-EW AGN. 

\begin{figure}[htbp]
\centering
\includegraphics[width=\textwidth]{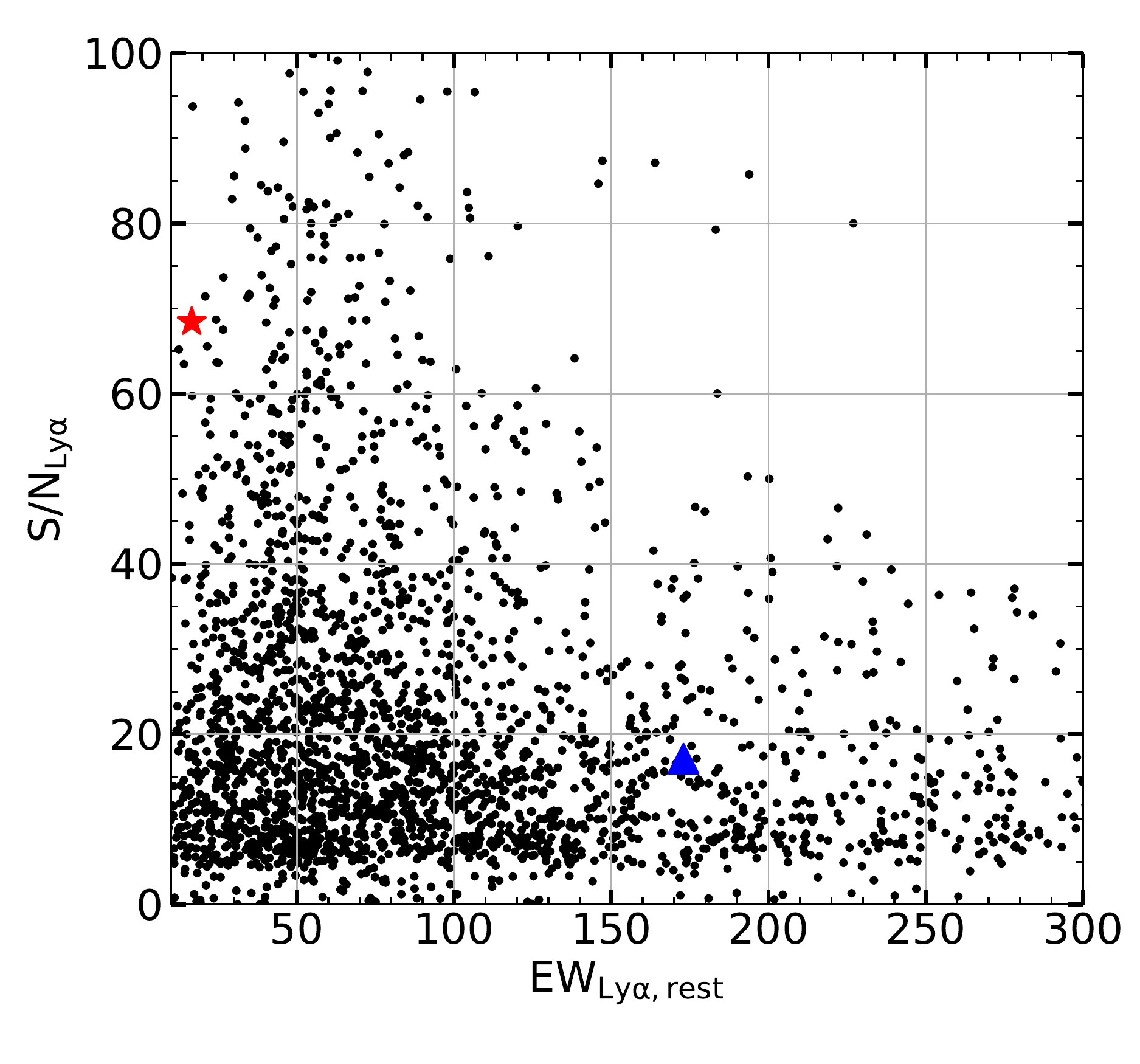}\\
\caption{Signal-to-noise ratio versus the EW of the $\rm Ly\alpha$ emission line of the HETDEX AGN sample. The red star is a random example AGN ($agnid=3118$) with not high $\rm EW_{Ly\alpha,rest}=16.6\,\AA$ and high $\rm S/N_{Ly\alpha}=68.4$. The blue triangle is another random example AGN ($agnid=849$) with high $\rm EW_{Ly\alpha,rest}=173\,\AA$ and not very high $\rm S/N_{Ly\alpha}=17.1$. We show their spectra in Figure \ref{f_example2}.}
\label{f_sn_ew}
\end{figure}

\begin{figure}[htbp]
\centering
\includegraphics[width=\textwidth]{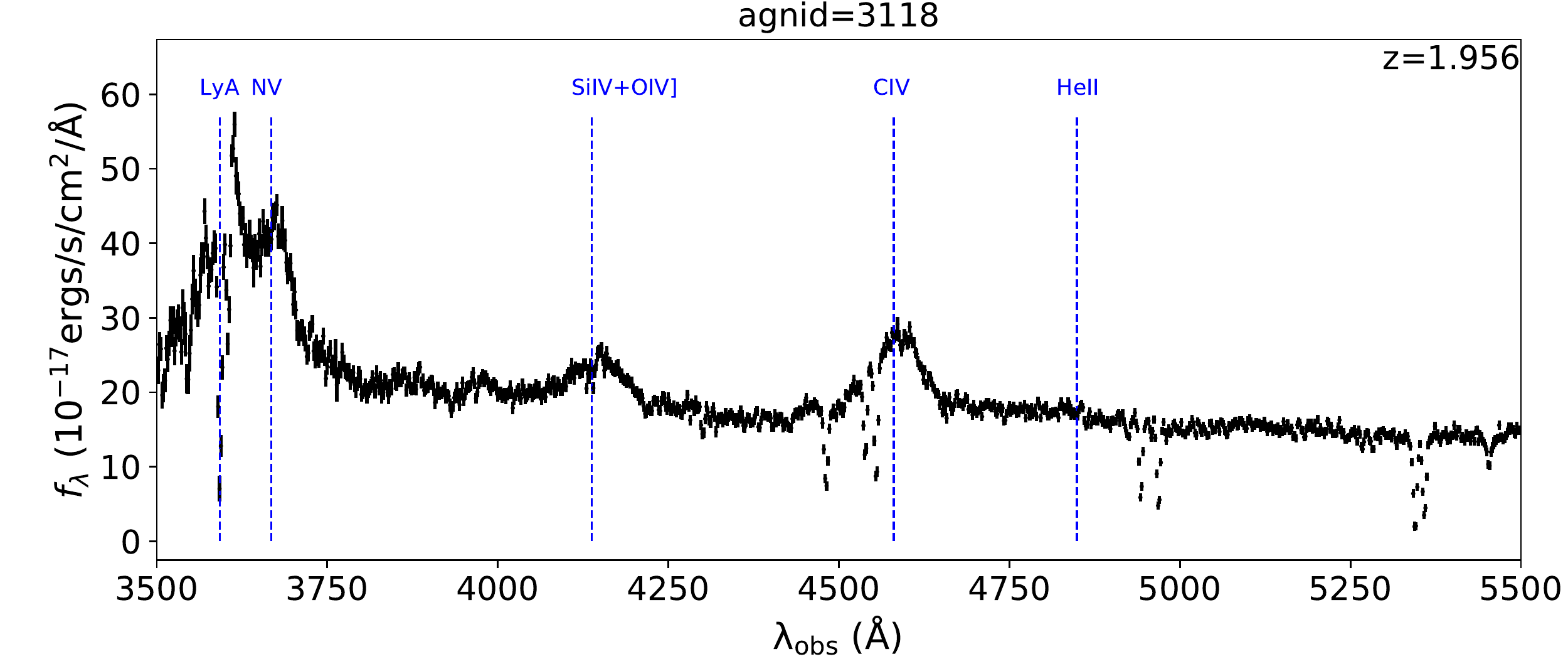}\\
\includegraphics[width=\textwidth]{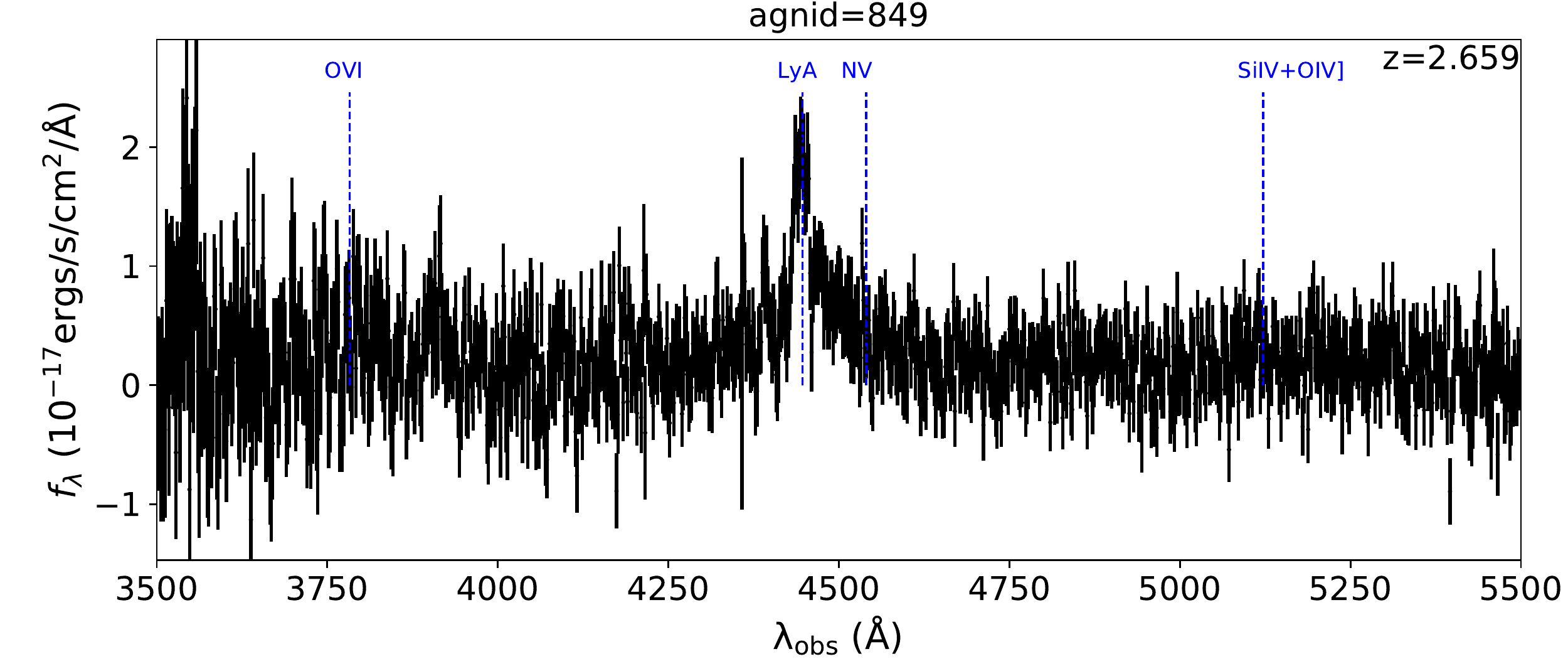}\\
\caption{Upper: The spectrum of the red star ($agnid=3118$) in Figure \ref{f_sn_ew}. Bottom: The spectrum of the blue triangle ($agnid=849$) in Figure \ref{f_sn_ew}.}
\label{f_example2}
\end{figure}

Our pure emission line based AGN selection is directly regulated by various signal-to-noise ratio of the emission lines as detailed in Section \ref{sec_completeness} rather than EW. Figure \ref{f_sn_ew} shows that there is no direct correlation between the signal-to-noise ratio and the EW using the $\rm Ly\alpha$ emission line as a typical example. In Figure \ref{f_example2}, we show $agnid=3118$ as a random example with not high $\rm EW_{Ly\alpha,rest}=16.6\,\AA$ and high $\rm S/N_{Ly\alpha}=68.4$, the red star in Figure \ref{f_sn_ew}. The $\rm EW_{Ly\alpha,rest}$ is low because the flux of $\rm Ly\alpha$ is heavily absorbed and the continuum at the line is very strong. $agnid=849$ is an opposite example with high $\rm EW_{Ly\alpha,rest}=173\,\AA$ and not very high $\rm S/N_{Ly\alpha}=17.1$. The very high $\rm EW_{Ly\alpha,rest}=173\,\AA$ is mainly resulted by its low continuum level.
Moreover, our LFs are carefully completeness corrected for their signal-to-noise ratio based on the simulations detailed in Section \ref{sec_completeness}. Thus, our completeness corrected LFs with our emission line selected HETDEX AGN sample in Section \ref{sec_LyA_LF} and Section \ref{sec_UV_LF_z} shouldn't be biased with the under-estimated incompleteness of the low-EW AGN.

Another concern is that the 2em selection could be biased by detecting more stronger emission lines compared to the weaker emission lines. For example, the Ly$\alpha$ emission line is stronger than all other emission lines, such as \ion{C}{4} $\lambda 1549$, \ion{C}{3}] $\lambda 1909$, \ion{Mg}{2} $\lambda 2799$, etc. The completeness of AGN identified by different emission lines should be different. Our completeness estimations for the 2em selected AGN properly correct this selection effect by using the composite spectrum of AGN as the template to make simulated spectra as detailed in Section \ref{sec_c_2em}. Stronger line pairs are of higher completeness (smaller completeness correction) in Figure \ref{f_c_2em}. For example, the (Ly$\alpha$ + \ion{C}{4} $\lambda 1549$) line pair, shown as the high completeness region at $z\sim2.3$, is stronger than the  (\ion{C}{4} $\lambda 1549$ + \ion{C}{3}] $\lambda 1909$) line pair, shown as the high completeness region at $z\sim1.6$. Therefore, the completeness of the (Ly$\alpha$ + \ion{C}{4} $\lambda 1549$) identified AGN is higher than that of the (\ion{C}{4} $\lambda 1549$ + \ion{C}{3}] $\lambda 1909$) identified AGN at low $\rm S/N_{spec}$ level ($\rm S/N_{spec}\lesssim80$). Therefore, our completeness correction for the 2em selected AGN ($\rm C_{2em}$) has already properly taken cares of the selection effects caused by different strengths of different emission lines.

\subsection{Line Width}
\label{sec_fwhm}

For the sBL selected AGN, one potential problem is that broader emission lines are less likely to be identified by the selection code than narrower emission lines at the same $\rm S/N_{em}$ level. This selection effect is properly modeled and completeness corrected based on our careful simulations for the sBL selected AGN as shown in Figure \ref{f_c_sBL}. However, there is one AGN population, the single narrow line AGN, that our completeness corrections do miss out. It would be hard to separate these from normal star-forming galaxies without detecting another emission line.
Both the 2em selection method and the sBL selection method are unable to identify the single narrow line AGN. Their completeness estimations are therefore not modeled with any simulations in Section \ref{sec_c_agn}. Narrow-line AGN are the ones whose broad-line region is obscured by the dusty torus. Their continuum can also be significantly obscured by the torus making them continuum faint AGN in the UV LF. Missing single narrow line AGN can result in the under-estimation of the space density of the faint AGN in the UV LF. Our study of the type-II AGN fraction in Section 7.4 of Paper I shows that the type-II AGN fraction is strongly correlated with the bolometric luminosity. In the faintest luminosity bins ($L_{\text{bol}}\lesssim10^{11}\ L_{\sun}$), this fraction is $\sim$ 50\% while the type-II AGN fraction is only $\sim$ 5\% for bright AGN ($L_{\text{bol}}\gtrsim10^{12}\ L_{\sun}$) as shown by Figure 17 of Paper I. Even if the space density of the faintest luminosity bins in the UV LFs is doubled, the turnover luminosity and the declined space density trend for fainter AGN would still be there, just the faint end slope would change.

\subsection{Host Galaxy Contamination}
\label{sec_host}

There are many ways that the host galaxy contamination can affect the LF of AGN. In this section, we briefly discuss the possible effects from the AGN-host decomposition in Section \ref{sec_AGN-host}, the potential contamination from star-forming galaxies in Section \ref{sec_SFgal}, and Seyferts dominated by host galaxies in Section \ref{sec_seyferts}.

\subsubsection{AGN-host decomposition}
\label{sec_AGN-host}

Proper AGN-host decomposition is important in evaluating the AGN contribution to the observed luminosity and calculating the intrinsic AGN LF.

For the $\rm Ly\alpha$ LF study, we do not perform AGN-host decomposition as the decomposition can introduce more problems than correcting the contribution from host galaxies. Therefore, the observed emission line luminosity is a combination of the contribution from the star formation of the host galaxies and the emission of AGN. This can cause the over-estimate of the completeness of AGN hosted by galaxies with high star formation rate. The space density of the low luminosity bins can be under-estimated due to the lack of the AGN-host decomposition of emission lines.

For the continuum LF study, we carry out the AGN-host decomposition for the continuum of the spectrum with the publicly available code \texttt{PyQSOFit}. However, such decomposition with HETDEX spectra of a short wavelength coverage of 3500 - 5500\,\AA\ can fail for sources with a faint continuum and the spectrum is then fit with a pure power-law continuum. The AGN contribution is then over-estimated in this way, causing over-estimated completeness and under-estimated space density at the faint end.

\subsubsection{Potential contamination from star-forming galaxies}
\label{sec_SFgal}
In this sub-section, we discuss the potential contamination from the star-forming galaxies to our AGN sample. We have two approaches to identify AGN from HETDEX - the 2em method that identifies the emission line pair characteristic of AGN and the sBL method that identifies the single broad emission line as broad-line AGN candidates. If the sBL identified AGN are matched with previous catalog(s), they are flagged with \texttt{zflag=1}, otherwise they are only broad-line AGN candidates with our best redshift estimates (details of the redshift estimates can be found in Paper I). Therefore, there could be some star-forming galaxies whose emission lines are broadened by strong stellar winds mis-identified as broad-line AGN for the \texttt{zflag=0} sBL identified AGN candidates. The difference between the red circles and the magenta stars in Figure \ref{f_LyA_LF} shows the contribution of the \texttt{zflag=0} AGN to the $\rm Ly\alpha$ LF. There is no difference at the bright end and the intermediate luminosities. The contamination from the mis-identified star-forming galaxies, if there is, can cause an over-estimated space density of the faint end of the LF.

\subsubsection{Seyferts dominated by host galaxies}
\label{sec_seyferts}
We have discussed that our emission-line identified AGN catalog is not biased against low-EW sources in Section \ref{sec_ew}. However, the catalog could possibly be biased against Seyferts dominated by host galaxies, because the broad line of Seyferts with low AGN contribution can hardly be resolved from the observed emission line and identified by our sBL selection. This can cause an under-estimate of the space density of AGN at the faint end.

\section{Summary}
\label{sec_summary}

In this paper, we present our study of the AGN LF of the HETDEX AGN sample, which is purely spectroscopically selected from the 3500\,\AA\ - 5500\,\AA\ spectra, and free of photometric preselections, such as broad band magnitude cuts, color selections, and/or point-like morphology. We explored the emission-line LF with the $\rm Ly\alpha$ emission (Section \ref{sec_LyA_LF}), the UV LF with the 1450\,\AA\ monochromatic luminosity of the power-law component of the continuum (Section \ref{sec_UV_LF_z}), and the bolometric LF (Section \ref{sec_Lbol}). In all three kinds of LFs, we find significant declination of the space density for fainter AGN at the faint end for our AGN sample. This suggests that there is a turnover luminosity ($L^*$) favored by the UV/optical AGN: Above $L^*$, the AGN would soon fade away and the number density of the bright AGN decreases at high luminosities at the bright end; Below $L^*$, the time scale of the UV/optical phase of AGN is shorter for fainter AGN. The very faint AGN can only be identified as X-ray AGN or mid-IR AGN. We studied the evolution of the turnover luminosity ($L^*$) and the turnover space density ($\rm \Phi^*$) of the double-power law fitted $\rm M_{1450}$ LF of AGN in Section \ref{sec_UV_LF_z}. We find that the evolution from $z\sim1.5$ to $z\sim2.6$ can be well fitted with a simple LEDE model, where $L^*$ and $\rm \Phi^*$ evolves with redshift independently. At lower redshifts, the number density of AGN peaks at lower luminosity.

\acknowledgments

HETDEX is led by the University of Texas at Austin McDonald Observatory and Department of Astronomy with participation from the Ludwig-Maximilians-Universit\"at M\"unchen, Max-Planck-Institut f\"ur Extraterrestrische Physik (MPE), Leibniz-Institut f\"ur Astrophysik Potsdam (AIP), Texas A\&M University, The Pennsylvania State University, Institut f\"ur Astrophysik G\"ottingen, The University of Oxford, Max-Planck-Institut f\"ur Astrophysik (MPA), The University of Tokyo, and Missouri University of Science and Technology. In addition to Institutional support, HETDEX is funded by the National Science Foundation (grant AST-0926815), the State of Texas, the US Air Force (AFRL FA9451-04-2-0355), and generous support from private individuals and foundations.

The Hobby-Eberly Telescope (HET) is a joint project of the University of Texas at Austin, the Pennsylvania State University, Ludwig-Maximilians-Universit\"at M\"unchen, and Georg-August-Universit\"at G\"ottingen. The HET is named in honor of its principal benefactors, William P. Hobby and Robert E. Eberly.

VIRUS is a joint project of the University of Texas at Austin, Leibniz-Institut f{\" u}r Astrophysik Potsdam (AIP), Texas A\&M University, Max-Planck-Institut f{\" u}r Extraterrestrische Physik, Ludwig-Maximilians-Universit{\" a}t M{\" u}nchen, Pennsylvania State University, Institut f{\" u}r Astrophysik G{\"o}ttingen, University of Oxford, and the Max-Planck-Institut f{\" u}r Astrophysik.

The authors acknowledge the Texas Advanced Computing Center (TACC) at The University of Texas at Austin for providing high performance computing, visualization, and storage resources that have contributed to the research results reported within this paper. URL: http://www.tacc.utexas.edu

The Institute for Gravitation and the Cosmos is supported by the Eberly College of Science and the Office of the Senior Vice President for Research at the Pennsylvania State University. \\


\bibliography{agn}

\begin{thebibliography}{}
\expandafter\ifx\csname natexlab\endcsname\relax\def\natexlab#1{#1}\fi

\bibitem[{{Aihara} {et~al.}(2018){Aihara}, {Arimoto}, {Armstrong}, {Arnouts},
  {Bahcall}, {Bickerton}, {Bosch}, {Bundy}, {Capak}, {Chan}, {Chiba}, {Coupon},
  {Egami}, {Enoki}, {Finet}, {Fujimori}, {Fujimoto}, {Furusawa}, {Furusawa},
  {Goto}, {Goulding}, {Greco}, {Greene}, {Gunn}, {Hamana}, {Harikane},
  {Hashimoto}, {Hattori}, {Hayashi}, {Hayashi}, {He{\l}miniak}, {Higuchi},
  {Hikage}, {Ho}, {Hsieh}, {Huang}, {Huang}, {Ikeda}, {Imanishi}, {Inoue},
  {Iwasawa}, {Iwata}, {Jaelani}, {Jian}, {Kamata}, {Karoji}, {Kashikawa},
  {Katayama}, {Kawanomoto}, {Kayo}, {Koda}, {Koike}, {Kojima}, {Komiyama},
  {Konno}, {Koshida}, {Koyama}, {Kusakabe}, {Leauthaud}, {Lee}, {Lin}, {Lin},
  {Lupton}, {Mandelbaum}, {Matsuoka}, {Medezinski}, {Mineo}, {Miyama},
  {Miyatake}, {Miyazaki}, {Momose}, {More}, {More}, {Moritani}, {Moriya},
  {Morokuma}, {Mukae}, {Murata}, {Murayama}, {Nagao}, {Nakata}, {Niida},
  {Niikura}, {Nishizawa}, {Obuchi}, {Oguri}, {Oishi}, {Okabe}, {Okamoto},
  {Okura}, {Ono}, {Onodera}, {Onoue}, {Osato}, {Ouchi}, {Price}, {Pyo}, {Sako},
  {Sawicki}, {Shibuya}, {Shimasaku}, {Shimono}, {Shirasaki}, {Silverman},
  {Simet}, {Speagle}, {Spergel}, {Strauss}, {Sugahara}, {Sugiyama}, {Suto},
  {Suyu}, {Suzuki}, {Tait}, {Takada}, {Takata}, {Tamura}, {Tanaka}, {Tanaka},
  {Tanaka}, {Tanaka}, {Terai}, {Terashima}, {Toba}, {Tominaga}, {Toshikawa},
  {Turner}, {Uchida}, {Uchiyama}, {Umetsu}, {Uraguchi}, {Urata}, {Usuda},
  {Utsumi}, {Wang}, {Wang}, {Wong}, {Yabe}, {Yamada}, {Yamanoi}, {Yasuda},
  {Yeh}, {Yonehara}, \& {Yuma}}]{Aihara2018}
{Aihara}, H., {Arimoto}, N., {Armstrong}, R., {et~al.} 2018, \pasj, 70, S4

\bibitem[{{Aihara} {et~al.}(2019){Aihara}, {AlSayyad}, {Ando}, {Armstrong},
  {Bosch}, {Egami}, {Furusawa}, {Furusawa}, {Goulding}, {Harikane}, {Hikage},
  {Ho}, {Hsieh}, {Huang}, {Ikeda}, {Imanishi}, {Ito}, {Iwata}, {Jaelani},
  {Kakuma}, {Kawana}, {Kikuta}, {Kobayashi}, {Koike}, {Komiyama}, {Li},
  {Liang}, {Lin}, {Luo}, {Lupton}, {Lust}, {MacArthur}, {Matsuoka}, {Mineo},
  {Miyatake}, {Miyazaki}, {More}, {Murata}, {Namiki}, {Nishizawa}, {Oguri},
  {Okabe}, {Okamoto}, {Okura}, {Ono}, {Onodera}, {Onoue}, {Osato}, {Ouchi},
  {Shibuya}, {Strauss}, {Sugiyama}, {Suto}, {Takada}, {Takagi}, {Takata},
  {Takita}, {Tanaka}, {Terai}, {Toba}, {Uchiyama}, {Utsumi}, {Wang}, {Wang}, \&
  {Yamada}}]{Aihara2019}
{Aihara}, H., {AlSayyad}, Y., {Ando}, M., {et~al.} 2019, \pasj, 71, 114

\bibitem[{{Alexander} \& {Hickox}(2012)}]{Alexander2012}
{Alexander}, D.~M., \& {Hickox}, R.~C. 2012, \nar, 56, 93

\bibitem[{{Blanc} {et~al.}(2011){Blanc}, {Adams}, {Gebhardt}, {Hill}, {Drory},
  {Hao}, {Bender}, {Ciardullo}, {Finkelstein}, {Fry}, {Gawiser}, {Gronwall},
  {Hopp}, {Jeong}, {Kelzenberg}, {Komatsu}, {MacQueen}, {Murphy}, {Roth},
  {Schneider}, \& {Tufts}}]{Blanc2011}
{Blanc}, G.~A., {Adams}, J.~J., {Gebhardt}, K., {et~al.} 2011, \apj, 736, 31

\bibitem[{{Bongiorno} {et~al.}(2007){Bongiorno}, {Zamorani}, {Gavignaud},
  {Marano}, {Paltani}, {Mathez}, {M{\o}ller}, {Picat}, {Cirasuolo},
  {Lamareille}, {Bottini}, {Garilli}, {Le Brun}, {Le F{\`e}vre}, {Maccagni},
  {Scaramella}, {Scodeggio}, {Tresse}, {Vettolani}, {Zanichelli}, {Adami},
  {Arnouts}, {Bardelli}, {Bolzonella}, {Cappi}, {Charlot}, {Ciliegi},
  {Contini}, {Foucaud}, {Franzetti}, {Guzzo}, {Ilbert}, {Iovino}, {McCracken},
  {Marinoni}, {Mazure}, {Meneux}, {Merighi}, {Pell{\`o}}, {Pollo}, {Pozzetti},
  {Radovich}, {Zucca}, {Hatziminaoglou}, {Polletta}, {Bondi}, {Brinchmann},
  {Cucciati}, {de la Torre}, {Gregorini}, {Mellier}, {Merluzzi}, {Temporin},
  {Vergani}, \& {Walcher}}]{Bongiorno2007A&A}
{Bongiorno}, A., {Zamorani}, G., {Gavignaud}, I., {et~al.} 2007, \aap, 472, 443

\bibitem[{{Boroson} \& {Green}(1992)}]{Boroson1992}
{Boroson}, T.~A., \& {Green}, R.~F. 1992, \apjs, 80, 109

\bibitem[{{Boyle} {et~al.}(2000){Boyle}, {Shanks}, {Croom}, {Smith}, {Miller},
  {Loaring}, \& {Heymans}}]{Boyle2000}
{Boyle}, B.~J., {Shanks}, T., {Croom}, S.~M., {et~al.} 2000, \mnras, 317, 1014

\bibitem[{{Brandt} \& {Alexander}(2015)}]{Brandt2015}
{Brandt}, W.~N., \& {Alexander}, D.~M. 2015, \aapr, 23, 1

\bibitem[{{Bruzual} \& {Charlot}(2003)}]{BC03}
{Bruzual}, G., \& {Charlot}, S. 2003, \mnras, 344, 1000

\bibitem[{{Cenarro} {et~al.}(2019){Cenarro}, {Moles},
  {Crist{\'o}bal-Hornillos}, {Mar{\'\i}n-Franch}, {Ederoclite}, {Varela},
  {L{\'o}pez-Sanjuan}, {Hern{\'a}ndez-Monteagudo}, {Angulo}, {V{\'a}zquez
  Rami{\'o}}, {Viironen}, {Bonoli}, {Orsi}, {Hurier}, {San Roman}, {Greisel},
  {Vilella-Rojo}, {D{\'\i}az-Garc{\'\i}a}, {Logro{\~n}o-Garc{\'\i}a},
  {Gurung-L{\'o}pez}, {Spinoso}, {Izquierdo-Villalba}, {Aguerri}, {Allende
  Prieto}, {Bonatto}, {Carvano}, {Chies-Santos}, {Daflon}, {Dupke},
  {Falc{\'o}n-Barroso}, {Gon{\c{c}}alves}, {Jim{\'e}nez-Teja}, {Molino},
  {Placco}, {Solano}, {Whitten}, {Abril}, {Ant{\'o}n}, {Bello}, {Bielsa de
  Toledo}, {Castillo-Ram{\'\i}rez}, {Chueca}, {Civera},
  {D{\'\i}az-Mart{\'\i}n}, {Dom{\'\i}nguez-Mart{\'\i}nez},
  {Garzar{\'a}n-Calderaro}, {Hern{\'a}ndez-Fuertes}, {Iglesias-Marzoa},
  {I{\~n}iguez}, {Jim{\'e}nez Ruiz}, {Kruuse}, {Lamadrid}, {Lasso-Cabrera},
  {L{\'o}pez-Alegre}, {L{\'o}pez-Sainz}, {Ma{\'\i}cas}, {Moreno-Signes},
  {Muniesa}, {Rodr{\'\i}guez-Llano}, {Rueda-Teruel}, {Rueda-Teruel},
  {Soriano-Lagu{\'\i}a}, {Tilve}, {Valdivielso}, {Yanes-D{\'\i}az}, {Alcaniz},
  {Mendes de Oliveira}, {Sodr{\'e}}, {Coelho}, {Lopes de Oliveira}, {Tamm},
  {Xavier}, {Abramo}, {Akras}, {Alfaro}, {Alvarez-Candal}, {Ascaso}, {Beasley},
  {Beers}, {Borges Fernandes}, {Bruzual}, {Buzzo}, {Carrasco}, {Cepa},
  {Cortesi}, {Costa-Duarte}, {De Pr{\'a}}, {Favole}, {Galarza}, {Galbany},
  {Garcia}, {Gonz{\'a}lez Delgado}, {Gonz{\'a}lez-Serrano},
  {Guti{\'e}rrez-Soto}, {Hernandez-Jimenez}, {Kanaan}, {Kuncarayakti},
  {Landim}, {Laur}, {Licandro}, {Lima Neto}, {Lyman}, {Ma{\'\i}z
  Apell{\'a}niz}, {Miralda-Escud{\'e}}, {Morate}, {Nogueira-Cavalcante},
  {Novais}, {Oncins}, {Oteo}, {Overzier}, {Pereira}, {Rebassa-Mansergas},
  {Reis}, {Roig}, {Sako}, {Salvador-Rusi{\~n}ol}, {Sampedro},
  {S{\'a}nchez-Bl{\'a}zquez}, {Santos}, {Schmidtobreick}, {Siffert}, {Telles},
  \& {Vilchez}}]{Cenarro2019}
{Cenarro}, A.~J., {Moles}, M., {Crist{\'o}bal-Hornillos}, D., {et~al.} 2019,
  \aap, 622, A176

\bibitem[{{Croom} {et~al.}(2004){Croom}, {Smith}, {Boyle}, {Shanks}, {Miller},
  {Outram}, \& {Loaring}}]{Croom2004}
{Croom}, S.~M., {Smith}, R.~J., {Boyle}, B.~J., {et~al.} 2004, \mnras, 349,
  1397

\bibitem[{{Croom} {et~al.}(2009{\natexlab{a}}){Croom}, {Richards}, {Shanks},
  {Boyle}, {Strauss}, {Myers}, {Nichol}, {Pimbblet}, {Ross}, {Schneider},
  {Sharp}, \& {Wake}}]{Croom2009}
{Croom}, S.~M., {Richards}, G.~T., {Shanks}, T., {et~al.} 2009{\natexlab{a}},
  \mnras, 399, 1755

\bibitem[{{Croom} {et~al.}(2009{\natexlab{b}}){Croom}, {Richards}, {Shanks},
  {Boyle}, {Sharp}, {Bland-Hawthorn}, {Bridges}, {Brunner}, {Cannon}, {Carson},
  {Chiu}, {Colless}, {Couch}, {De Propris}, {Drinkwater}, {Edge}, {Fine},
  {Loveday}, {Miller}, {Myers}, {Nichol}, {Outram}, {Pimbblet}, {Roseboom},
  {Ross}, {Schneider}, {Smith}, {Stoughton}, {Strauss}, \&
  {Wake}}]{Croom2009survey}
---. 2009{\natexlab{b}}, \mnras, 392, 19

\bibitem[{{Davies} {et~al.}(2018){Davies}, {Robotham}, {Driver}, {Lagos},
  {Cortese}, {Mannering}, {Foster}, {Lidman}, {Hashemizadeh}, {Koushan},
  {O'Toole}, {Baldry}, {Bilicki}, {Bland-Hawthorn}, {Bremer}, {Brown},
  {Bryant}, {Catinella}, {Croom}, {Grootes}, {Holwerda}, {Jarvis}, {Maddox},
  {Meyer}, {Moffett}, {Phillipps}, {Taylor}, {Windhorst}, \&
  {Wolf}}]{Davies2018}
{Davies}, L.~J.~M., {Robotham}, A.~S.~G., {Driver}, S.~P., {et~al.} 2018,
  \mnras, 480, 768

\bibitem[{{Di Matteo} {et~al.}(2005){Di Matteo}, {Springel}, \&
  {Hernquist}}]{DiMatteo2005}
{Di Matteo}, T., {Springel}, V., \& {Hernquist}, L. 2005, \nat, 433, 604

\bibitem[{{Finkelstein} {et~al.}(2009){Finkelstein}, {Rhoads}, {Malhotra}, \&
  {Grogin}}]{Finkelstein2009}
{Finkelstein}, S.~L., {Rhoads}, J.~E., {Malhotra}, S., \& {Grogin}, N. 2009,
  \apj, 691, 465

\bibitem[{{Finkelstein} {et~al.}(2007){Finkelstein}, {Rhoads}, {Malhotra},
  {Pirzkal}, \& {Wang}}]{Finkelstein2007}
{Finkelstein}, S.~L., {Rhoads}, J.~E., {Malhotra}, S., {Pirzkal}, N., \&
  {Wang}, J. 2007, \apj, 660, 1023

\bibitem[{{Gavignaud} {et~al.}(2006){Gavignaud}, {Bongiorno}, {Paltani},
  {Mathez}, {Zamorani}, {M{\o}ller}, {Picat}, {Le Brun}, {Marano}, {Le
  F{\`e}vre}, {Bottini}, {Garilli}, {Maccagni}, {Scaramella}, {Scodeggio},
  {Tresse}, {Vettolani}, {Zanichelli}, {Adami}, {Arnaboldi}, {Arnouts},
  {Bardelli}, {Bolzonella}, {Cappi}, {Charlot}, {Ciliegi}, {Contini},
  {Foucaud}, {Franzetti}, {Guzzo}, {Ilbert}, {Iovino}, {McCracken}, {Marinoni},
  {Mazure}, {Meneux}, {Merighi}, {Pell{\`o}}, {Pollo}, {Pozzetti}, {Radovich},
  {Zucca}, {Bondi}, {Busarello}, {Cucciati}, {de la Torre}, {Gregorini},
  {Lamareille}, {Mellier}, {Merluzzi}, {Ripepi}, {Rizzo}, \&
  {Vergani}}]{Gavignaud2006}
{Gavignaud}, I., {Bongiorno}, A., {Paltani}, S., {et~al.} 2006, \aap, 457, 79

\bibitem[{{Gawiser} {et~al.}(2006){Gawiser}, {van Dokkum}, {Gronwall},
  {Ciardullo}, {Blanc}, {Castander}, {Feldmeier}, {Francke}, {Franx},
  {Haberzettl}, {Herrera}, {Hickey}, {Infante}, {Lira}, {Maza}, {Quadri},
  {Richardson}, {Schawinski}, {Schirmer}, {Taylor}, {Treister}, {Urry}, \&
  {Virani}}]{Gawiser2006}
{Gawiser}, E., {van Dokkum}, P.~G., {Gronwall}, C., {et~al.} 2006, \apjl, 642,
  L13

\bibitem[{{Gebhardt} {et~al.}(2000){Gebhardt}, {Bender}, {Bower}, {Dressler},
  {Faber}, {Filippenko}, {Green}, {Grillmair}, {Ho}, {Kormendy}, {Lauer},
  {Magorrian}, {Pinkney}, {Richstone}, \& {Tremaine}}]{Gebhardt2000}
{Gebhardt}, K., {Bender}, R., {Bower}, G., {et~al.} 2000, \apjl, 539, L13

\bibitem[{{Gebhardt} {et~al.}(2021){Gebhardt}, {Mentuch Cooper}, {Ciardullo},
  {Acquaviva}, {Bender}, {Bowman}, {Castanheira}, {Dalton}, {Davis}, {de Jong},
  {DePoy}, {Devarakonda}, {Dongsheng}, {Drory}, {Fabricius}, {Farrow},
  {Feldmeier}, {Finkelstein}, {Froning}, {Gawiser}, {Gronwall}, {Herold},
  {Hill}, {Hopp}, {House}, {Janowiecki}, {Jarvis}, {Jeong}, {Jogee}, {Kakuma},
  {Kelz}, {Kollatschny}, {Komatsu}, {Krumpe}, {Landriau}, {Liu}, {Niemeyer},
  {MacQueen}, {Marshall}, {Mawatari}, {McLinden}, {Mukae}, {Nagaraj}, {Ono},
  {Ouchi}, {Papovich}, {Sakai}, {Saito}, {Schneider}, {Schulze},
  {Shanmugasundararaj}, {Shetrone}, {Sneden}, {Snigula}, {Steinmetz}, {Thomas},
  {Thomas}, {Tuttle}, {Urrutia}, {Wisotzki}, {Wold}, {Zeimann}, \&
  {Zhang}}]{Gebhardt2021}
{Gebhardt}, K., {Mentuch Cooper}, E., {Ciardullo}, R., {et~al.} 2021, \apj,
  923, 217

\bibitem[{{Giallongo} {et~al.}(2015){Giallongo}, {Grazian}, {Fiore}, {Fontana},
  {Pentericci}, {Vanzella}, {Dickinson}, {Kocevski}, {Castellano}, {Cristiani},
  {Ferguson}, {Finkelstein}, {Grogin}, {Hathi}, {Koekemoer}, {Newman}, \&
  {Salvato}}]{Giallongo2015}
{Giallongo}, E., {Grazian}, A., {Fiore}, F., {et~al.} 2015, \aap, 578, A83

\bibitem[{{Glikman} {et~al.}(2011){Glikman}, {Djorgovski}, {Stern}, {Dey},
  {Jannuzi}, \& {Lee}}]{Glikman2011}
{Glikman}, E., {Djorgovski}, S.~G., {Stern}, D., {et~al.} 2011, \apjl, 728, L26

\bibitem[{{Gronwall} {et~al.}(2007){Gronwall}, {Ciardullo}, {Hickey},
  {Gawiser}, {Feldmeier}, {van Dokkum}, {Urry}, {Herrera}, {Lehmer}, {Infante},
  {Orsi}, {Marchesini}, {Blanc}, {Francke}, {Lira}, \&
  {Treister}}]{Gronwall2007}
{Gronwall}, C., {Ciardullo}, R., {Hickey}, T., {et~al.} 2007, \apj, 667, 79

\bibitem[{{Guo} {et~al.}(2018){Guo}, {Shen}, \& {Wang}}]{Guo2018}
{Guo}, H., {Shen}, Y., \& {Wang}, S. 2018, {PyQSOFit: Python code to fit the
  spectrum of quasars}, Astrophysics Source Code Library, ascl:1809.008

\bibitem[{{Haardt} \& {Salvaterra}(2015)}]{Haardt2015}
{Haardt}, F., \& {Salvaterra}, R. 2015, \aap, 575, L16

\bibitem[{{Hao} {et~al.}(2005){Hao}, {Strauss}, {Fan}, {Tremonti}, {Schlegel},
  {Heckman}, {Kauffmann}, {Blanton}, {Gunn}, {Hall}, {Ivezi{\'c}}, {Knapp},
  {Krolik}, {Lupton}, {Richards}, {Schneider}, {Strateva}, {Zakamska},
  {Brinkmann}, \& {Szokoly}}]{Hao2005}
{Hao}, L., {Strauss}, M.~A., {Fan}, X., {et~al.} 2005, \aj, 129, 1795

\bibitem[{{Herenz} {et~al.}(2019){Herenz}, {Wisotzki}, {Saust}, {Kerutt},
  {Urrutia}, {Diener}, {Schmidt}, {Marino}, {de la Vieuville}, {Boogaard},
  {Schaye}, {Guiderdoni}, {Richard}, \& {Bacon}}]{Herenz2019}
{Herenz}, E.~C., {Wisotzki}, L., {Saust}, R., {et~al.} 2019, \aap, 621, A107

\bibitem[{{Hill} {et~al.}(2008){Hill}, {Gebhardt}, {Komatsu}, {Drory},
  {MacQueen}, {Adams}, {Blanc}, {Koehler}, {Rafal}, {Roth}, {Kelz}, {Gronwall},
  {Ciardullo}, \& {Schneider}}]{Hill08}
{Hill}, G.~J., {Gebhardt}, K., {Komatsu}, E., {et~al.} 2008, Astronomical
  Society of the Pacific Conference Series, Vol. 399, {The Hobby-Eberly
  Telescope Dark Energy Experiment (HETDEX): Description and Early Pilot Survey
  Results}, ed. T.~{Kodama}, T.~{Yamada}, \& K.~{Aoki}, 115

\bibitem[{{Hill} {et~al.}(2021){Hill}, {Lee}, {MacQueen}, {Kelz}, {Drory},
  {Vattiat}, {Good}, {Ramsey}, {Kriel}, {Peterson}, {DePoy}, {Gebhardt},
  {Marshall}, {Tuttle}, {Bauer}, {Chonis}, {Fabricius}, {Froning},
  {H{\"a}user}, {Indahl}, {Jahn}, {Landriau}, {Leck}, {Montesano}, {Prochaska},
  {Snigula}, {Zeimann}, {Bryant}, {Damm}, {Fowler}, {Janowiecki}, {Martin},
  {Mrozinski}, {Odewahn}, {Rostopchin}, {Shetrone}, {Spencer}, {Mentuch
  Cooper}, {Armandroff}, {Bender}, {Dalton}, {Hopp}, {Komatsu}, {Nicklas},
  {Ramsey}, {Roth}, {Schneider}, {Sneden}, \& {Steinmetz}}]{Hill2021}
{Hill}, G.~J., {Lee}, H., {MacQueen}, P.~J., {et~al.} 2021, \aj, 162, 298

\bibitem[{{Hopkins} {et~al.}(2008){Hopkins}, {Hernquist}, {Cox}, \&
  {Kere{\v{s}}}}]{Hopkins2008}
{Hopkins}, P.~F., {Hernquist}, L., {Cox}, T.~J., \& {Kere{\v{s}}}, D. 2008,
  \apjs, 175, 356

\bibitem[{{Hopkins} {et~al.}(2007){Hopkins}, {Richards}, \&
  {Hernquist}}]{Hopkins2007}
{Hopkins}, P.~F., {Richards}, G.~T., \& {Hernquist}, L. 2007, \apj, 654, 731

\bibitem[{{Hunt} {et~al.}(2004){Hunt}, {Steidel}, {Adelberger}, \&
  {Shapley}}]{Hunt2004}
{Hunt}, M.~P., {Steidel}, C.~C., {Adelberger}, K.~L., \& {Shapley}, A.~E. 2004,
  \apj, 605, 625

\bibitem[{{Johnston}(2011)}]{Johnston2011}
{Johnston}, R. 2011, \aapr, 19, 41

\bibitem[{{Kormendy} \& {Ho}(2013)}]{Kormendy2013}
{Kormendy}, J., \& {Ho}, L.~C. 2013, \araa, 51, 511

\bibitem[{{Kulkarni} {et~al.}(2019){Kulkarni}, {Worseck}, \&
  {Hennawi}}]{Kulkarni2019}
{Kulkarni}, G., {Worseck}, G., \& {Hennawi}, J.~F. 2019, \mnras, 488, 1035

\bibitem[{{Liu} {et~al.}(2022){Liu}, {Gebhardt}, {Mentuch Cooper}, {Davis},
  {Schneider}, {Ciardullo}, {Farrow}, {Finkelstein}, {Gronwall}, {Guo}, {Hill},
  {House}, {Jeong}, {Jogee}, {Kollatschny}, {Krumpe}, {Landriau}, {Chavez
  Ortiz}, \& {Zhang}}]{Liu2022}
{Liu}, C., {Gebhardt}, K., {Mentuch Cooper}, E., {et~al.} 2022, arXiv e-prints,
  arXiv:2204.13658

\bibitem[{{Palanque-Delabrouille} {et~al.}(2016){Palanque-Delabrouille},
  {Magneville}, {Y{\`e}che}, {P{\^a}ris}, {Petitjean}, {Burtin}, {Dawson},
  {McGreer}, {Myers}, {Rossi}, {Schlegel}, {Schneider}, {Streblyanska}, \&
  {Tinker}}]{PD2016}
{Palanque-Delabrouille}, N., {Magneville}, C., {Y{\`e}che}, C., {et~al.} 2016,
  \aap, 587, A41

\bibitem[{{P{\^a}ris} {et~al.}(2018){P{\^a}ris}, {Petitjean}, {Aubourg},
  {Myers}, {Streblyanska}, {Lyke}, {Anderson}, {Armengaud}, {Bautista},
  {Blanton}, {Blomqvist}, {Brinkmann}, {Brownstein}, {Brandt}, {Burtin},
  {Dawson}, {de la Torre}, {Georgakakis}, {Gil-Mar{\'\i}n}, {Green}, {Hall},
  {Kneib}, {LaMassa}, {Le Goff}, {MacLeod}, {Mariappan}, {McGreer}, {Merloni},
  {Noterdaeme}, {Palanque-Delabrouille}, {Percival}, {Ross}, {Rossi},
  {Schneider}, {Seo}, {Tojeiro}, {Weaver}, {Weijmans}, {Y{\`e}che}, {Zarrouk},
  \& {Zhao}}]{Paris2018}
{P{\^a}ris}, I., {Petitjean}, P., {Aubourg}, {\'E}., {et~al.} 2018, \aap, 613,
  A51

\bibitem[{{Puchwein} {et~al.}(2019){Puchwein}, {Haardt}, {Haehnelt}, \&
  {Madau}}]{Puchwein2019}
{Puchwein}, E., {Haardt}, F., {Haehnelt}, M.~G., \& {Madau}, P. 2019, \mnras,
  485, 47

\bibitem[{{Rakshit} {et~al.}(2020){Rakshit}, {Stalin}, \&
  {Kotilainen}}]{Rakshit2020}
{Rakshit}, S., {Stalin}, C.~S., \& {Kotilainen}, J. 2020, \apjs, 249, 17

\bibitem[{{Ramsey} {et~al.}(1998){Ramsey}, {Adams}, {Barnes}, {Booth},
  {Cornell}, {Fowler}, {Gaffney}, {Glaspey}, {Good}, {Hill}, {Kelton},
  {Krabbendam}, {Long}, {MacQueen}, {Ray}, {Ricklefs}, {Sage}, {Sebring},
  {Spiesman}, \& {Steiner}}]{Ramsey98}
{Ramsey}, L.~W., {Adams}, M.~T., {Barnes}, T.~G., {et~al.} 1998, in Society of
  Photo-Optical Instrumentation Engineers (SPIE) Conference Series, Vol. 3352,
  Advanced Technology Optical/IR Telescopes VI, ed. L.~M. {Stepp}, 34--42

\bibitem[{{Richards} {et~al.}(2006){Richards}, {Strauss}, {Fan}, {Hall},
  {Jester}, {Schneider}, {Vanden Berk}, {Stoughton}, {Anderson}, {Brunner},
  {Gray}, {Gunn}, {Ivezi{\'c}}, {Kirkland}, {Knapp}, {Loveday}, {Meiksin},
  {Pope}, {Szalay}, {Thakar}, {Yanny}, {York}, {Barentine}, {Brewington},
  {Brinkmann}, {Fukugita}, {Harvanek}, {Kent}, {Kleinman}, {Krzesi{\'n}ski},
  {Long}, {Lupton}, {Nash}, {Neilsen}, {Nitta}, {Schlegel}, \&
  {Snedden}}]{Richards2006}
{Richards}, G.~T., {Strauss}, M.~A., {Fan}, X., {et~al.} 2006, \aj, 131, 2766

\bibitem[{{Ross} {et~al.}(2013){Ross}, {McGreer}, {White}, {Richards}, {Myers},
  {Palanque-Delabrouille}, {Strauss}, {Anderson}, {Shen}, {Brandt},
  {Y{\`e}che}, {Swanson}, {Aubourg}, {Bailey}, {Bizyaev}, {Bovy}, {Brewington},
  {Brinkmann}, {DeGraf}, {Di Matteo}, {Ebelke}, {Fan}, {Ge}, {Malanushenko},
  {Malanushenko}, {Mandelbaum}, {Maraston}, {Muna}, {Oravetz}, {Pan},
  {P{\^a}ris}, {Petitjean}, {Schawinski}, {Schlegel}, {Schneider}, {Silverman},
  {Simmons}, {Snedden}, {Streblyanska}, {Suzuki}, {Weinberg}, \&
  {York}}]{Ross2013}
{Ross}, N.~P., {McGreer}, I.~D., {White}, M., {et~al.} 2013, \apj, 773, 14

\bibitem[{{Salviander} {et~al.}(2007){Salviander}, {Shields}, {Gebhardt}, \&
  {Bonning}}]{Salviander2007}
{Salviander}, S., {Shields}, G.~A., {Gebhardt}, K., \& {Bonning}, E.~W. 2007,
  \apj, 662, 131

\bibitem[{{Schmidt}(1968)}]{Schmidt1968}
{Schmidt}, M. 1968, \apj, 151, 393

\bibitem[{{Schneider} {et~al.}(2010){Schneider}, {Richards}, {Hall}, {Strauss},
  {Anderson}, {Boroson}, {Ross}, {Shen}, {Brandt}, {Fan}, {Inada}, {Jester},
  {Knapp}, {Krawczyk}, {Thakar}, {Vanden Berk}, {Voges}, {Yanny}, {York},
  {Bahcall}, {Bizyaev}, {Blanton}, {Brewington}, {Brinkmann}, {Eisenstein},
  {Frieman}, {Fukugita}, {Gray}, {Gunn}, {Hibon}, {Ivezi{\'c}}, {Kent}, {Kron},
  {Lee}, {Lupton}, {Malanushenko}, {Malanushenko}, {Oravetz}, {Pan}, {Pier},
  {Price}, {Saxe}, {Schlegel}, {Simmons}, {Snedden}, {SubbaRao}, {Szalay}, \&
  {Weinberg}}]{Schneider2010}
{Schneider}, D.~P., {Richards}, G.~T., {Hall}, P.~B., {et~al.} 2010, \aj, 139,
  2360

\bibitem[{{Shankar} {et~al.}(2009){Shankar}, {Weinberg}, \&
  {Miralda-Escud{\'e}}}]{Shankar2009}
{Shankar}, F., {Weinberg}, D.~H., \& {Miralda-Escud{\'e}}, J. 2009, \apj, 690,
  20

\bibitem[{{Shen} {et~al.}(2020){Shen}, {Hopkins}, {Faucher-Gigu{\`e}re},
  {Alexander}, {Richards}, {Ross}, \& {Hickox}}]{Shen2020}
{Shen}, X., {Hopkins}, P.~F., {Faucher-Gigu{\`e}re}, C.-A., {et~al.} 2020,
  \mnras, 495, 3252

\bibitem[{{Spinoso} {et~al.}(2020){Spinoso}, {Orsi}, {L{\'o}pez-Sanjuan},
  {Bonoli}, {Viironen}, {Izquierdo-Villalba}, {Sobral}, {Gurung-L{\'o}pez},
  {Hern{\'a}n-Caballero}, {Ederoclite}, {Varela}, {Overzier},
  {Miralda-Escud{\'e}}, {Muniesa}, {V{\'\i}lchez}, {Alcaniz}, {Angulo},
  {Cenarro}, {Crist{\'o}bal-Hornillos}, {Dupke}, {Hern{\'a}ndez-Monteagudo},
  {Mar{\'\i}n-Franch}, {Moles}, {Sodr{\'e}}, \&
  {V{\'a}zquez-Rami{\'o}}}]{Spinoso2020}
{Spinoso}, D., {Orsi}, A., {L{\'o}pez-Sanjuan}, C., {et~al.} 2020, \aap, 643,
  A149

\bibitem[{{Steidel} {et~al.}(2003){Steidel}, {Adelberger}, {Shapley},
  {Pettini}, {Dickinson}, \& {Giavalisco}}]{Steidel2003}
{Steidel}, C.~C., {Adelberger}, K.~L., {Shapley}, A.~E., {et~al.} 2003, \apj,
  592, 728

\bibitem[{{Thorne} {et~al.}(2022){Thorne}, {Robotham}, {Davies}, {Bellstedt},
  {Brown}, {Croom}, {Delvecchio}, {Groves}, {Jarvis}, {Shabala}, {Seymour},
  {Whittam}, {Bravo}, {Cook}, {Driver}, {Holwerda}, {Phillipps}, \&
  {Siudek}}]{Thorne2022}
{Thorne}, J.~E., {Robotham}, A. S.~G., {Davies}, L. J.~M., {et~al.} 2022,
  \mnras, 509, 4940

\bibitem[{{Tremaine} {et~al.}(2002){Tremaine}, {Gebhardt}, {Bender}, {Bower},
  {Dressler}, {Faber}, {Filippenko}, {Green}, {Grillmair}, {Ho}, {Kormendy},
  {Lauer}, {Magorrian}, {Pinkney}, \& {Richstone}}]{Tremaine2002}
{Tremaine}, S., {Gebhardt}, K., {Bender}, R., {et~al.} 2002, \apj, 574, 740

\bibitem[{{Tsuzuki} {et~al.}(2006){Tsuzuki}, {Kawara}, {Yoshii}, {Oyabu},
  {Tanab{\'e}}, \& {Matsuoka}}]{Tsuzuki06}
{Tsuzuki}, Y., {Kawara}, K., {Yoshii}, Y., {et~al.} 2006, \apj, 650, 57

\bibitem[{{Vestergaard} \& {Wilkes}(2001)}]{Vestergaard2001}
{Vestergaard}, M., \& {Wilkes}, B.~J. 2001, \apjs, 134, 1

\bibitem[{{Zhang} {et~al.}(2021){Zhang}, {Ouchi}, {Gebhardt}, {Mentuch Cooper},
  {Liu}, {Davis}, {Jeong}, {Farrow}, {Finkelstein}, {Gawiser}, {Hill},
  {Harikane}, {Kakuma}, {Acquaviva}, {Casey}, {Fabricius}, {Hopp}, {Jarvis},
  {Landriau}, {Mawatari}, {Mukae}, {Ono}, {Sakai}, \& {Schneider}}]{Zhang2021}
{Zhang}, Y., {Ouchi}, M., {Gebhardt}, K., {et~al.} 2021, \apj, 922, 167

\end{thebibliography}

\end{document}